\documentclass[aps,prb,twocolumn]{revtex4-2}
\usepackage{amssymb}
\usepackage{graphicx,amsmath}
\usepackage{hyperref}
\allowdisplaybreaks
\setcitestyle{numbers,square}

\newcommand{\bea}{\begin{eqnarray}}
\newcommand{\eea}{\end{eqnarray}}
\newcommand{\beq}{\begin{equation}}
\newcommand{\eeq}{\end{equation}}
\newcommand{\la}{\langle}
\newcommand{\ra}{\rangle}
\newcommand{\al}{\alpha}
\newcommand{\be}{\beta}
\newcommand{\ga}{\gamma}
\newcommand{\om}{\omega}
\newcommand{\Om}{\Omega}
\newcommand{\ep}{\epsilon}

\newcommand{\si}{\sigma}
\newcommand{\dl}{\delta}

\newcommand{\ord}{\mathcal{O}}

\newcommand{\vol}{\mathcal{V}}

\newcommand{\ptl}{\partial}
\newcommand{\sgn}{{\rm Sgn}}
\newcommand{\cda}{c^{\dagger}}
\newcommand{\bk}{{\bf k}}
\newcommand{\bq}{{\bf q}}
\newcommand{\bp}{{\bf p}}
\newcommand{\br}{{\bf r}}

\newcommand{\bE}{{\bf E}}
\newcommand{\bA}{{\bf A}}

\newcommand{\bj}{{\bf j}}
\newcommand{\hham}{\hat{\mathcal{H}}}
\newcommand{\hV}{\hat{V}}
\newcommand{\hO}{\hat{O}}
\newcommand{\hU}{\hat{U}}
\newcommand{\hv}{\hat{v}}
\newcommand{\hj}{\hat{j}}
\newcommand{\hro}{\hat{\rho}}
\newcommand{\tp}{t^{\prime}}
\newcommand{\omp}{\omega^{+}}

\begin{document}

\title{Nonlinear terahertz electro-optical responses in centrosymmetric electronic systems}
\date{\today}
\author{I. Paul}
\affiliation{
Laboratoire Mat\'{e}riaux et Ph\'{e}nom\`{e}nes Quantiques, Universit\'{e} Paris Cit\'{e}, CNRS,
75205 Paris, France
}

\begin{abstract}
Motivated by the recent developments in terahertz spectroscopy using pump-probe setups
to study correlated electronic materials, we review the field theoretical formalism to compute
finite frequency nonlinear electro-optical responses in centrosymmetric systems starting from
basic time dependent perturbation theory.
We express the nonlinear current kernel as a sum of several causal response functions.
These causal functions cannot be evaluated using perturbative field theory methods, since they are
not contour ordered. Consequently, we associate
each response function with a corresponding imaginary time ordered current correlation function,
since the latter can be factorized using Wick's theorem. The mapping between the response functions and the correlation functions,
suitably analytically continued to real frequencies, is proven
exactly. We derive constraints satisfied by the nonlinear current kernel and we prove
a generalized $f$-sum rule for the nonlinear conductivity, all of which are consequences of particle
number conservation. The constraints guarantee that the nonlinear static responses are free from
spurious divergences. We apply the theory to compute the gauge invariant nonlinear conductivity
of a system of noninteracting electrons in the presence of weak disorder.
As special cases of this generalized nonlinear response, we discuss its third harmonic and its instantaneous terahertz Kerr signals. The formalism can be
used to compute the nonlinear conductivity in symmetry broken phases of electronic systems
such as superconductors, density waves and nematic states.
\end{abstract}

%\pacs{
%}

\maketitle
%%%%%%%%%%%%%%%%%%%%%%%%%%%%%%%%%%%%%%%%%%%%%%%%%%%%%%%%%%%%%%%%%%%%%

\section{Introduction}
Pump-probe spectroscopy has emerged as an important experimental method to probe and manipulate
correlated electronic
matter~\cite{reviews,okamoto07,fausti11,liu12,Zong,mitrano16, Kogar,Buzzi,Thewalt18}.
In this technique the system is first subjected to an intense laser pump, and then
the reaction of the system is probed with a weaker laser pulse at a later time.
Traditionally, this technique has been used mostly with pumps in optical frequency
range, and with pulse durations that are shorter
than the relaxation time scale of the system. Such setups probe the
nonequilibrium dynamics of the system.
More recently, with the development of terahertz lasers
it has become possible to excite systems at milli-electron volt scale, which is an energy range of
great importance for correlated electron systems with interesting low temperature quantum
phases~\cite{Hebling08,Kampfrath13,Junginger12,Matsunaga12,Matsunaga13,Wu17,Wu18,Nakamura20}.
Simultaneously, it has become possible to generate
pump pulses that are long compared to the system's relaxation time.
In this case the
system stays in equilibrium in the presence of the pump, and one can probe the finite frequency
nonlinear electro-optical response of the system~\cite{Shen,Boyd}.
The purpose of this paper is to review the theoretical framework for such nonlinear finite frequency
electro-optical responses
in centrosymmetric electronic systems starting from basic time dependent Hamiltonian formalism.

In the context of terahertz spectroscopy there are two types of nonlinear responses that are
currently being discussed the most.
(i) Third harmonic generation, which is a measurement in frequency domain where the system is
excited with a pump electric field with frequency $\om$, and a response at $3\om$ is
detected~\cite{Matsunaga14,Matsunaga17,Kaiser20,ZheWang20}.
(ii) The terahertz Kerr effect where the optical property (such as the optical
conductivity or equivalently, the refractive index)
of the system is transiently modified in the presence of the pump, the change being
proportional to second order in the pump electric
field~\cite{Hoffmann09,Freysz10,Yada13,Cornet14,Sajadi17,Katsumi18,Katsumi20}. Of special
interest is the
instantaneous Kerr effect, where the probe measures a time~($t$) dependent
response that is proportional to the square of the pump electric field
$\bE_{pp}(t)^2$.

In the last few years, in parallel with the experimental
developments~\cite{Matsunaga14,Matsunaga17,Kaiser20,ZheWang20,Hoffmann09,Katsumi18,
Katsumi20,Grasset18,Grasset19},
a lot of theoretical effort has been
put to study and interpret Kerr and third harmonic responses of
superconductors~\cite{Tsuji15,Cea16,Murotani17,Cea18,Jujo18,Murotani19,Silaev19,Udina19,Schwarz20,
Tsuji20,Shimano20,Seibold20,Schwarz20b}.
The motivation has been to study exotic collective modes of electrons that exist only
in broken symmetry phases, such as the superconducting gap amplitude mode or the so-called Higgs
mode~\cite{Tsuji15,Cea16,Murotani17,Cea18,Jujo18,Murotani19,Silaev19,Udina19,Schwarz20,
Tsuji20,Shimano20,Seibold20,Schwarz20b,Mueller19,Kumar19,Gabrielle20,Sun20,Yang20,Golez20}.
Recent theory works have also studied nonlinear electro-optical responses of topological
metals, see,
e.g.~\cite{Mikhailov08,Moore10,Morimoto16,Parker19,Cheng20,Ahn20,Villegas20,Rostami20}.
From a technical point of view, the topic has been widely
studied in the past using semiclassical equation of motion approach within band theory, see, e.g., Refs~\cite{SipeI,SipeII,SipeIII,SipeIV,Juan20}. The equation of motion formalism, however, is
not particularly suited to treat electron-electron interaction in a systematic way.

The aim of the current work is to discuss a field theoretical formalism.
The advantage of a field theoretic treatment
is that, in principle, it allows a systematic way to include electron-electron interaction effects.
In this review we pay particular attention to the following two aspects that
have been mostly glossed over in the recent literature.

Firstly, there is a basic dichotomy between what is experimentally measured, and what can be
computed using perturbative field theoretic techniques. As we show below,
the measured nonlinear current involves the sum of several
response functions that obey causality. However, the causal functions are not contour ordered
and, as such, they cannot be factorized using Wick's theorem. The latter is crucial in
order to include electron-electron interaction in a perturbative fashion. Instead, what
can be calculated using field theory are contour ordered correlators that can be factorized
using Wick's theorem. Thus, one goal of this work is to establish an exact mapping between the
causal response functions and the correlators.

The contour can be ordered either
in real time using Keldysh's two-time formalism~\cite{Kamenev},
or it can be ordered in imaginary time using Matsubara technique~\cite{Mahan}.
The advantage of the former is that, at the end of the book keeping, one can avoid
an additional step which is necessary in the Matsubara method, namely having to make
analytic continuations from imaginary to real axes. The advantage
of the latter is that, in the intermediate steps, the typical expressions in the Matsubara formalism
are more compact.

Besides the technical intricacies, the careful extraction of the response functions from the correlation
functions is important to keep track of finite temperature effects
in the nonlinear signals.
The above dichotomy between causal response functions and contour ordered correlation functions
is already present at a linear response level. The only difference here is that the causality structure, or
equivalently, the analytic continuations are more complex in the case of a nonlinear response.
Importantly, we show that the basic intuition concerning finite temperature effects remain the same
here as in linear response. Namely, thermal factors are not
important if the dominant relaxation process is elastic scattering, just as in linear Drude conductivity.
While, inelastic scattering (not discussed in this work) leads to nontrivial temperature dependencies.

Secondly, the importance of obtaining results that are consistent with particle number conservation,
that can be expressed in terms of a global $U(1)$ symmetry. This
conservation leads to the generalization of the $f$-sum rule,
and it also ensures that the response is zero for constant time independent vector potentials.
Physically, such a vector potential
implies zero electric field in the bulk, and consequently such a potential does not affect the system,
provided the electromagnetic response at the boundary is unremarkable.
In terms of the $U(1)$ symmetry, a constant vector potential can be absorbed, and
therefore ``gauged out'', in a global redefinition of the phases of the single particle
wavefunctions, provided the system is in a non-superconducting phase.
From a diagrammatic point
of view, this implies that keeping only an arbitrary subset of diagrams in the calculation of $\bj_{nl}$,
the nonlinear current, will often lead to unphysical answers.

More concretely, a second goal of this work is to examine how
particle number conservation or gauge invariance
imposes constraints on the nonlinear responses.
One such constraint involves the vanishing of the nonlinear current response
for non-superconducting phases if the external vector potential is time independent.
This property guarantees the absence of spurious divergences and that the static nonlinear
responses remain finite.
A second constraint is a generalization of the familiar $f$-sum rule
that is invoked in the context of linear conductivity.

Traditionally, in nonlinear optics the quantity of central interest is the electric
polarization ${\bf P}$~\cite{Boyd}. However, in the context of
metals we find it more convenient, and physically more intuitive, to develop the theory
in terms of the electrical current $\bj$ and the associated nonlinear
conductivity $\si^{(3)}$.  This choice is not fundamental,
and is more a matter of taste, since the two are related by
$\bj(t) = \bj_{DC} + \partial {\bf P}(t)/\partial t$~\cite{Shen}. Thus,
one can extract time dependent polarization ${\bf P}(t)$ from $\bj(t)$ and vice versa.

The rest of the paper is organized as follows.
In section~\ref{sec2} we derive formal expressions for the
nonlinear current $\left(j_{NL}\right)_{\al}(\om)$ in terms of the nonlinear current
kernel $\Pi_{\al \be \ga \dl}^{(3)}(\om_1, \om_2, \om_3)$, see Eq.~(\ref{eq:2.20}),
or equivalently in terms of the nonlinear conductivity $\si_{\al \be \ga \dl}^{(3)}(\om_1, \om_2, \om_3)$,
see Eqs.~(\ref{eq:2.21-bis}) and (\ref{eq:2.21-bis2}).
The nonlinear kernel and the conductivity are rank-four tensors, and the
indices $(\al, \be, \ga, \dl)$ denote photon polarizations. The arguments
$(\om_1, \om_2, \om_3)$ denote incoming photon
frequencies, with polarizations $(\be, \ga, \dl)$, respectively. The outgoing photon has polarization
$\al$ and carries frequency $\om = \om_1 + \om_2 + \om_3$.
The nonlinear kernel itself is expressed as a sum of several current correlation functions, suitably analytically
continued from imaginary to real frequencies, see Eq.~(\ref{eq:2.21}). The mapping between the
causal response functions and the imaginary time ordered correlation functions is proven using
the Lehmann representation.
This mapping is exact to all orders in the interaction strength, and it also 
holds if the electrons are in a random potential due to the presence of impurities.
In section~\ref{sec3} we prove
the following two properties of the kernel stemming from particle number
conservation. First,  in non superconducting phases the nonlinear kernel vanishes 
if any one of the three incoming
photon frequencies is set to zero, see Eq.~(\ref{eq:3.1}).
This ensures that there is no nonlinear diamagnetic response in non-superconducting phases, and 
that the static nonlinear responses are free from spurious divergences. Second, a generalization
of the $f$-sum rule which shows that the nonlinear conductivity integrated over the three incoming
frequencies is a constant that depends only the electronic spectrum, and is independent of the electron
lifetime, see Eq.~(\ref{eq:3.15}). This sum rule has been noted earlier~\cite{Watanabe20}.
We use the first property to express the nonlinear conductivity in a manifestly gauge invariant form.
It is this gauge invariant response that is studied in the remaining sections.
In section~\ref{sec4} we use diagrammatic method to calculate the gauge invariant
$\Pi_{\al \be \ga \dl}^{(3)}(\om_1, \om_2, \om_3)$ of a Drude metal,
namely a system of non-interacting electrons in the presence of weak disorder. The third harmonic
and the terahertz instantaneous Kerr
signals are special cases of the generalized nonlinear response, and these quantities for a Drude system
are computed in sections~\ref{sec5} and~\ref{sec6}, respectively. In the concluding section~\ref{sec7}
we give a summary of the main results, and we mention the various settings where the field theoretic 
formalism is useful.

\section{Derivation of the nonlinear electro-optical response}
\label{sec2}

We consider an electronic system in a crystalline environment described by the Hamiltonian
\beq
\label{eq:2.1}
\hham_{\rm tot} = \hham +\hV(t),
\eeq
where $\hham$ is a time independent Hamiltonian that describes the system in the absence of external
time-dependent perturbations. Depending on the context, $\hham$ can include electron-electron
interaction and scattering of electrons due to disorder.
To simplify the discussion we assume that only one band is relevant.
The multiband generalization of the formalism is straightforward.
Thus, the part of $\hham$ that describes the band dispersion
is given by $\hham_0 = \sum_{\bk} \ep_{\bk} \cda_{\bk} c_{\bk}$, where
$(\cda_{\bk}, c_{\bk})$ are creation and annihilation operators of electrons with wavevector $\bk$,
and $\ep_{\bk}$ is the band dispersion. We take the electrons to be spinless, since it does not play any
role in the following.

$\hV(t)$ is a time-dependent potential that the electrons experience
due to the electric field $\bE(t)$ of the pump and the probe lasers. Since the typical
photon wavelength is much longer than the Fermi wavelength $1/k_F$, where $k_F$ is the Fermi wavevector,
the electric field can be taken as spatially uniform.
We describe the light-matter coupling by Peierls substitution, such that
$\ep_{\bk} \rightarrow \ep_{\bk - e \bA}$ in the presence of the electromagnetic field.
Here $e$ is the electron charge
and $\bA(t)$ is the vector potential to which the electric field is related by
$\bE(t) = -\ptl_t \bA(t)$. Expanding $\ep_{\bk - e \bA}$
in powers of the vector potential we get
\begin{align}
\label{eq:2.2}
\hV(t) =& \left[ -e \hv_{\al}(t) + \frac{e^2}{2!}\hv_{\al \be}(t) A_{\be}(t)
- \frac{e^3}{3!}\hv_{\al \be \ga}(t) A_{\be}(t)
\right. \nonumber \\
 &\left.  \times  A_{\ga}(t)
 + \frac{e^4}{4!}\hv_{\al \be \ga \dl}(t) A_{\be}(t) A_{\ga}(t) A_{\dl}(t) \right] A_{\al}(t),
\end{align}
where
\begin{subequations}
\label{eq:2.3}
\begin{align}
\hv_{\al} &= \sum_{\bk} \frac{\ptl \ep_{\bk}}{\ptl k_{\al}} \cda_{\bk} c_{\bk},
\label{eq:2.3a}\\
\hv_{\al \be} &= \sum_{\bk} \frac{\ptl^2 \ep_{\bk}}{\ptl k_{\al} \ptl k_{\be}} \cda_{\bk} c_{\bk},
\label{eq:2.3b}\\
\hv_{\al \be \ga} &= \sum_{\bk} \frac{\ptl^3 \ep_{\bk}}
{\ptl k_{\al} \ptl k_{\be} \ptl k_{\ga}} \cda_{\bk} c_{\bk},
\label{eq:2.3c}\\
\hv_{\al \be \ga \dl} &= \sum_{\bk} \frac{\ptl^4 \ep_{\bk}}
{\ptl k_{\al} \ptl k_{\be} \ptl k_{\ga} \ptl k_{\dl}} \cda_{\bk} c_{\bk},
\label{eq:2.3d}
\end{align}
\end{subequations}
and $(\al, \be, \ga, \dl)$ denote spatial indices $(x, y, z)$. In Eq.~(\ref{eq:2.2}) and in the rest of the paper
summation over repeated indices is implied, unless the contrary is explicitly mentioned.
The associated charge current operator
$\hj_{\al} \equiv - \dl \hham[\bA]/\dl A_{\al}$, is given by
\begin{align}
\label{eq:2.4}
\hj_{\al}(t) =& e \hv_{\al}(t) - e^2 \hv_{\al \be}(t) A_{\be}(t) +
\frac{e^3}{2}\hv_{\al \be \ga}(t) A_{\be}(t) A_{\ga}(t)
\nonumber \\
-&  \frac{e^4}{6}\hv_{\al \be \ga \dl}(t) A_{\be}(t) A_{\ga}(t) A_{\dl}(t).
\end{align}

The next step is to calculate the average current of the system which is defined as
\beq
\label{eq:2.5}
j_{\al}(t) \equiv \frac{1}{Z} \sum_n e^{-\be E_n} \la n(t) | \hj_{\al}(t) | n(t) \ra,
\eeq
where $|n\ra$ are the eigenstates of $\hham$ with $\hham | n \ra = E_n | n \ra$. Thus,
we assume that the perturbation due to the electromagnetic field does not take the system out of
equilibrium, and that the system remains in thermal equilibrium with temperature $T$. Therefore, a
measurable quantity is simply the thermal average of the associated operator. Following the usual
rules of equilibrium statistical mechanics, the eigenstate $|n\ra$ has Boltzmann weight $e^{-\be E_n}$,
the partition function $Z \equiv \sum_n e^{-\be E_n}$, and $\be \equiv 1/(k_B T)$
with $k_B$ the Boltzmann
constant. In other words, in the following the role of the time-dependent perturbation $\hV(t)$ is simply
to modify the time evolution of the states and/or the operators depending on the picture (Schroedinger,
Heisenberg or interaction).

In practice, the pump-probe experiments typically measure not just an equilibrium nonlinear response,
but also a nonequilibrium response where the system relaxes back to equilibrium after having put
out-of-equilibrium by the pump. Thus, how the nonlinear signal gets modified due to the simultaneous
presence of an inequilibrium component is a question that will be both interesting and relevant to address
in the future. In the current treatment we simply assume that the out of equilibrium component is absent.

In the following we use the operator formalism to compute the current $j_{\al}(t)$, while the same can be
done using the effective action principle, see, e.g.~\cite{Udina19,benfatto04}.
We adopt the interaction picture in which the time evolution of an operator $\hO(t)$ is given by
$
\hO(t) = e^{i \hham t} \hO(0) e^{-i \hham t},
$
and that of a state by
$
| n(t) \ra = \hU (t, t_0) | n(t_0) \ra,
$
where the time evolution operator is
\beq
\label{eq:2.6}
 \hU (t, t_0) = \hat{T}_+ \exp [-i \int_{t_0}^t d \tp \hV(\tp)],
 \eeq
 and $\hat{T}_+$ is the time ordering operator. The reference time $t_0$ is an instant before the introduction
 of the perturbation $\hV(t)$. It will be convenient later to set $t_0 \rightarrow -\infty$.

We assume the system to be centrosymmetric for which the lowest order nonlinear current is cubic in the
vector potential. Consequently, the operators $\hj_{\al}(t)$ and $\hU(t,t_0)$
need to be expanded to third order  in the vector potential.
For convenience we define the quantity $\hat{J}_{\al}^{(3)}(t) = [\hU^{\dagger}(t, t_0)
\hj_{\al}(t) \hU(t, t_0)]_{\ord(A^3)}$, and after collecting terms we get

\begin{widetext}
\begin{align}
\label{eq:2.7}
&\hat{J}_{\al}^{(3)}(t) = -e^4 \left[
\frac{1}{6} \hv_{\al \be \ga \dl}(t) A_{\be}(t) A_{\ga}(t) A_{\dl}(t)
- \frac{i}{2} \int_{t_0}^t dt_1 \left[ \hv_{\al \be}(t), \hv_{\ga \dl}(t_1) \right]
A_{\be}(t) A_{\ga}(t_1) A_{\dl}(t_1)
- \frac{i}{2} \int_{t_0}^t dt_1 \left[ \hv_{\al \be \ga}(t), \hv_{\dl}(t_1) \right]
\right.
\nonumber \\
&\times A_{\be}(t) A_{\ga}(t) A_{\dl}(t_1)
- \frac{i}{6} \int_{t_0}^t dt_1 \left[ \hv_{\al}(t), \hv_{\be \ga \dl}(t_1) \right]
A_{\be}(t_1) A_{\ga}(t_1) A_{\dl}(t_1)
+ \left\{ \frac{1}{2} \int_{t_0}^t dt_1 \int_{t_0}^t dt_2 \left(
\hv_{\ga}(t_1) \hv_{\al \be}(t) \hv_{\dl}(t_2) + {\rm h.c.} \right) \right.
\nonumber \\
&-  \left. \int_{t_0}^t dt_1 \int_{t_0}^{t_1} dt_2 \left( \hv_{\al \be}(t) \hv_{\ga}(t_1) \hv_{\dl}(t_2)
+ {\rm h.c.} \right) \right\}
A_{\be}(t) A_{\ga}(t_1) A_{\dl}(t_2)
+ \left\{ \frac{1}{2} \int_{t_0}^t dt_1 \int_{t_0}^t dt_2 \left(
\hv_{\be}(t_1) \hv_{\al}(t) \hv_{\ga \dl}(t_2) + {\rm h.c.} \right) \right.
\nonumber\\
&-  \left. \frac{1}{2}  \int_{t_0}^t dt_1 \int_{t_0}^{t_1} dt_2
\left( \hv_{\al}(t) \hv_{\be}(t_1) \hv_{\ga \dl}(t_2) + {\rm h.c.} \right)
-  \frac{1}{2}  \int_{t_0}^t dt_2 \int_{t_0}^{t_2} dt_1
\left( \hv_{\al}(t) \hv_{\ga \dl}(t_2) \hv_{\be}(t_1) + {\rm h.c.} \right)
\right\}  A_{\be}(t_1) A_{\ga}(t_2) A_{\dl}(t_2)
\nonumber \\
&+ \left\{  i \int_{t_0}^t dt_1 \int_{t_0}^{t_1} dt_2 \int_{t_0}^{t_2} dt_3
\left(\hv_{\al}(t) \hv_{\be}(t_1) \hv_{\ga}(t_2) \hv_{\dl}(t_3) - {\rm h.c.} \right)
- i  \int_{t_0}^t dt_1 \int_{t_0}^{t} dt_2  \int_{t_0}^{t_2} dt_3
 \left(\hv_{\be}(t_1) \hv_{\al}(t) \hv_{\ga}(t_2) \hv_{\dl}(t_3) - {\rm h.c.} \right)
\right\} \nonumber \\
&\left. \times A_{\be}(t_1) A_{\ga}(t_2) A_{\dl}(t_3) \right].
\end{align}
\end{widetext}

In the above there are seven different types of terms which can be distinguished from the different ways
in which the time arguments of the three factors of the vector potential appear. Therefore, using
Eqs.~(\ref{eq:2.5}) and (\ref{eq:2.7}) the measured nonlinear current, proportional to $A^3$,
can be expressed as a sum of seven terms as
\begin{align}
\label{eq:2.8}
\left(j_{NL}\right)_{\al}(t) &= j_{\al}(t)^{1p} + j_{\al}(t)^{2p,a} + j_{\al}(t)^{2p,b} + j_{\al}(t)^{2p,c}
\nonumber \\
&+ j_{\al}(t)^{3p,a} + j_{\al}(t)^{3p,b} + j_{\al}(t)^{4p},
\end{align}
where, after taking $t_0 \rightarrow -\infty$ for convenience,
\begin{subequations}
\label{eq:2.9}
\begin{align}
 j_{\al}(t)^{1p} &= - e^4 R_{\al \be \ga \dl}^{(1p)}(t) A_{\be}(t) A_{\ga}(t) A_{\dl}(t),
\label{eq:2.9a} \\
j_{\al}(t)^{2p,a} &= -\frac{e^4}{2} \int_{-\infty}^{\infty} dt_1 R_{\al \be, \ga \dl}^{(2p,a)}(t, t_1)
A_{\be}(t) A_{\ga}(t_1) A_{\dl}(t_1),  \label{eq:2.9b} \\
j_{\al}(t)^{2p,b} &= -\frac{e^4}{2} \int_{-\infty}^{\infty} dt_1 R_{\al \be \ga, \dl}^{(2p,b)}(t, t_1)
A_{\be}(t) A_{\ga}(t) A_{\dl}(t_1),  \label{eq:2.9c} \\
j_{\al}(t)^{2p,c} &= -\frac{e^4}{6} \int_{-\infty}^{\infty} dt_1 R_{\al, \be \ga \dl}^{(2p,c)}(t, t_1)
A_{\be}(t_1) A_{\ga}(t_1) A_{\dl}(t_1),  \label{eq:2.9d} \\
j_{\al}(t)^{3p,a} &= - e^4 \int_{-\infty}^{\infty} dt_1  \int_{-\infty}^{\infty} dt_2
R_{\al \be, \ga, \dl}^{(3p,a)}(t, t_1, t_2)
\nonumber \\
&\times A_{\be}(t) A_{\ga}(t_1) A_{\dl}(t_2),  \label{eq:2.9e} \\
j_{\al}(t)^{3p,b} &= - e^4 \int_{-\infty}^{\infty} dt_1  \int_{-\infty}^{\infty} dt_2
R_{\al, \be, \ga \dl}^{(3p,b)}(t, t_1, t_2)
\nonumber \\
&\times
A_{\be}(t_1) A_{\ga}(t_2) A_{\dl}(t_2),  \label{eq:2.9f} \\
j_{\al}(t)^{4p} &= - e^4 \int_{-\infty}^{\infty} dt_1  \int_{-\infty}^{\infty} dt_2
\int_{-\infty}^{\infty} dt_3
\nonumber \\
&\times R_{\al, \be, \ga, \dl}^{(4p)}(t, t_1, t_2, t_3)
A_{\be}(t_1) A_{\ga}(t_2) A_{\dl}(t_3),  \label{eq:2.9g}
\end{align}
\end{subequations}
and the response functions are defined by
\begin{subequations}
\label{eq:2.10}
\begin{align}
&R_{\al \be \ga \dl}^{(1p)}(t) \equiv \la \hv_{\al \be \ga \dl}(t) \ra,
\label{eq:2.10a}\\
&R_{\al \be, \ga \dl}^{(2p,a)}(t, t_1) \equiv -i \theta(t - t_1)
\la \left[ \hv_{\al \be}(t) , \hv_{\ga \dl}(t_1) \right] \ra, \label{eq:2.10b}\\
&R_{\al \be \ga, \dl}^{(2p,b)}(t, t_1) \equiv -i \theta(t - t_1)
\la \left[ \hv_{\al \be \ga}(t) , \hv_{\dl}(t_1) \right] \ra, \label{eq:2.10c}\\
&R_{\al, \be \ga \dl}^{(2p,c)}(t, t_1) \equiv -i \theta(t - t_1)
\la \left[ \hv_{\al}(t) , \hv_{\be \ga \dl}(t_1) \right] \ra, \label{eq:2.10d}\\
&R_{\al \be, \ga, \dl}^{(3p,a)}(t, t_1, t_2) \equiv \theta(t - t_1)  \theta(t - t_2)
\la \hv_{\ga}(t_1) \hv_{\al \be}(t)   \nonumber \\
&\times \hv_{\dl}(t_2)/2 + {\rm h.c.} \ra
- \theta(t - t_1)  \theta(t_1 - t_2) \la \hv_{\al \be}(t) \hv_{\ga}(t_1) \nonumber\\
&\times \hv_{\dl}(t_2)  + {\rm h.c.} \ra,
\label{eq:2.10e}\\
&R_{\al, \be, \ga \dl}^{(3p,b)}(t, t_1, t_2) \equiv \theta(t - t_1)  \theta(t - t_2)
\la \hv_{\be}(t_1) \hv_{\al}(t)    \nonumber \\
&\times \hv_{\ga \dl}(t_2)/2 + {\rm h.c.} \ra - \theta(t - t_1)  \theta(t_1 - t_2) \la \hv_{\al}(t)
\hv_{\be}(t_1)  \nonumber \\
&\times  \hv_{\ga \dl}(t_2)/2  + {\rm h.c.} \ra - \theta(t - t_2)  \theta(t_2 - t_1)
\la \hv_{\al}(t)  \hv_{\ga \dl}(t_2)  \nonumber\\
&\times \hv_{\be}(t_1)/2  + {\rm h.c.} \ra,
\label{eq:2.10f}\\
&R_{\al, \be, \ga, \dl}^{(4p)}(t, t_1, t_2, t_3) \equiv i \theta(t - t_1)  \theta(t_1 - t_2) \theta(t_2 - t_3)
\nonumber \\
&\times \la \hv_{\al}(t) \hv_{\be}(t_1) \hv_{\ga}(t_2) \hv_{\dl}(t_3)  - {\rm h.c.} \ra
- i \theta(t - t_1)  \theta(t - t_2) \nonumber \\
&\times   \theta(t_2 -t_3) \la \hv_{\be}(t_1)  \hv_{\al}(t)
\hv_{\ga}(t_2) \hv_{\dl}(t_3)   + {\rm h.c.} \ra.
\label{eq:2.10g}
\end{align}
\end{subequations}
Here the average $\la \hO \ra$ of an operator $\hO$ is defined as
\[
\la \hO \ra \equiv (1/Z) \sum_n \exp(-\be E_n) \la n| \hO | n \ra,
\]
with $| n \ra = | n(t_0 \rightarrow - \infty) \ra$.
In the above the indices $(1p, 2p, 3p, 4p)$ imply that the corresponding response functions are related to
1-point, 2-point, 3-point and 4-point contour ordered current-current correlators, respectively. This link
between the response functions and the correlators will be demonstrated below. For the moment it is
obvious from the definitions of each of the response functions in Eq.~(\ref{eq:2.10})
that a $n$-point response function involves $n$ number of current operators.
Thus, there are three types of 2-point response functions that are distinguished by the labels
$(a, b, c)$, and there are
two types of 3-point response functions that are denoted by labels $(a, b)$. Also, since the trace involves the
energy eigenstates of the time translation invariant Hamiltonian $\hham$, it is clear that
the 1-point response function is a $t$-independent constant, the 2-point responses
are functions of the single variable $(t-t_1)$, the 3-point responses are functions of the two variables
$(t-t_1)$ and $(t-t_2)$, and the 4-point response is a function of the three variables $(t-t_1)$, $(t-t_2)$
and $(t-t_3)$.
Finally, from the presence of the
$\theta$-functions in Eq.~(\ref{eq:2.10}), it is clear that the response functions are causal.

The next step is to express the nonlinear response in the frequency domain. Accordingly, we
define the Fourier transform of the nonlinear current as
\beq
\label{eq:2.11}
\left(j_{NL}\right)_{\al}(\om) \equiv \int_{-\infty}^{\infty} dt e^{i \om t} \left(j_{NL}\right)_{\al}(t),
\eeq
and likewise the Fourier transforms of the seven components $ j_{\al}(\om)^{1p}$, $ j_{\al}(\om)^{2p,a}$,
$\cdots$, $ j_{\al}(\om)^{4p}$ such that
\begin{align}
\label{eq:2.12}
\left(j_{NL}\right)_{\al}(\om) &= j_{\al}(\om)^{1p} + j_{\al}(\om)^{2p,a} + j_{\al}(\om)^{2p,b} +
 j_{\al}(\om)^{2p,c}
\nonumber \\
&+ j_{\al}(\om)^{3p,a} + j_{\al}(\om)^{3p,b} + j_{\al}(\om)^{4p}.
\end{align}
Simultaneously, we define the Fourier transforms of the response functions by
\begin{subequations}
\label{eq:2.13}
\begin{align}
&R_{\al \be, \ga \dl}^{(2p,a)}(\Om) \equiv \int_{-\infty}^{\infty} d(t-t_1)
e^{i (\Om + i \eta)(t-t_1)} R_{\al \be, \ga \dl}^{(2p,a)}(t- t_1),
\label{eq:2.13a}\\
&R_{\al \be \ga, \dl}^{(2p,b)}(\Om) \equiv \int_{-\infty}^{\infty} d(t-t_1)
e^{i (\Om + i \eta)(t-t_1)} R_{\al \be \ga, \dl}^{(2p,b)}(t- t_1),
\label{eq:2.13b}\\
&R_{\al, \be \ga \dl}^{(2p,c)}(\Om) \equiv \int_{-\infty}^{\infty} d(t-t_1)
e^{i (\Om + i \eta)(t-t_1)} R_{\al, \be \ga \dl}^{(2p,c)}(t- t_1),
\label{eq:2.13c}\\
&R_{\al \be, \ga, \dl}^{(3p,a)}(\Om_1, \Om_2) \equiv
\int_{-\infty}^{\infty} d(t-t_1) e^{i (\Om_1 + i \eta)(t-t_1)} \nonumber \\
&\times \int_{-\infty}^{\infty} d(t-t_2) e^{i (\Om_2 + i \eta)(t-t_2)}
R_{\al \be, \ga, \dl}^{(3p,a)}(t- t_1, t-t_2),
\label{eq:2.13d}\\
&R_{\al, \be, \ga \dl}^{(3p,b)}(\Om_1, \Om_2) \equiv
\int_{-\infty}^{\infty} d(t-t_1) e^{i (\Om_1 + i \eta)(t-t_1)} \nonumber \\
&\times \int_{-\infty}^{\infty} d(t-t_2) e^{i (\Om_2 + i \eta)(t-t_2)}
R_{\al, \be, \ga \dl}^{(3p,b)}(t- t_1, t-t_2),
\label{eq:2.13e}\\
&R_{\al, \be, \ga, \dl}^{(4p)}(\Om_1, \Om_2, \Om_3) \equiv
\int_{-\infty}^{\infty} d(t-t_1) e^{i (\Om_1 + i \eta)(t-t_1)} \nonumber \\
&\times \int_{-\infty}^{\infty} d(t-t_2) e^{i (\Om_2 + i \eta)(t-t_2)}
\int_{-\infty}^{\infty} d(t-t_3) e^{i (\Om_3 + i \eta)(t-t_3)} \nonumber \\
&\times R_{\al, \be, \ga, \dl}^{(4p)}(t- t_1, t-t_2, t-t_3).
\label{eq:2.13f}
\end{align}
\end{subequations}
Using these definitions it is straightforward to check that the nonlinear current is given by
\begin{widetext}
\begin{align}
\label{eq:2.14}
&\left(j_{NL}\right)_{\al}(\om)  = - \frac{e^4}{6} \int_{-\infty}^{\infty} \int_{-\infty}^{\infty}
\int_{-\infty}^{\infty}  \frac{d \om_1 d \om_2 d \om_3}{(2\pi)^2}
\dl(\om - \om_1 - \om_2 - \om_3)  A_{\be}(\om_1) A_{\ga}(\om_2) A_{\dl}(\om_3)
\left[R_{\al \be \ga \dl}^{(1p)} + \left\{ R_{\al \be, \ga \dl}^{(2p,a)}(\om_2 + \om_3)
\right. \right. \nonumber\\
&+ \left. R_{\al \ga, \be \dl}^{(2p,a)}(\om_1 + \om_3)  + R_{\al \dl, \be \ga}^{(2p,a)}(\om_1 + \om_2)
\right\} + \left\{ R_{\al \be \ga, \dl}^{(2p,b)}(\om_3)  + R_{\al \be \dl, \ga}^{(2p,b)}(\om_2)
+ R_{\al \ga \dl, \be}^{(2p,b)}(\om_1) \right\} + R_{\al, \be \ga \dl}^{(2p,c)}(\om_1 + \om_2 + \om_3)
\nonumber \\
&+ \left\{ R_{\al \be, \ga, \dl}^{(3p,a)}(\om_2, \om_3) + R_{\al \be, \dl, \ga}^{(3p,a)}(\om_3, \om_2)
+ R_{\al \ga, \be, \dl}^{(3p,a)}(\om_1, \om_3) + R_{\al \ga, \dl, \be}^{(3p,a)}(\om_3, \om_1)
+ R_{\al \dl, \be, \ga}^{(3p,a)}(\om_1, \om_2) + R_{\al \dl, \ga, \be}^{(3p,a)}(\om_2, \om_1)
\right\} \nonumber\\
&+ \left\{ 2R_{\al, \be, \ga \dl}^{(3p,b)}(\om_1, \om_2 + \om_3)
+ 2R_{\al, \ga, \be \dl}^{(3p,b)}(\om_2, \om_1 + \om_3)
+ 2R_{\al, \dl, \be \ga}^{(3p,b)}(\om_3, \om_1 + \om_2) \right\}
+ \left\{ R_{\al, \be, \ga, \dl}^{(4p)}(\om_1, \om_2, \om_3) \right. \nonumber\\
&+ \left. \left.
+ R_{\al, \be, \dl, \ga}^{(4p)}(\om_1, \om_3, \om_2)  + R_{\al, \ga, \be, \dl}^{(4p)}(\om_2, \om_1, \om_3)
+ R_{\al, \ga, \dl, \be}^{(4p)}(\om_2, \om_3, \om_1)  + R_{\al, \dl, \be, \ga}^{(4p)}(\om_3, \om_1, \om_2)
+ R_{\al, \dl, \ga, \be}^{(4p)}(\om_3, \om_2, \om_1) \right\} \right].
\end{align}
\end{widetext}
In the above the total nonlinear response kernel, given by the expression within the
square bracket  $[ \cdots ]$, is symmetric with respect to
all permutations of the running variables $(\be, \om_1)$, $(\ga, \om_2)$ and $(\dl, \om_3)$. The various
terms within each curly bracket $\{ \cdots \}$ are equal since they differ only in dummy variables, and they
appear in the process of symmetrization. This also ensures that all the terms within $[ \cdots ]$ have the
same symmetry factor of $1/(3!)$.

The difficulty with Eq.~(\ref{eq:2.14}) is that the response functions, defined in Eq.~(\ref{eq:2.13}),
are not contour-ordered objects, and therefore they cannot be evaluated using the standard tools of
manybody field theory. Formally, the response functions can be expressed using the Lehmann representation,
and this is done in Appendix~\ref{appA}.
However, to evaluate such expressions one needs the exact eigenstates of
$\hham$ which are not known in most cases of interest. To circumvent this difficulty we need to identify
each response function with a contour-ordered correlation function.

With the above motivation we define the following imaginary time ordered current correlation functions.
\begin{subequations}
\label{eq:2.15}
\begin{align}
C_{\al \be, \ga \dl}^{(2p,a)}(\tau, \tau_1) &\equiv - T_{\tau} \la \hv_{\al \be}(\tau)
\hv_{\ga \dl}(\tau_1) \ra,
\label{eq:2.15a}\\
C_{\al \be \ga, \dl}^{(2p,b)}(\tau, \tau_1) &\equiv - T_{\tau} \la \hv_{\al \be \ga}(\tau)
\hv_{\dl}(\tau_1) \ra,
\label{eq:2.15b}\\
C_{\al, \be \ga \dl}^{(2p,c)}(\tau, \tau_1) &\equiv - T_{\tau} \la \hv_{\al}(\tau)
\hv_{\be \ga \dl}(\tau_1) \ra,
\label{eq:2.15c}\\
C_{\al \be, \ga, \dl}^{(3p,a)}(\tau, \tau_1,\tau_2) &\equiv + T_{\tau} \la \hv_{\al \be}(\tau)
\hv_{\ga}(\tau_1) \hv_{\dl}(\tau_2) \ra,
\label{eq:2.15d}\\
C_{\al, \be, \ga \dl}^{(3p,b)}(\tau, \tau_1,\tau_2) &\equiv + T_{\tau} \la \hv_{\al}(\tau)
\hv_{\be}(\tau_1) \hv_{\ga \dl}(\tau_2) \ra,
\label{eq:2.15e}\\
C_{\al, \be, \ga, \dl}^{(4p)}(\tau, \tau_1,\tau_2,\tau_3) &\equiv - T_{\tau} \la \hv_{\al}(\tau)
\hv_{\be}(\tau_1) \hv_{\ga}(\tau_2) \hv_{\dl}(\tau_3)\ra,
\label{eq:2.15f}
\end{align}
\end{subequations}
where $T_{\tau}$ is the imaginary time ordering operator. Note, when $n$ is odd it is convenient
to define the $n$-point correlator with an overall sign which is opposite to the case when $n$ is even.
Next we define the Fourier transforms of the correlators as functions of bosonic Matsubara frequencies as
follows.
\begin{subequations}
\label{eq:2.16}
\begin{align}
&C_{\al \be, \ga \dl}^{(2p,a)}(i \Om_{1n}) \equiv \int_0^{\be} d(\tau - \tau_1)
e^{i \Om_{1n}(\tau - \tau_1)} C_{\al \be, \ga \dl}^{(2p,a)}(\tau, \tau_1)
\label{eq:2.16a}\\
&C_{\al \be \ga, \dl}^{(2p,b)}(i \Om_{1n}) \equiv \int_0^{\be} d(\tau - \tau_1)
e^{i \Om_{1n}(\tau - \tau_1)} C_{\al \be \ga, \dl}^{(2p,b)}(\tau, \tau_1)
\label{eq:2.16b}\\
&C_{\al, \be \ga \dl}^{(2p,c)}(i \Om_{1n}) \equiv \int_0^{\be} d(\tau - \tau_1)
e^{i \Om_{1n}(\tau - \tau_1)}  C_{\al, \be \ga \dl}^{(2p,c)}(\tau, \tau_1)
\label{eq:2.16c}\\
&C_{\al \be, \ga, \dl}^{(3p,a)}(i \Om_{1n}, i \Om_{2n}) \equiv \frac{1}{\be}
\int_0^{\be} d \tau \int_0^{\be} d \tau_1 e^{i \Om_{1n}(\tau - \tau_1)}
\nonumber\\
&\times \int_0^{\be} d \tau_2 e^{i \Om_{2n}(\tau - \tau_2)}
C_{\al \be, \ga, \dl}^{(3p,a)}(\tau, \tau_1, \tau_2)
\label{eq:2.16d}\\
&C_{\al, \be, \ga \dl}^{(3p,b)}(i \Om_{1n}, i \Om_{2n}) \equiv \frac{1}{\be}
\int_0^{\be} d \tau \int_0^{\be} d \tau_1 e^{i \Om_{1n}(\tau - \tau_1)}
\nonumber\\
&\times \int_0^{\be} d \tau_2 e^{i \Om_{2n}(\tau - \tau_2)}
C_{\al, \be, \ga \dl}^{(3p,b)}(\tau, \tau_1, \tau_2)
\label{eq:2.16e}\\
&C_{\al, \be, \ga, \dl}^{(4p)}(i \Om_{1n}, i\Om_{2n}, i \Om_{3n}) \equiv \frac{1}{\be}
\int_0^{\be} d \tau \int_0^{\be} d \tau_1 e^{i \Om_{1n}(\tau - \tau_1)}
\nonumber\\
&\times \int_0^{\be} d \tau_2 e^{i \Om_{2n}(\tau - \tau_2)}
\int_0^{\be} d \tau_3 e^{i \Om_{3n}(\tau - \tau_3)}
\nonumber\\
&\times C_{\al, \be, \ga, \dl}^{(4p)}(\tau, \tau_1, \tau_2, \tau_3).
\label{eq:2.16f}
\end{align}
\end{subequations}
In the above the structure of the 2-point functions is familiar from linear response theory.
For example, due to time translation symmetry $C_{\al \be, \ga \dl}^{(2p,a)}(\tau, \tau_1)$
is a function of $s_1 = \tau - \tau_1$, and consequently, there is only one way its Fourier transform
in imaginary frequency space can be defined. Furthermore, it satisfies bosonic periodicity with
$C_{\al \be, \ga \dl}^{(2p,a)}(s_1) = C_{\al \be, \ga \dl}^{(2p,a)}(s_1 + \beta)$ for $-\be < s_1 <0$,
which further simplifies the structure of the correlator in the frequency space. By contrast,
the nonlinear correlators are functions of more than one imaginary time variable. For example,
$C_{\al \be, \ga, \dl}^{(3p,a)}(\tau, \tau_1, \tau_2)$ is a function of two variables $s_1 = \tau - \tau_1$
and $s_2 = \tau - \tau_2$. Consequently, there are more than one way to take Fourier transforms,
and the appropriate one has to be chosen with care. Moreover, unlike the 2-point functions,
$C_{\al \be, \ga, \dl}^{(3p,a)}(s_1, s_2)$ and the other nonlinear correlators
do not have the property of $\be$-periodicity. As a
consequence, in Eqs.~(\ref{eq:2.16d}), (\ref{eq:2.16e}), (\ref{eq:2.16f}) there are additional
$\tau$-integrals which are crucial to obtain the correct quantities in Matsubara space.

The next step is to express the response functions defined by Eq.~(\ref{eq:2.13}) and the correlation
functions defined by Eq.~(\ref{eq:2.16}) using Lehmann representation, and to compare them. The
procedure is somewhat long, but straightforward, and the details of this step are given in
Appendix~\ref{appA}. Based on it, we find that the 2-point functions are related by
\begin{subequations}
\label{eq:2.17}
\begin{align}
&C_{\al \be, \ga \dl}^{(2p,a)}(i \Om_{n} \rightarrow \Om^+)
= R_{\al \be, \ga \dl}^{(2p,a)}(\Om),
\label{eq:2.17a}\\
&C_{\al \be \ga, \dl}^{(2p,b)}(i \Om_{n} \rightarrow \Om^+)
= R_{\al \be \ga, \dl}^{(2p,b)}(\Om),
\label{eq:2.17b}\\
&C_{\al, \be \ga \dl}^{(2p,c)}(i \Om_{n} \rightarrow \Om^+)
= R_{\al, \be \ga \dl}^{(2p,c)}(\Om),
\label{eq:2.17c}
\end{align}
\end{subequations}
where $\Om^+ \equiv \Om + i\eta$. This mapping
is well-known from linear response theory. Next, the 3-point functions are related by
\begin{subequations}
\label{eq:2.18}
\begin{align}
&C_{\al \be, \ga, \dl}^{(3p,a)}(i \Om_{1n} \rightarrow \Om_1^+,
i \Om_{2n} \rightarrow \Om_2^+) = R_{\al \be, \ga, \dl}^{(3p,a)}(\Om_{1}, \Om_{2})
\nonumber\\
&+ R_{\al \be, \dl, \ga}^{(3p,a)}(\Om_{2}, \Om_{1}),
\label{eq:2.18a}\\
&C_{\al, \be, \ga \dl}^{(3p,b)}(i \Om_{1n} \rightarrow \Om_1^+,
i \Om_{2n}  \rightarrow \Om_2^+)
= 2 R_{\al, \be, \ga \dl}^{(3p,b)}(\Om_{1}, i \Om_{2}),
\label{eq:2.18b}
\end{align}
\end{subequations}
and the 4-point functions by
\begin{align}
\label{eq:2.19}
&C_{\al, \be, \ga, \dl}^{(4p)}(i \Om_{1n} \rightarrow \Om_1^+,
i\Om_{2n} \rightarrow \Om_2^+, i \Om_{3n} \rightarrow \Om_3^+)
\nonumber\\
&= R_{\al, \be, \ga, \dl}^{(4p)}(\Om_1, \Om_2, \Om_3)
+ R_{\al, \be, \dl, \ga}^{(4p)}(\Om_1, \Om_3, \Om_2)  \nonumber\\
&+ R_{\al, \ga, \be, \dl}^{(4p)}(\Om_2, \Om_1, \Om_3)
+ R_{\al, \ga, \dl, \be}^{(4p)}(\Om_2, \Om_3, \Om_1)
\nonumber\\
&+ R_{\al, \dl, \be, \ga}^{(4p)}(\Om_3, \Om_1, \Om_2)
+ R_{\al, \dl, \ga, \be}^{(4p)}(\Om_3, \Om_2, \Om_1).
\end{align}
Using Eqs.~(\ref{eq:2.17}), (\ref{eq:2.18}) and (\ref{eq:2.19}) the nonlinear current in Eq.~(\ref{eq:2.14})
can be re-written as
\begin{align}
\label{eq:2.20}
&\left(j_{NL}\right)_{\al}(\om)  = \int_{-\infty}^{\infty} \int_{-\infty}^{\infty}
\int_{-\infty}^{\infty}  \frac{d \om_1 d \om_2 d \om_3}{(2\pi)^2}
\nonumber\\
&\times
\dl(\om - \om_1 - \om_2 - \om_3)  A_{\be}(\om_1) A_{\ga}(\om_2) A_{\dl}(\om_3)
\nonumber\\
&\times \Pi_{\al \be \ga \dl}^{(3)}(\om_1, \om_2, \om_3),
\end{align}
where the nonlinear current kernel $\Pi_{\al \be \ga \dl}^{(3)}(\om_1, \om_2, \om_3)$
is given by
\begin{widetext}
\begin{align}
\label{eq:2.21}
&\Pi_{\al \be \ga \dl}^{(3)}(\om_1, \om_2, \om_3) =  - \frac{e^4}{6}
\left[C_{\al \be \ga \dl}^{(1p)} + \left\{ C_{\al \be, \ga \dl}^{(2p,a)}(\om_2 + \om_3 + i\eta)
+ C_{\al \ga, \be \dl}^{(2p,a)}(\om_1 + \om_3 + i\eta)
+ C_{\al \dl, \be \ga}^{(2p,a)}(\om_1 + \om_2 + i\eta) \right\}
\right. \nonumber\\
&+ \left\{ C_{\al \be \ga, \dl}^{(2p,b)}(\om_3 + i\eta)
+ C_{\al \be \dl, \ga}^{(2p,b)}(\om_2 + i\eta)
+ C_{\al \ga \dl, \be}^{(2p,b)}(\om_1 + i\eta) \right\}
+ C_{\al, \be \ga \dl}^{(2p,c)}(\om_1 + \om_2 + \om_3 + i\eta)
\nonumber \\
&+ \left\{ C_{\al \be, \ga, \dl}^{(3p,a)}(\om_2 + i\eta, \om_3 + i\eta)
+ C_{\al \ga, \be, \dl}^{(3p,a)}(\om_1 + i\eta, \om_3 + i\eta)
+ C_{\al \dl, \be, \ga}^{(3p,a)}(\om_1 + i\eta, \om_2 + i\eta)
\right\} \nonumber\\
&+ \left\{ C_{\al, \be, \ga \dl}^{(3p,b)}(\om_1 + i\eta, \om_2 + \om_3 + i\eta)
+ C_{\al, \ga, \be \dl}^{(3p,b)}(\om_2 + i\eta, \om_1 + \om_3 + i\eta)
+ C_{\al, \dl, \be \ga}^{(3p,b)}(\om_3 + i\eta, \om_1 + \om_2 + i\eta) \right\}
\nonumber\\
&+ \left.
C_{\al, \be, \ga, \dl}^{(4p)}(\om_1 + i\eta, \om_2 + i\eta, \om_3 + i\eta) \right].
\end{align}
\end{widetext}
In the above $C_{\al \be \ga \dl}^{(1p)} \equiv R_{\al \be \ga \dl}^{(1p)}$
is a frequency independent constant. Note, the nonlinear current kernel
$\Pi_{\al \be \ga \dl}^{(3)}(\om_1, \om_2, \om_3)$ is fully symmetric with respect to permutations of the
variables $(\be, \om_1)$, $(\ga, \om_2)$ and $(\dl, \om_3)$.
The advantage of Eq.~(\ref{eq:2.20}), compared to Eq.~(\ref{eq:2.14}), is that the nonlinear
current response is now given in terms of current-current correlators. Being
contour-ordered objects, the correlators can be factorized using Wick's theorem, and therefore expressed as
products of single particle Green's function. In other words, standard techniques of manybody field theory
and controlled approximation schemes can be used to compute the nonlinear current response.

The current response in Eq.~(\ref{eq:2.20}) can be expressed alternatively in terms of the external electric
field $\bE(\om) = i \om \bA(\om)$ and the third order nonlinear conductivity as
\begin{align}
\label{eq:2.21-bis}
&\left(j_{NL}\right)_{\al}(\om)  = \int_{-\infty}^{\infty} \int_{-\infty}^{\infty}
\int_{-\infty}^{\infty}  \frac{d \om_1 d \om_2 d \om_3}{(2\pi)^2}
\nonumber\\
&\times
\dl(\om - \om_1 - \om_2 - \om_3)  E_{\be}(\om_1) E_{\ga}(\om_2) E_{\dl}(\om_3)
\nonumber\\
&\times \si_{\al \be \ga \dl}^{(3)}(\om_1, \om_2, \om_3),
\end{align}
where the third order nonlinear conductivity is defined as
\begin{align}
\label{eq:2.21-bis2}
\si_{\al \be \ga \dl}^{(3)}(\om_1, \om_2, \om_3)
\equiv \frac{i \Pi_{\al \be \ga \dl}^{(3)}(\om_1, \om_2, \om_3)}
{(\om_1 + i\eta)(\om_2 + i\eta)(\om_3 + i\eta)}.
\end{align}
Note, as we show in Sec.~\ref{sec3}, particle number conservation, or gauge invariance, ensures
that in non-superconducting systems
$\Pi_{\al \be \ga \dl}^{(3)}(\om_1, \om_2, \om_3) \sim \om_1 \om_2 \om_3$ for
$(\om_1, \om_2,\om_3) \rightarrow 0$. Consequently, the nonlinear conductivity remains finite
even when one or more of the external frequencies are set to zero.

Eqs.~(\ref{eq:2.20}) - (\ref{eq:2.21-bis2}) constitute the main results of this section.
They express the nonlinear electro-optical response of an electronic system in terms of gauge
invariant quantities, see Sec.~\ref{sec3} for further discussion.
Thus, the nonlinear current is expressed in terms of the nonlinear conductivity,
and the latter in terms of a sum of several current correlators. These relations are quite general and
they are relevant not only for metallic phases, but for superconducting ones as well.
Note, since the Eqs.~(\ref{eq:2.17}), (\ref{eq:2.18}) and (\ref{eq:2.19}) relating
the response functions with the correlators
is proven using the Lehmann representation and the exact
eigenstates of $\hham$ (see Appendix~\ref{appA}),
Eqs.~(\ref{eq:2.20}) - (\ref{eq:2.21-bis2}) are formally exact to all orders
in interaction and disorder strengths.

\section{Gauge invariance and sum rule}
\label{sec3}

In this section we discuss certain general properties of the nonlinear kernel
$\Pi_{\al \be \ga \dl}^{(3)}(\om_1, \om_2, \om_3)$
that follow from particle number conservation.

\subsection{Gauge invariance}
\label{subsec:3.1}

A vector potential that is constant in time $\bA(t) = \bA_0$ is equivalent to zero electric
field in the bulk.
Such a potential should not affect the system, provided the electromagnetic response
of the boundary is trivial, which is the case of non-superconducting phases.
Since in frequency space such a vector
potential is $\bA(\om) = \bA_0 \dl(\om)$, we expect that for such phases, and for any given set of
polarizations $(\al, \be, \ga, \dl)$
\begin{align}
\label{eq:3.1}
&\Pi_{\al \be \ga \dl}^{(3)}(\om_1=0, \om_2, \om_3) =
\Pi_{\al \be \ga \dl}^{(3)}(\om_1, \om_2=0, \om_3)
\nonumber\\
&= \Pi_{\al \be \ga \dl}^{(3)}(\om_1, \om_2, \om_3=0)
=0.
\end{align}
Taken together, the above three relations imply that
$\Pi_{\al \be \ga \dl}^{(3)}(\om_1, \om_2, \om_3) \sim \om_1 \om_2 \om_3$ for
$(\om_1, \om_2,\om_3) \rightarrow 0$, such that the responses stay finite even if one or more of 
the external photon frequencies are set to zero.
Below we provide a proof of these relations.

The first step is to express the current operators defined in Eq.~(\ref{eq:2.3}) in terms of the generalized
density operator
\beq
\label{eq:3.2}
\hro_{\bq} \equiv \sum_{\bk} \cda_{\bk + \bq} c_{\bk}.
\eeq
The paramagnetic current operator, defined in Eq.~(\ref{eq:2.3a}) can be written as
\beq
\label{eq:3.3}
\hv_{\al} = \lim_{\bq \to 0} \frac{1}{q_{\al}} \left[ \hham , \hro_{q_{\al}} \right].
\eeq
The above relation follows from the continuity equation, which itself is a consequence of particle number
conservation. Alternately, it can be verified explicitly for an interacting electron Hamiltonian of the
form $\hham = \sum_{\bk} \ep_{\bk} \cda_{\bk} c_{\bk} + \sum_{\bq} V(\bq) \hro_{\bq} \hro_{-\bq}$,
where $V(\bq)$ is the interaction potential. For unscreened Coulomb potential
$V(\bq) \propto 1/q^2$ and for screened Coulomb $V(\bq) \propto 1/(q^2 + q_{0}^2)$, with
$1/q_{0}$ the Thomas-Fermi screening length. Likewise, the remaining current operators defined in
Eqs.~(\ref{eq:2.3b}) - (\ref{eq:2.3d}) can be written as
\beq
\label{eq:3.4}
\hv_{\al \be} = \lim_{\bq \to 0} \frac{1}{q_{\al} q_{\be}}
\left[\left[ \hham , \hro_{q_{\al}} \right] , \hro_{q_{\be}} \right],
\eeq
\beq
\label{eq:3.5}
\hv_{\al \be \ga} = \lim_{\bq \to 0} \frac{1}{q_{\al} q_{\be} q_{\ga}}
\left[ \left[\left[ \hham , \hro_{q_{\al}} \right] , \hro_{q_{\be}} \right] , \hro_{q_{\ga}} \right],
\eeq
\beq
\label{eq:3.6}
\hv_{\al \be \ga \dl} = \lim_{\bq \to 0} \frac{1}{q_{\al} q_{\be} q_{\ga} q_{\dl}}
\left[ \left[ \left[\left[ \hham , \hro_{q_{\al}} \right] , \hro_{q_{\be}} \right] , \hro_{q_{\ga}} \right] ,
\hro_{q_{\dl}} \right].
\eeq

The second step is to convert, using Eqs.~(\ref{eq:3.3}) - (\ref{eq:3.6}),
the various current matrix elements, that enter in the definition of the various correlators
in Appendix~\ref{appA}, into equivalent density matrix elements. For this purpose we
define the following matrix elements involving the density operators in the Lehmann basis.
\begin{subequations}
\label{eq:3.7}
\begin{align}
\left(T_1^{\bq} \right)_{nmpl} &\equiv (\hro_{q_{\al}})_{nm}(\hro_{q_{\be}})_{mp}
(\hro_{q_{\ga}})_{pl}(\hro_{q_{\dl}})_{ln},
\label{eq:3.7a}\\
\left(T_2^{\bq} \right)_{nmpl} &\equiv (\hro_{q_{\al}})_{nm}(\hro_{q_{\be}})_{mp}
(\hro_{q_{\dl}})_{pl}(\hro_{q_{\ga}})_{ln},
\label{eq:3.7b}\\
\left(T_3^{\bq} \right)_{nmpl} &\equiv (\hro_{q_{\al}})_{nm}(\hro_{q_{\ga}})_{mp}
(\hro_{q_{\be}})_{pl}(\hro_{q_{\dl}})_{ln},
\label{eq:3.7c}\\
\left(T_4^{\bq} \right)_{nmpl} &\equiv (\hro_{q_{\al}})_{nm}(\hro_{q_{\ga}})_{mp}
(\hro_{q_{\dl}})_{pl}(\hro_{q_{\be}})_{ln},
\label{eq:3.7d}\\
\left(T_5^{\bq} \right)_{nmpl} &\equiv (\hro_{q_{\al}})_{nm}(\hro_{q_{\dl}})_{mp}
(\hro_{q_{\ga}})_{pl}(\hro_{q_{\be}})_{ln},
\label{eq:3.7e}\\
\left(T_6^{\bq} \right)_{nmpl} &\equiv (\hro_{q_{\al}})_{nm}(\hro_{q_{\dl}})_{mp}
(\hro_{q_{\be}})_{pl}(\hro_{q_{\ga}})_{ln}.
\label{eq:3.7f}
\end{align}
\end{subequations}
In the above no summation over repeated indices is implied, and $(\hO)_{nm} \equiv \la n | \hO | m\ra$,
where $(n, m, p, l)$ are indices associated with the energy eigenstates of $\hham$ such that
$\hham | n \ra = E_n | n \ra$.

The third step of the proof is to express $\Pi_{\al \be \ga \dl}^{(3)}(\om_1, \om_2, \om_3)$ in terms of
the matrix elements introduced in Eq.~(\ref{eq:3.7}). As an example of a two-point function,
$C_{\al \be, \ga \dl}^{(2p,a)}(i\Om)$ given in Eq.~(\ref{eq:A4}) can be re-expressed as
$C_{\al \be, \ga \dl}^{(2p,a)}(i\Om) = \lim_{\bq \to 0}
C_{\al \be, \ga \dl}^{(2p,a)}(i\Om,\bq)/(q_{\al} q_{\be} q_{\ga} q_{\dl})$, where
\begin{align}
&Z C_{\al \be, \ga \dl}^{(2p,a)}(i\Om,\bq) =
\frac{ e^{-\be E_n} - e^{- \be E_p}}{i \Om_n + E_{np}}
\left[ \left(T_1^{\bq} \right)_{nmpl} E_{nm} E_{pl}
\right. \nonumber\\
&- \left.
\left(T_2^{\bq} \right)_{nmpl} E_{nm} E_{ln} \right]
- \frac{ e^{-\be E_l} - e^{- \be E_m}}{i \Om_n + E_{lm}}
\nonumber\\
&\times
\left[ \left(T_4^{\bq} \right)_{nmpl} E_{nm} E_{mp}
- \left(T_5^{\bq} \right)_{nmpl} E_{nm} E_{pl} \right].
\nonumber
\end{align}
Likewise, as an example of a three-point function,
\[
C_{\al \be, \ga, \dl}^{(3p,a)}(i \Om_{1n}, i \Om_{2n}) = \lim_{\bq \to 0}
\frac{C_{\al \be, \ga, \dl}^{(3p,a)}(i \Om_{1n}, i \Om_{2n}, \bq)}{q_{\al} q_{\be} q_{\ga} q_{\dl}},
\]
where
\begin{widetext}
\begin{align}
& Z C_{\al \be, \ga, \dl}^{(3p,a)}(i \Om_{1n}, i \Om_{2n}, \bq)
= - \frac{\left(T_1^{\bq} \right)_{nmpl} E_{nm} E_{pl} E_{ln}}{i \Om_{12n} + E_{np}}
\left[ \frac{e^{-\be E_l} - e^{-\be E_p}}{i\Om_{1n} + E_{lp}} +
\frac{e^{-\be E_l} - e^{-\be E_n}}{i\Om_{2n} + E_{nl}} \right]
+  \frac{\left(T_4^{\bq} \right)_{nmpl} E_{nm} E_{mp} E_{pl}}{i \Om_{12n} + E_{lm}}
\nonumber\\
&\times
\left[ \frac{e^{-\be E_p} - e^{-\be E_m}}{i\Om_{1n} + E_{pm}} +
\frac{e^{-\be E_p} - e^{-\be E_l}}{i\Om_{2n} + E_{lp}} \right]
-  \frac{\left(T_2^{\bq} \right)_{nmpl} E_{nm} E_{pl} E_{ln}}{i \Om_{12n} + E_{np}}
\left[ \frac{e^{-\be E_l} - e^{-\be E_p}}{i\Om_{2n} + E_{lp}} +
\frac{e^{-\be E_l} - e^{-\be E_n}}{i\Om_{1n} + E_{nl}} \right]
\nonumber\\
&+  \frac{\left(T_5^{\bq} \right)_{nmpl} E_{nm} E_{mp} E_{pl}}{i \Om_{12n} + E_{lm}}
\left[ \frac{e^{-\be E_p} - e^{-\be E_m}}{i\Om_{2n} + E_{pm}} +
\frac{e^{-\be E_p} - e^{-\be E_l}}{i\Om_{1n} + E_{lp}} \right].
\nonumber
\end{align}
\end{widetext}
In order to re-express the four-point function we use relations such as
\begin{align}
\left(W_1 \right)_{nmpl} &\equiv (\hv_{\al})_{nm}(\hv_{\be})_{mp}(\hv_{\ga})_{pl}(\hv_{\dl})_{ln},
\nonumber\\
&= \lim_{\bq \to 0} \frac{\left(T_1^{\bq} \right)_{nmpl} E_{nm} E_{mp} E_{pl} E_{ln}}
{q_{\al} q_{\be} q_{\ga} q_{\dl}},
\nonumber
\end{align}
and so on, and also Eq.~(\ref{eq:A21}).

From the above discussion it is clear that the nonlinear susceptibility can be expressed as a limit
in the form
\beq
\label{eq:3.8}
\Pi_{\al \be \ga \dl}^{(3)}(\om_1, \om_2, \om_3) =
\lim_{\bq \to 0}
\frac{\Pi_{\al \be \ga \dl}^{(3)}(\om_1, \om_2, \om_3, \bq)}{q_{\al} q_{\be} q_{\ga} q_{\dl}}
\eeq
where $\Pi_{\al \be \ga \dl}^{(3)}(\om_1, \om_2, \om_3, \bq)$ has the structure
\begin{align}
\label{eq:3.9}
&\Pi_{\al \be \ga \dl}^{(3)}(\om_1, \om_2, \om_3, \bq)
= - \left[ \left(T_1^{\bq} \right)_{nmpl} Q_1(\om_1, \om_2, \om_3)_{nmpl} \right.
\nonumber\\
&+ \left. \cdots + \left(T_6^{\bq} \right)_{nmpl} Q_6(\om_1, \om_2, \om_3)_{nmpl} \right] e^4 E_{nm}/6.
\end{align}
The coefficients $Q_i(\om_1, \om_2, \om_3)_{nmpl}$, $i= 1, \cdots, 6$, are given
in Appendix~\ref{appB}, see Eqs.~(\ref{eq:B1})-(\ref{eq:B6}).

Now we set $\om_3=0$. It is simple to check using  Eqs.~(\ref{eq:B1})-(\ref{eq:B6}) that
$Q_i(\om_1, \om_2, \om_3=0)_{nmpl} = 0$, $\forall i$. Thus,
\beq
\label{eq:3.10}
\Pi_{\al \be \ga \dl}^{(3)}(\om_1, \om_2, \om_3 =0, \bq) =0.
\eeq
Since the above relation holds for all sets of polarizations $(\al, \be, \ga, \dl)$,
it is clear from the cyclic property of the kernel that
\[
\Pi_{\al \be \ga \dl}^{(3)}(\om_1=0, \om_2, \om_3, \bq) =
\Pi_{\al \be \ga \dl}^{(3)}(\om_1, \om_2=0, \om_3, \bq)  = 0
\]
will hold as well. These two relations can also be shown from the following arguments.

We set $\om_2=0$ in Eq.~(\ref{eq:3.9}), and we get $Q_3(\om_1, \om_2=0, \om_3)_{nmpl}
= Q_6(\om_1, \om_2=0, \om_3)_{nmpl} = 0$, while
\begin{align}
\label{eq:3.11}
&Q_1(\om_1, \om_2=0, \om_3)_{nmpl}= - Q_2(\om_1, \om_2=0, \om_3)_{nmpl}
\nonumber\\
&= \frac{\om_3(\om_{13} + E_{np}) e^{-\be E_n}}{(\om_{13} + E_{nm})(\om_3 + E_{np})}
+ \frac{\om_3 E_{mp} e^{-\be E_m}}{(\om_{13} + E_{nm})(\om_1 + E_{pm})}
\nonumber\\
&-\frac{\om_1 \om_3  e^{-\be E_p}}{(\om_{3} + E_{np})(\om_1 + E_{pm})},
\end{align}
and
\begin{align}
\label{eq:3.12}
&Q_4(\om_1, \om_2=0, \om_3)_{nmpl}= - Q_5(\om_1, \om_2=0, \om_3)_{nmpl}
\nonumber\\
&= \frac{\om_3 E_{ln} e^{-\be E_n}}{(\om_{13} + E_{nm})(\om_1 + E_{nl})}
+ \frac{\om_3 (\om_{13} + E_{lm}) e^{-\be E_m}}{(\om_{13} + E_{nm})(\om_3 + E_{lm})}
\nonumber\\
&-\frac{\om_1 \om_3  e^{-\be E_l}}{(\om_{3} + E_{lm})(\om_1 + E_{nl})}.
\end{align}
In the above equations $\om_{13} \equiv \om_1 + \om_3$. Importantly,
$Q_1(\om_1, \om_2=0, \om_3)_{nmpl}$ and $Q_2(\om_1, \om_2=0, \om_3)_{nmpl}$ are independent
of the Lehmann basis index $l$. Thus, once $\om_2=0$, in Eq.~(\ref{eq:3.9}) the summation over the
index $l$ can be performed for $\left(T_1^{\bq} \right)_{nmpl}$ and for $\left(T_2^{\bq} \right)_{nmpl}$.
Using $|l \ra \la l | = 1$, and the fact that $[ \hro_{q_{\ga}} , \hro_{q_{\dl}}] =0$ we conclude that
\[
\sum_l \left(T_1^{\bq} \right)_{nmpl} = \sum_l \left(T_2^{\bq} \right)_{nmpl}.
\]
In other words, the coefficients
$Q_1(\om_1, \om_2=0, \om_3)_{nmpl}$ and $Q_2(\om_1, \om_2=0, \om_3)_{nmpl}$ add up
to zero in Eq.~(\ref{eq:3.9}). Likewise,
$Q_4(\om_1, \om_2=0, \om_3)_{nmpl}$ and $Q_5(\om_1, \om_2=0, \om_3)_{nmpl}$ are independent
of the Lehmann basis index $p$.
Using the same argument we conclude that
\[
\sum_p \left(T_4^{\bq} \right)_{nmpl} = \sum_p \left(T_5^{\bq} \right)_{nmpl}.
\]
Thus, the coefficients $Q_4(\om_1, \om_2=0, \om_3)_{nmpl}$ and $Q_5(\om_1, \om_2=0, \om_3)_{nmpl}$
also add up to zero in Eq.~(\ref{eq:3.9}), and we find
\beq
\label{eq:3.13}
\Pi_{\al \be \ga \dl}^{(3)}(\om_1, \om_2=0, \om_3, \bq) =0.
\eeq

Lastly, we set $\om_1=0$ in Eq.~(\ref{eq:3.9}).
In this case the argument is similar. First, we find that
$Q_1(\om_1=0, \om_2, \om_3)_{nmpl}$ is independent of the index $p$, which allows
 $\left(T_1^{\bq} \right)_{nmpl}$ to be written as $\left(T_3^{\bq} \right)_{nmpl}$ in
 Eq.~(\ref{eq:3.9}). Next, we find that $Q_4(\om_1=0, \om_2, \om_3)_{nmpl}$ is
 independent of the index $l$, which allows $\left(T_4^{\bq} \right)_{nmpl}$ to be written as
 $\left(T_3^{\bq} \right)_{nmpl}$ as well. Likewise, both $\left(T_2^{\bq} \right)_{nmpl}$
 and $\left(T_5^{\bq} \right)_{nmpl}$ can be written as $\left(T_6^{\bq} \right)_{nmpl}$ once
 $\om_1=0$. Finally, using the fact that
 \begin{align}
 &Q_1(\om_1=0, \om_2, \om_3)_{nmpl} + Q_3(\om_1=0, \om_2, \om_3)_{nmpl}
 \nonumber\\
 &+ Q_4(\om_1=0, \om_2, \om_3)_{nmpl} =0,
 \nonumber\\
 &Q_2(\om_1=0, \om_2, \om_3)_{nmpl} + Q_5(\om_1=0, \om_2, \om_3)_{nmpl}
 \nonumber\\
 &+ Q_6(\om_1=0, \om_2, \om_3)_{nmpl} =0,
 \nonumber
 \end{align}
 we conclude that
 \beq
\label{eq:3.14}
\Pi_{\al \be \ga \dl}^{(3)}(\om_1=0, \om_2, \om_3, \bq) =0.
\eeq

As equations~(\ref{eq:3.10}), (\ref{eq:3.13}) and (\ref{eq:3.14}) hold for general wavevector $\bq$, it also
holds in the limit $\bq \rightarrow 0$. Thus, we conclude that the kernel vanishes
in the limit where the frequency $\om_i$,
$i= (1, 2,3)$, is first set to zero, and then the wavevector $\bq \rightarrow 0$ (quasistatic limit). However, the
quantity of interest in Eq.~(\ref{eq:3.1}) is the one for which first the wavevector is set to zero, and then
the frequency $\om_i \rightarrow 0$ (quasidynamic limit). Consequently, the question is whether
the two ways of taking limits commute.

In general, the non-commutation of the two ways of taking limits
signify the presence of non-analytic terms in the kernel
$\Pi_{\al \be \ga \dl}^{(3)}(\om_1, \om_2, \om_3, \bq)$, and there are two potential sources of
non-analyticity that need to be considered here. (i) In metals there are gapless
excitations close to the Fermi surface that can lead to non-analytic response.
However, one can show that, in the presence of a
finite elastic scattering lifetime, such non-analytic terms
are absent. This point has been discussed recently in the context of quadrupolar charge susceptibility of
metals~\cite{Gallais16}. (ii) The above proof is only a statement about the longitudinal response for which
$\nabla \times \bj_{NL}(\br) =0$. This follows from
Eq.~(\ref{eq:3.8}) which shows that the kernel considered here has the structure
\[
\lim_{\bq \to 0} \Pi_{\al \be \ga \dl}^{(3)}(\om_1, \om_2, \om_3, \bq)
= q_{\al} q_{\be} q_{\ga} q_{\dl}  \Pi^{(3L)}(\om_1, \om_2, \om_3, q) ,
\]
where $\Pi^{(3L)}(\om_1, \om_2, \om_3, q)$ is a scalar function independent of the direction of $\bq$.
On the other hand, in superconductors the transverse response is non-zero in the quasistatic limit
(Meissner effect). This finite transverse response also shows up, and gives a nonzero
contribution in the quasidynamic limit, and
consequently  Eq.~(\ref{eq:3.1})  does not hold for superconductors. But for non superconducting
phases no such
transverse response is expected, and therefore switching the two limits is justified.

This completes the proof of the assertion in Eq.~(\ref{eq:3.1}).
Note, since the proof uses the exact eigenstates of the Hamiltonian $\hham$, it is nonperturbative, and it
holds to all orders in electron-electron interaction and disorder strengths.

\subsection{Sum rule}
\label{subsec:3.2}

The nonlinear conductivity satisfies a generalization of the $f$-sum rule which can be expressed as
\begin{align}
\label{eq:3.15}
&\int_{-\infty}^{\infty} \int_{-\infty}^{\infty}
\int_{-\infty}^{\infty}  \frac{d \om_1 d \om_2 d \om_3}{(\pi)^3}
\si_{\al \be \ga \dl}^{(3)}(\om_1, \om_2, \om_3)
\nonumber\\
&= \frac{e^4}{6} \la \sum_{\bk} \frac{\ptl^4 \ep_{\bk}}
{\ptl k_{\al} \ptl k_{\be} \ptl k_{\ga} \ptl k_{\dl}} \cda_{\bk} c_{\bk} \ra.
\end{align}

The above relation follows simply from the causal structure of the response which guarantees that, as a
function of the three frequencies, $\si_{\al \be \ga \dl}^{(3)}(\om_1, \om_2, \om_3)$ has poles only on
the lower half planes, and is analytic in the upper half planes. Thus, all the frequency-dependent terms in
Eq.~(\ref{eq:2.21}) necessarily have an integral of the type
\[
\int_{-\infty}^{\infty} d \om_i \frac{1}{(\om_i + i\eta)(\om_i + E_0 + i\eta)} =0,
\]
where $E_0$ is an energy scale. The above integral vanishes since the contour can be completed in the
upper half plane where the integrand is analytic. Thus, the only term that survives the frequency integrals
is the constant $C_{\al \be \ga \dl}^{(1p)}$, and the above sum rule is established using
Eq.~(\ref{eq:2.3d}). The sum rule and its generalization to higher order nonlinear conductivities
was discussed earlier~\cite{Watanabe20}. Note, since the sum rule is proven using causality and the
general expression of the current kernel [Eq.~(\ref{eq:2.21})] which holds for all phases, in particular,
it is valid for superconductors as well.

\section{Nonlinear Drude response}
\label{sec4}
%====================
\begin{figure*}
\begin{center}
\includegraphics[width=15cm,trim=0 0 0 0]{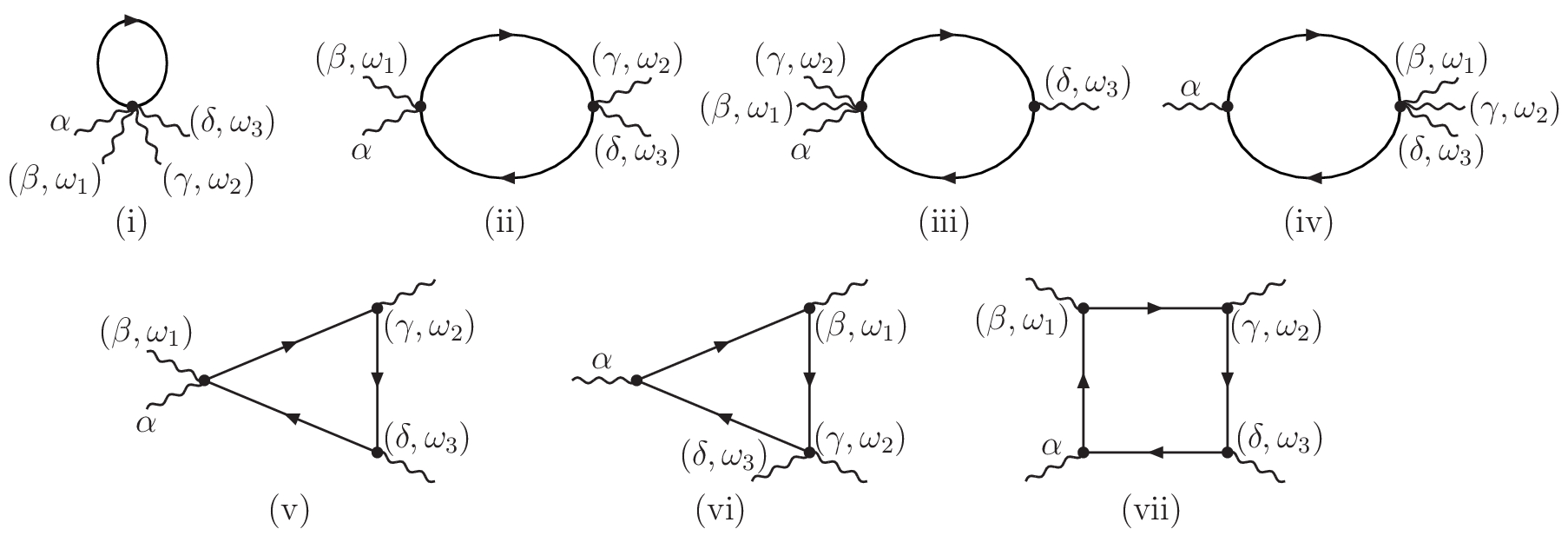}
\caption{Diagrams without vertex corrections for the nonlinear electro-optical kernel
$\Pi_{\al \be \ga \dl}^{(3)}(\om_1, \om_2, \om_3)$, see Eq.~(\ref{eq:2.21}).
The solid lines are electron Green's functions, and the wiggly lines are one outgoing and three incoming
photons with polarizations $(\al, \be, \ga, \dl)$ and frequencies $(\om, \om_1, \om_2, \om_3)$,
respectively, with  $\om = \om_1 + \om_2 + \om_3$.  (i) is the one-point function
$C_{\al \be \ga \dl}^{(1p)}$. (ii), (iii) and (iv) are the two-point functions
$C_{\al \be, \ga \dl}^{(2p,a)}(\om_{2} + \om_{3})$, $C_{\al \be \ga, \dl}^{(2p,b)}(\om_{3})$,
and $C_{\al, \be \ga \dl}^{(2p,c)}(\om_{1} + \om_{2} + \om_{3})$, respectively.
Diagram (v) plus that obtained by interchanging the positions of the $(\ga, \om_2)$
and the $(\dl, \om_3)$ photons give the three-point function
$C_{\al \be, \ga, \dl}^{(3p,a)}(\om_{2}, \om_{3})$. The diagram (vi) and
that obtained by interchanging the positions of the $(\al, \om)$ and $(\be, \om_1)$
photons together give the three-point function
$C_{\al, \be, \ga \dl}^{(3p,b)}(\om_{1}, \om_{2} + \om_{3})$.
Diagram (vii) and five others obtained by permuting the indices $(\be, \om_1)$, $(\ga, \om_2)$
and $(\dl, \om_3)$ together give the four-point function
$C_{\al, \be, \ga, \dl}^{(4p)}(\om_{1}, \om_{2}, \om_{3})$.
}
\label{fig1}
\end{center}
\end{figure*}
%====================
In this section we calculate the nonlinear electro-optical response of the simplest nontrivial system,
namely noninteracting electrons in the presence of weak disorder, using the formalism developed in
section~\ref{sec2}. Accordingly, we take
\beq
\label{eq:4.1}
\hham = \sum_{\bk} \ep_{\bk} \cda_{\bk} c_{\bk} + \frac{1}{\vol} \sum_{\bk, \bq} V_{\bq} \cda_{\bk + \bq}
c_{\bk}.
\eeq
In the above $\vol$ is the system volume, and $V_{\bq}$ is the disorder potential which obeys Gaussian
distribution, such that disorder average leads to
\[
\langle V_{\bq} V_{-\bq^{\prime}} \rangle_{\rm dis} = \dl_{\bq, \bq^{\prime}} \frac{\vol}{2\pi \nu_0 \tau}.
\]
Here $\nu_0$ is the electron density of states at the Fermi level, and $\tau$ is the elastic scattering
lifetime. The effect of impurity scattering can be taken into account perturbatively where the small
parameter is $1/(E_F \tau)$, $E_F$ being the Fermi energy. In this case the various correlation functions that
enter in the definition of the nonlinear current-current susceptibility
$\Pi_{\al \be \ga \dl}^{(3)}(\om_1, \om_2, \om_3)$ given by Eq.~(\ref{eq:2.21}) can be evaluated using diagrammatic
perturbation theory. The basic building block of such a calculation is the disorder averaged single electron
Green's function which is given by
\beq
\label{eq:4.2}
G^{-1}_{\bk} (i\om_n) = i \om_n - \ep_{\bk} + i/(2\tau) \sgn(\om_n),
\eeq
where $\sgn$ is the sign function.

The set of diagrams for computing $\Pi_{\al \be \ga \dl}^{(3)}(\om_1, \om_2, \om_3)$,
ignoring vertex corrections
for the moment, are given in Fig.~\ref{fig1}. They have been discussed earlier in the
literature, see. e.g.~\cite{Udina19,benfatto04}.
The solid lines indicate disorder averaged single electron
Green's function given by Eq.~(\ref{eq:4.2}), and photons indicated by wiggly lines.
The various current vertices involving $n= 1, \cdots, 4$ photons are given by Eq.~(\ref{eq:2.3}). The
diagram (i) represents the one-point function $C_{\al \be \ga \dl}^{(1p)}$. The diagram (ii) gives
the two-point function $C_{\al \be, \ga \dl}^{(2p,a)}(i\om_{2n} + i\om_{3n})$.
The diagram (iii) gives the two-point function $C_{\al \be \ga, \dl}^{(2p,b)}(i\om_{3n})$.
The diagram (iv) gives
$C_{\al, \be \ga \dl}^{(2p,c)}(i\om_{1n} + i\om_{2n} + i\om_{3n})$. The diagram (v) and that obtained by
interchanging the positions of the $(\ga, \om_2)$ and the $(\dl, \om_3)$ photons give the three-point function
$C_{\al \be, \ga, \dl}^{(3p,a)}(i\om_{2n}, i\om_{3n})$. The diagram (vi) and
that obtained by interchanging the positions of the $(\al, \om)$ and $(\be, \om_1)$
photons together give the three-point function
$C_{\al, \be, \ga \dl}^{(3p,b)}(i\om_{1n}, i\om_{2n} + i\om_{3n})$.
Finally, diagram (vii) and five others obtained by permuting the indices $(\be, \om_1)$, $(\ga, \om_2)$
and $(\dl, \om_3)$ give the four-point function
$C_{\al, \be, \ga, \dl}^{(4p)}(i\om_{1n}, i\om_{2n}, i\om_{3n})$.

In the following we consider only the contribution of the low-energy electrons, for which the wavevector
sum can be replaced by an angular integral around the Fermi surface followed by an energy integral,
\beq
\label{eq:4.2-bis}
(1/\vol) \sum_{\bk} \rightarrow \nu_0 \int_{-\infty}^{\infty} d \ep_{\bk} \oint_{FS} d \Om_k.
\eeq
In this approximation one can show that the three-point and four-point functions, as well as the vertex
correction terms, do not contribute. This is demonstrated in Appendix~\ref{appC}. Thus, we need to consider
only the one- and two-point functions.

We denote the various current vertices by
\begin{align}
&(v_{\bk})_{\al} \equiv \frac{\ptl \ep_{\bk}}{\ptl k_{\al}},
\quad \quad
(v_{\bk})_{\al \be} \equiv \frac{\ptl^2 \ep_{\bk}}{\ptl k_{\al} \ptl k_{\be}},
\nonumber \\
&(v_{\bk})_{\al \be \ga} \equiv \frac{\ptl^3 \ep_{\bk}}{\ptl k_{\al} \ptl k_{\be} \ptl k_{\ga}},
\quad
(v_{\bk})_{\al \be \ga \dl} \equiv \frac{\ptl^4 \ep_{\bk}}{\ptl k_{\al} \ptl k_{\be} \ptl k_{\ga} \ptl k_{\dl}}.
\nonumber
\end{align}
Using integration by parts, and setting boundary terms to zero we write $C_{\al \be \ga \dl}^{(1p)}$ as
\[
C_{\al \be \ga \dl}^{(1p)} = - \frac{1}{\be \vol} \sum_{\bk, \nu_n}
(v_{\bk})_{\al} (v_{\bk})_{\be \ga \dl} G_{\bk}^2(i\nu_n).
\]
The above term can be evaluated together with
$C_{\al, \be \ga \dl}^{(2p,c)}(i\om_{1n} + i\om_{2n} + i\om_{3n})$. We take the external photon
frequencies $(\om_{1n}, \om_{2n}, \om_{3n}) >0$, since the eventual analytic continuation is to be
performed from the upper complex frequency plane. The $\ep_{\bk}$ integral can be performed using
the method of contours. After analytic continuation we get
\begin{align}
\label{eq:4.3}
 &C_{\al \be \ga \dl}^{(1p)} + C_{\al, \be \ga \dl}^{(2p,c)}(\om_1 + \om_2 + \om_3 + i\eta)
 \nonumber \\
 &= \nu_0 \la (v_{\bk})_{\al} (v_{\bk})_{\be \ga \dl} \ra_{FS}
 \left[\frac{\om_1 + \om_2 + \om_3 }{\om_1 + \om_2 + \om_3  + i/\tau}\right]
 \end{align}
 Next, we consider the correlation functions of the type $(2p,a)$. Using integration by parts we get
 \begin{align}
 &C_{\al \be, \ga \dl}^{(2p,a)}(i\Om_n) =   - \frac{1}{\be \vol} \sum_{\bk, \nu_n}
(v_{\bk})_{\al} (v_{\bk})_{\be \ga \dl} G_{\bk}(i\nu_n)
\nonumber \\
&\times  G_{\bk}(i\nu_n + i\Om_n) - \frac{1}{\be \vol} \sum_{\bk, \nu_n}
(v_{\bk})_{\al} (v_{\bk})_{\be}  (v_{\bk})_{\ga \dl}
\nonumber \\
&\times [ G_{\bk}^2(i\nu_n)G_{\bk}(i\nu_n + i\Om_n) + G_{\bk}(i\nu_n)G_{\bk}^2(i\nu_n + i\Om_n) ].
\nonumber
\end{align}
In the above the second term can be set to zero since
\begin{align}
&\int_{-\infty}^{\infty} d \ep_{\bk}
[ G_{\bk}^2(i\nu_n)G_{\bk}(i\nu_n + i\Om_n)  +G_{\bk}(i\nu_n)
\nonumber \\
& \times  G_{\bk}^2(i\nu_n + i\Om_n) ]
=0.
\nonumber
\end{align}
For the same reason, after two integration by parts the correlation function $(2p,b)$ can be expressed as
\begin{align}
&C_{\al \be \ga, \ga \dl}^{(2p,b)}(i\Om_n) =   \frac{1}{\be \vol} \sum_{\bk, \nu_n}
(v_{\bk})_{\al} (v_{\bk})_{\be \ga \dl} G_{\bk}(i\nu_n)
\nonumber \\
&\times  G_{\bk}(i\nu_n + i\Om_n) + \cdots,
\nonumber
\end{align}
where the terms in the ellipsis can be set to zero after the energy integral. To each of the three terms
involving the correlation functions $(2p,a)$ the constant $C_{\al \be \ga \dl}^{(1p)}$ can be subtracted,
and to each of the three terms involving the correlation functions $(2p,b)$ the constant
$C_{\al \be \ga \dl}^{(1p)}$ can be added. This makes the frequency momentum sums in these
correlation functions fully convergent. Eventually we get
\begin{align}
\label{eq:4.4}
 &C_{\al \be, \ga \dl}^{(2p,a)}(\om_2 + \om_3 + i\eta) - C_{\al \be \ga \dl}^{(1p)}
 \nonumber \\
 &= -\nu_0 \la (v_{\bk})_{\al} (v_{\bk})_{\be \ga \dl} \ra_{FS}
 \left[\frac{\om_2 + \om_3 }{\om_2 + \om_3  + i/\tau}\right],
 \end{align}
and
\begin{align}
\label{eq:4.5}
 &C_{\al \be \ga, \dl}^{(2p,b)}(\om_3 + i\eta) + C_{\al \be \ga \dl}^{(1p)}
 \nonumber \\
 &= \nu_0 \la (v_{\bk})_{\al} (v_{\bk})_{\be \ga \dl} \ra_{FS}
 \left[\frac{\om_3 }{\om_3  + i/\tau}\right].
 \end{align}
 Finally, using Eqs.~(\ref{eq:4.3}), (\ref{eq:4.4}), and (\ref{eq:4.5}) the nonlinear current
 kernel, defined in Eq.~(\ref{eq:2.21}), of a Drude metal is given by
 \begin{align}
 \label{eq:4.6}
 &\Pi_{\al \be \ga \dl}^{(3)}(\om_1, \om_2, \om_3) =
 - \frac{e^4 \nu_0 \la (v_{\bk})_{\al} (v_{\bk})_{\be \ga \dl} \ra_{FS}}{6}
 \nonumber \\
 &\left[
 \frac{\om_1 + \om_2 + \om_3 }{\om_1 + \om_2 + \om_3  + i/\tau}
 - \frac{\om_1 + \om_2 }{\om_1 + \om_2  + i/\tau} - \frac{\om_2 + \om_3 }{\om_2 + \om_3  + i/\tau}
 \right. \nonumber \\
 & \left.
 - \frac{\om_3 + \om_1 }{\om_3 + \om_1  + i/\tau} + \frac{\om_1 }{\om_1  + i/\tau}
 + \frac{\om_2 }{\om_2  + i/\tau} + \frac{\om_3 }{\om_3  + i/\tau} \right].
 \end{align}
 Note, this result is consistent with the constraints imposed in Eq.~(\ref{eq:3.1}) by
 gauge invariance, since $\Pi_{\al \be \ga \dl}^{(3)}(\om_1, \om_2, \om_3) \sim \om_1 \om_2 \om_3$
 for $(\om_1, \om_2, \om_3) \rightarrow 0$. 
 Finally, the nonlinear conductivity can be readily obtained from the above by using
 Eq.~(\ref{eq:2.21-bis2}). Note also, the above Eq.~(\ref{eq:4.6}) is relevant as a low energy asymptotic
 behavior also for non-superconducting
 symmetry broken states such as nematic and density wave phases, as long as such phases stay 
 metallic.

 Alternatively, the above result can be derived by considering the manifestly gauge invariant
 susceptibility
 \begin{align}
 \label{eq:4.7}
 &\Pi_{\al \be \ga \dl}^{(3)}(\om_1, \om_2, \om_3)_{\rm inv} \equiv
 \Pi_{\al \be \ga \dl}^{(3)}(\om_1, \om_2, \om_3)
 \nonumber \\
 &- \Pi_{\al \be \ga \dl}^{(3)}(0, \om_2, \om_3)
 - \Pi_{\al \be \ga \dl}^{(3)}(\om_1, 0, \om_3) - \Pi_{\al \be \ga \dl}^{(3)}(\om_1, \om_2, 0)
 \nonumber \\
 &+ \Pi_{\al \be \ga \dl}^{(3)}(0, 0, \om_3) + \Pi_{\al \be \ga \dl}^{(3)}(0, \om_2, 0)
 + \Pi_{\al \be \ga \dl}^{(3)}(\om_1, 0, 0)
 \nonumber \\
 &- \Pi_{\al \be \ga \dl}^{(3)}(0, 0, 0).
 \end{align}
 In the above zeroes have been added and subtracted using the gauge invariance condition
 of Eq.~(\ref{eq:3.1}).
 It is simple to check that, for the gauge invariant quantity, the correlation functions $(1p)$,
 $(2p,a)$ and $(2p,b)$ vanish identically and only the correlation function $(2p,c)$ contribute.

 Next, we show that the result expressed in Eq.~(\ref{eq:4.6}) is consistent with the sum rule discussed
 in Section~\ref{subsec:3.2}. It is simple to perform the three frequency integrals in Eq.~(\ref{eq:3.15}),
 and the left hand side gives
\begin{align}
&\int_{-\infty}^{\infty} \int_{-\infty}^{\infty}
\int_{-\infty}^{\infty}  \frac{d \om_1 d \om_2 d \om_3}{(\pi)^3}
\si_{\al \be \ga \dl}^{(3)}(\om_1, \om_2, \om_3)
\nonumber\\
&= \frac{e^4}{6} \nu_0 \la (v_{\bk})_{\al} (v_{\bk})_{\be \ga \dl} \ra_{FS}.
\nonumber
\end{align}
Simultaneously the right hand side can be written as
\begin{align}
&\frac{e^4}{6 \vol} \sum_{\bk} (v_{\bk})_{\al \be \ga \dl} n_F(\ep_{\bk})
= - \frac{e^4}{6 \vol} \sum_{\bk} (v_{\bk})_{\al} (v_{\bk})_{\be \ga \dl} n_F^{\prime}(\ep_{\bk})
\nonumber \\
&= \frac{e^4}{6} \nu_0 \la (v_{\bk})_{\al} (v_{\bk})_{\be \ga \dl} \ra_{FS},
\nonumber
\end{align}
where $n_F(\ep_{\bk})$ is the Fermi function, and prime denotes its derivative with respect to
energy. Thus, the sum rule is indeed verified.

\section{Third harmonic generation}
\label{sec5}

In third harmonic generation the system is perturbed by a monochromatic light pulse of frequency $\nu$,
and the nonlinear response at frequency $3\nu$ is studied. Below we describe the theory of the third harmonic
signal of a Drude metal.

We consider the perturbing electric field to be of the form $\bE(t) = \bE_{in} e^{- i \nu t}$,
which in Fourier space is $\bE (\om) = 2 \pi \bE_{in} \dl (\om - \nu)$. Using Eq.~(\ref{eq:2.21-bis}) we
find that the third harmonic current density is given by
\beq
\label{eq:5.1}
\left(j_{TH}\right)_{\al}(t) = \left[ \si_{\al \be \ga \dl}^{(3)}(\nu, \nu, \nu)
E_{in, \be} E_{in, \ga} E_{in, \dl} \right] e^{-3 i \nu t}.
\eeq
In frequency space this corresponds to
\beq
\label{eq:5.1-bis}
\left(j_{TH}\right)_{\al}(\om) = 2\pi \dl(\om - 3\nu)
\si_{\al \be \ga \dl}^{(3)}(\nu, \nu, \nu)
E_{in, \be} E_{in, \ga} E_{in, \dl}.
\eeq
In turn, the above current can be associated with a third harmonic electric field
$\left(E_{TH}\right)_{\al}(\om)$, where $\om = 3\nu$. The computation of the third harmonic
field $\bE_{TH}$ from the third harmonic current $\bj_{TH}$ involves solving nonlinear Maxwell
equations within the material with appropriate boundary conditions,
see e.g., chapter 2 of Ref.~\cite{Boyd}.

Next, we discuss how the third harmonic response depends upon the pump and the probe
polarizations~\cite{Matsunaga17,Cea18}.
We consider the pump electric field to be $\bE_{in} = E_0 (\hat{x} \cos \phi  + \hat{y} \sin \phi )$.  For
a centrosymmetric system the third harmonic currents generated are (suppressing frequency indices)
\begin{align}
\left(j_{TH}\right)_{x} &\propto \si_{x x x x}^{(3)} \cos^3 \phi +
3 \si_{x x y y}^{(3)} \cos \phi \sin^2 \phi,
\nonumber\\
\left(j_{TH}\right)_{y} &\propto \si_{y y y y}^{(3)} \sin^3 \phi +
3 \si_{y y x x}^{(3)} \sin \phi \cos^2 \phi.
\nonumber
\end{align}
In the above we used the property
$\si_{x x y y}^{(3)}(\nu,\nu,\nu) = \si_{x y x y}^{(3)}(\nu,\nu,\nu) = \si_{x y y x}^{(3)}(\nu,\nu,\nu)$, and
so on. Furthermore, for a system with tetragonal or higher symmetry
$\si_{x x x x}^{(3)} = \si_{y y y y}^{(3)}$, and
$\si_{x x y y}^{(3)} = \si_{y y x x}^{(3)}$. Then, depending on whether the probe polarization is parallel
or perpendicular to the pump polarization, the third harmonic responses are
\begin{subequations}
\label{eq:5.7}
\begin{align}
\left(j_{TH}\right)_{\parallel}(\om) &=
2\pi \dl(\om - 3\nu) E_0^3 \left[ A(\nu) + 2 B(\nu) \sin^2(2\phi) \right],
\label{eq:5.7a}\\
\left(j_{TH}\right)_{\perp}(\om) &= 2\pi \dl(\om - 3\nu) E_0^3 B(\nu) \sin (4 \phi),
\label{eq:5.7b}
\end{align}
\end{subequations}
respectively, where $A(\nu) = \si_{x x x x}^{(3)}(\nu,\nu,\nu)$, and
$B(\nu) = [ 3 \si_{x x y y}^{(3)}(\nu,\nu,\nu) - \si_{x x x x}^{(3)}(\nu,\nu,\nu)]/4$.

\section{Terahertz Kerr effect}
\label{sec6}

We consider measurement of electro-optical Kerr effect that
involves perturbing the system with a pump
electric field $\bE_{pp}(t)$ in the  terahertz range, and then to probe
the system with a field $\bE_{pb}(t)$ which is at a much higher frequency, typically in the
optical range. The instantaneous
Kerr signal is the response of the system which is proportional to the
square of the pump field $\bE_{pp}(t)^2$.

In this setup the system is probed in the presence of the pump, and therefore the total nonlinear
current is proportional to $(\bE_{pp}(t) + \bE_{pb}(t))^3$. In this expansion there are three
terms that are of the type $\bE_{pp}(t)^2 \bE_{pb}(t)$, which contribute to the Kerr signal. It
is simple to check that these three terms contribute equally. Thus, using Eq.~\ref{eq:2.21-bis}
the nonlinear current associated with Kerr effect can be expressed as
\begin{align}
\label{eq:6.1}
&\left(j_{NL}\right)_{\al}(\om)  = 3 \int_{-\infty}^{\infty} \int_{-\infty}^{\infty}
\int_{-\infty}^{\infty}  \frac{d \om_1 d \om_2 d \om_3}{(2\pi)^2}
\nonumber\\
&\times
\dl(\om - \om_1 - \om_2 - \om_3)  E_{pb,\be}(\om_1) E_{pp, \ga}(\om_2) E_{pp, \dl}(\om_3)
\nonumber\\
&\times \si_{\al \be \ga \dl}^{(3)}(\om_1, \om_2, \om_3).
\end{align}
In the above $(\al, \be)$ are fixed by the probe polarization, and $(\ga, \dl)$ are fixed by the
pump polarization.
Since the pump frequencies $(\om_2, \om_3)$ are much smaller compared to the typical probe
frequency $\om_1$, we can Taylor expand
\beq
\label{eq:6.1-bis}
\si_{\al \be \ga \dl}^{(3)}(\om_1, \om_2, \om_3)
= \si_{\al \be \ga \dl}^{(3)}(\om_1, 0, 0) + \cdots.
\eeq
The first term above gives the instantaneous Kerr response, while the ellipsis denote terms that
lead to retarded Kerr response. Since in a typical pump-probe setup the overall nonlinear response is also
accompanied by an out of equilibrium relaxational dynamics, it is nontrivial to distinguish the retarded
Kerr response from the nonequilibrium component.
Note, in setups where both the pump and the probe frequencies are in the terahertz range,
the retarded Kerr response can dominate the overall nonlinear response.

\begin{figure}[!!t]
\begin{center}
\includegraphics[width=7.0cm,trim=0 0 0 0]{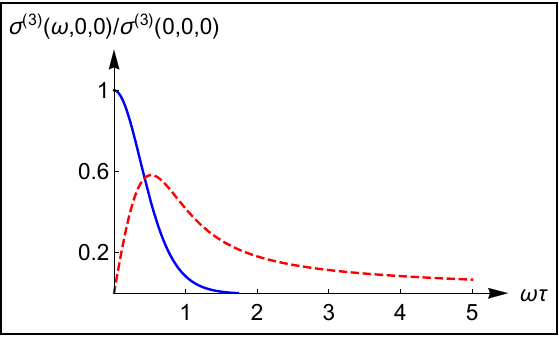}
\caption{
(color online) The real (solid, blue) and the imaginary (dashed, red) parts of the nonlinear conductivity
(polarization indices suppressed for clarity)
associated with the Kerr signal as a function of probe frequency $\om$,
see Eqs.~(\ref{eq:6.4}) and~(\ref{eq:6.5}). $\tau$ is the elastic scattering
lifetime of the electrons.}
\label{fig3}
\end{center}
\end{figure}

Keeping only the instantaneous Kerr component in Eq.~(\ref{eq:6.1}),
the nonlinear current in the time domain can be written as
\begin{align}
\label{eq:6.2}
\left(j_{NL}\right)_{\al}(t) &= \int_{-\infty}^{\infty} \frac{d \om}{2\pi}
\left[3 \si_{\al \be \ga \dl}^{(3)}(\om, 0, 0) E_{pp, \ga}(t) E_{pp, \dl}(t) \right]
\nonumber \\
&\times E_{pb,\be}(\om) e^{-i \om t}.
\end{align}
This expression is to be compared with the linear current response to the probe field which is
\[
\left(j_{L}\right)_{\al}(t) = \int_{-\infty}^{\infty} \frac{d \om}{2\pi}
\si_{\al \be}^{(1)}(\om) E_{pb,\be}(\om) e^{-i \om t}.
\]
Since the total current in the presence of the pump is $\bj_L + \bj_{NL}$, the
instantaneous Kerr response
can be expressed as a time and frequency dependent shift of the linear conductivity tensor
$\si_{\al \be}^{(1)}(\om) \rightarrow \si_{\al \be}^{(1)}(\om)
+ \Delta \si_{\al \be}^{(1)}(\om,t)$ that is due to the presence of the pump, where
\beq
\label{eq:6.3}
\Delta \si_{\al \be}^{(1)}(\om,t) = 3  \si_{\al \be \ga \dl}^{(3)}(\om, 0, 0)
E_{pp, \ga}(t) E_{pp, \dl}(t).
\eeq
Thus, if the Kerr signal is measured as a change in the reflectivity $R$, then
\begin{align}
\label{eq:6.4}
\left( \Delta R \right)_{\al \be} &= 3 \left[
\left( \frac{\ptl R}{\ptl \si_1} \right) {\rm Re} \si_{\al \be \ga \dl}^{(3)}(\om, 0, 0)
\right. \nonumber \\
&+ \left.  \left( \frac{\ptl R}{\ptl \si_2} \right) {\rm Im} \si_{\al \be \ga \dl}^{(3)}(\om, 0, 0)
\right] E_{pp, \ga}(t) E_{pp, \dl}(t).
\end{align}
Here $\si_{1,2}$ are the real and imaginary parts of the complex linear conductivity, respectively.
From Eq.~(\ref{eq:4.6}) the relevant nonlinear conductivity for a Drude metal is given by
\begin{align}
\label{eq:6.5}
\si_{\al \be \ga \dl}^{(3)}(\om, 0, 0)  &=
\nu_0 e^4 \tau^3 \la (v_{\bk})_{\al} (v_{\bk})_{\be \ga \dl} \ra_{FS}
\nonumber\\
&\times
\frac{(3 - 3i \om \tau - \om^2 \tau^2)}{3 (1 - i \om \tau)^3}.
\end{align}
The real and imaginary parts of the above are shown in Fig.~\ref{fig3} as a function of the probe
frequency. Note, in the frequency range $\om \sim 1/\tau$, both the real and the imaginary parts of
$\si_{\al \be \ga \dl}^{(3)}(\om, 0, 0)$ contribute to the instantaneous Kerr response.
%====================
%====================

Next, we discuss how the instantaneous Kerr signal depends upon the pump and the probe
polarizations~\cite{Katsumi18,Katsumi20}. In a typical reflectivity measurement of the Kerr signal the
probe field is incident normally on the surface of the system. The quantity of interest
is the change in the reflectivity $( \Delta R )_{\al \al}$, where $\hat{\al}$ denotes the direction
of the probe polarization. We characterize $\hat{\al}$ by an angle $\phi_{pb}$ such that
$\hat{\al} = \hat{x} \cos \phi_{pb} + \hat{y} \sin \phi_{pb}$. Then,
\begin{align}
\label{eq:6.6}
2 \left( \Delta R \right)_{\al \al} &= [(\Delta R)_{xx} + (\Delta R)_{yy}]
+ \cos (2\phi_{pb}) [(\Delta R)_{xx}
\nonumber \\
&- (\Delta R)_{yy}] + \sin (2\phi_{pb}) [(\Delta R)_{xy} + (\Delta R)_{yx}],
\end{align}
which follows from standard transformation of a rank two tensor under rotation.

Next, we take the pump polarization to be also in the $xy$-plane, making an 
angle $\phi_{pp}$ with $\hat{x}$.
Using Eq.~(\ref{eq:6.3}), for a centrosymmetric system  we get
\begin{align}
\label{eq:6.7}
\Delta \si_{xx}^{(1)}(\om,t) &= 3  E^2_{pp}(t) \left[\si_{xxxx}^{(3)}(\om, 0, 0) \cos^2 \phi_{pp} \right.
\nonumber\\
&\left. + \si_{xxyy}^{(3)}(\om, 0, 0) \sin^2 \phi_{pp} \right].
\end{align}
One can write similar expressions for $\Delta \si_{yy}^{(1)}(\om,t)$, $\Delta \si_{xy}^{(1)}(\om,t)$ and
$\Delta \si_{yx}^{(1)}(\om,t)$. Using Eqs.~(\ref{eq:6.4}), (\ref{eq:6.6}) and (\ref{eq:6.7}), for a system with
tetragonal or higher symmetry, the polarization dependencies of the instantaneous Kerr response can be written as
\begin{align}
\label{eq:6.8}
&\left( \Delta R \right)_{\al \al}(\om, t) =
3 E^2_{pp}(t) \left[ K_{A_{1g}}(\om) + \cos (2\phi_{pp}) \cos (2\phi_{pb})
\right. \nonumber\\
&\left. \times K_{B_{1g}}(\om) + \sin (2\phi_{pp}) \sin (2\phi_{pb}) K_{B_{2g}}(\om) \right],
\end{align}
where
\begin{subequations}
\label{eq:6.9}
\begin{align}
K_{A_{1g}}(\om)  &\equiv \left( \frac{\ptl R}{\ptl \si_1} \right)_{xx} {\rm Re} \si^{(3)}_{A_{1g}}(\om)
+ \left( \frac{\ptl R}{\ptl \si_2} \right)_{xx} {\rm Im} \si^{(3)}_{A_{1g}}(\om),
\label{eq:6.9a}\\
K_{B_{1g}}(\om)  &\equiv \left( \frac{\ptl R}{\ptl \si_1} \right)_{xx} {\rm Re} \si^{(3)}_{B_{1g}}(\om)
+ \left( \frac{\ptl R}{\ptl \si_2} \right)_{xx} {\rm Im} \si^{(3)}_{B_{1g}}(\om),
\label{eq:6.9b}\\
K_{B_{2g}}(\om)  &\equiv \left( \frac{\ptl R}{\ptl \si_1} \right)_{x^{\prime}x^{\prime}} \! \! \!
{\rm Re} \si^{(3)}_{B_{2g}}(\om)
+ \left( \frac{\ptl R}{\ptl \si_2} \right)_{x^{\prime}x^{\prime}} \! \!  \! {\rm Im} \si^{(3)}_{B_{2g}}(\om),
\label{eq:6.9c}
\end{align}
\end{subequations}
with $x^{\prime} \equiv (x+y)/\sqrt{2}$, and
\begin{subequations}
\label{eq:6.10}
\begin{align}
\si^{(3)}_{A_{1g}}(\om) &\equiv [ \si_{xxxx}^{(3)}(\om, 0, 0) + \si_{xxyy}^{(3)}(\om, 0, 0)]/2,
\label{eq:6.10a}\\
\si^{(3)}_{B_{1g}}(\om) &\equiv [ \si_{xxxx}^{(3)}(\om, 0, 0) - \si_{xxyy}^{(3)}(\om, 0, 0)]/2,
\label{eq:6.10b}\\
\si^{(3)}_{B_{2g}}(\om) &\equiv \si_{xyxy}^{(3)}(\om, 0, 0).
\label{eq:6.10c}
\end{align}
\end{subequations}
For a tight binding model with nearest and next nearest neighbor hoppings $t$ and $t^{\prime}$,
respectively, we expect $\si_{xxxx}^{(3)} \sim t^2$, $\si_{xxyy}^{(3)} \sim t t^{\prime}$, and
$\si_{xyxy}^{(3)} \sim t t^{\prime}$.

\section{Conclusion}
\label{sec7}

To summarize, in this work we reviewed the field theoretical framework to compute
the nonlinear electro-optical
responses of centrosymmetric electronic systems.
The formalism itself, starting from standard time dependent
perturbation theory,  is described in section~\ref{sec2}. We showed that the nonlinear current can be
expressed in terms of a sum of several response functions that are causal. However, the
response functions do not obey Wick's theorem and, therefore, they cannot be computed directly
using perturbative field theory methods. Consequently, we associated each response function with a
corresponding
imaginary time ordered correlation function that can be factorized by means of Wick's theorem.
Using the Lehmann representation we
showed that the correlation functions, analytically continued to real frequencies, map on to the
response functions. This mapping is exact to all orders in the interaction strength, and it also 
holds if the electrons are in a random potential due to the presence of impurities.
This mapping leads to formal expressions for the
nonlinear current $\left(j_{NL}\right)_{\al}(\om)$ in terms of the nonlinear current
kernel $\Pi_{\al \be \ga \dl}^{(3)}(\om_1, \om_2, \om_3)$, see Eq.~(\ref{eq:2.20}),
or equivalently in terms of the nonlinear
conductivity $\si_{\al \be \ga \dl}^{(3)}(\om_1, \om_2, \om_3)$,
see Eqs.~~(\ref{eq:2.21-bis}) and (\ref{eq:2.21-bis2}).
The nonlinear kernel and the conductivity are rank-four tensors, and the
indices $(\al, \be, \ga, \dl)$ denote spatial directions (photon polarizations). The arguments
$(\om_1, \om_2, \om_3)$ denote the incoming photon frequencies,
with $\om = \om_1 + \om_2 + \om_3$.
In section~\ref{sec3} we showed that the nonlinear kernel satisfy certain constraints,
namely that, for non superconducting phases, it vanishes
if either one of the three incoming photon frequencies is set to zero, see Eq.~(\ref{eq:3.1}).
These constraints ensure that there are no spurious divergences in the static limit, and that
the static nonlinear responses are finite.
We also showed that the nonlinear conductivity satisfies a generalized $f$-sum rule.
Thus, the nonlinear conductivity integrated over the three external
frequencies is a constant that depends only on 
the electronic spectrum, and is independent of the electron
lifetime, see Eq.~(\ref{eq:3.15}).
The constraints and the sum rule are consequences of gauge
invariance, or particle number conservation. In section~\ref{sec4} we applied the theory to compute the
gauge invariant nonlinear kernel for a Drude metal, i.e., a system of noninteracting electrons in the
presence of weak disorder, see Eq.~(\ref{eq:4.6}). As special cases of the generalized response, we derived
expressions for the third harmonic and the instantaneous terahertz
Kerr signals in sections~\ref{sec5} and~\ref{sec6}, respectively.

The field theoretic formalism reviewed here is ideal to include effects of
electron-electron interaction. It particular, it can be used to describe the nonlinear
electro-optical responses of broken symmetry states of electronic systems, such as
superconductors, nematic states and density waves. In such systems there will be
contributions to the current correlation functions due to coupling of the carriers with
the collective modes of the broken symmetry states.  Furthermore, effects of inelastic
lifetimes of the electrons, and their temperature dependencies can be addressed using the
current formalism.

\acknowledgments
The author is grateful to Yann Gallais and Ryo Shimano for illuminating discussions.
\appendix
\section{}
\label{appA}
In this Appendix we provide the technical details of the results obtained in Section~\ref{sec2}. In particular,
we show how the response functions can be mapped on to the correlation functions
by comparing their expressions in the Lehmann basis.

\subsection{Two-point functions}
The structure of the two-point functions is well-known from linear response theory, and it has been discussed
in standard textbooks. Here we discuss it for the sake of completeness.

Using Lehmann representation the real time response function $R_{\al \be, \ga \dl}^{(2p,a)}(t, t_1)$,
defined in Eq.~(\ref{eq:2.10b}), can be written as
\begin{align}
\label{eq:A1}
&R_{\al \be, \ga \dl}^{(2p,a)}(t, t_1) = -i \theta(t - t_1) \frac{e^{-\be E_n}}{Z}
\left[ \left(\hv_{\al \be}\right)_{nm}  \left(\hv_{\ga \dl}\right)_{mn}
\right. \nonumber\\
&\times e^{i E_{nm}(t-t_1)}
-\left. \left(\hv_{\ga \dl}\right)_{nm} \left(\hv_{\al \be}\right)_{mn} e^{-i E_{nm}(t-t_1)} \right],
\end{align}
where $(\hO)_{nm} \equiv \la n | \hO | m\ra$, and $E_{nm} \equiv E_n - E_m$.
Also, summation over repeated Lehmann basis indices $(n, m)$ is implied. Its Fourier transform
$R_{\al \be, \ga \dl}^{(2p,a)}(\Om)$, defined in Eq.~(\ref{eq:2.13a}), is given by
\begin{align}
\label{eq:A2}
R_{\al \be, \ga \dl}^{(2p,a)}(\Om) = \frac{1}{Z} \left( e^{-\be E_n} - e^{-\be E_m} \right)
\frac{\left(\hv_{\al \be}\right)_{nm}  \left(\hv_{\ga \dl}\right)_{mn}}
{\Om + i\eta + E_{nm}}.
\end{align}

Next we express the imaginary time ordered correlation function $C_{\al \be, \ga \dl}^{(2p,a)}(\tau, \tau_1)$
given by Eq.~(\ref{eq:2.15a}). We get
\begin{align}
\label{eq:A3}
&C_{\al \be, \ga \dl}^{(2p,a)}(\tau, \tau_1) = - \frac{e^{-\be E_n}}{Z}
\left[ \theta(\tau - \tau_1) \left(\hv_{\al \be}\right)_{nm}  \left(\hv_{\ga \dl}\right)_{mn}
\right. \nonumber\\
&\times e^{E_{nm}(\tau -\tau_1)}
+\left. \theta(\tau_1 - \tau) \left(\hv_{\ga \dl}\right)_{nm} \left(\hv_{\al \be}\right)_{mn}
e^{E_{nm}(\tau_1 - \tau)} \right].
\end{align}
Note, as a function of $s_1 \equiv \tau - \tau_1$, the correlation function satisfies bosonic periodicity
$C_{\al \be, \ga \dl}^{(2p,a)}(s < 0) = C_{\al \be, \ga \dl}^{(2p,a)}(s+ \be > 0)$. This property, unique
to two-point functions, considerably simplifies the structure of the Fourier transform
$C_{\al \be, \ga \dl}^{(2p,a)}(i \Om_{1n}) $ that is needed for the mapping.
From its definition in Eq.~(\ref{eq:2.16a}) we get
\begin{align}
\label{eq:A4}
C_{\al \be, \ga \dl}^{(2p,a)}(i\Om) = \frac{1}{Z} \left( e^{-\be E_n} - e^{-\be E_m} \right)
\frac{\left(\hv_{\al \be}\right)_{nm}  \left(\hv_{\ga \dl}\right)_{mn}}
{i\Om + E_{nm}}.
\end{align}
Thus, comparing Eqs.~(\ref{eq:A2}) and Eq.~(\ref{eq:A4}) we conclude
\[
C_{\al \be, \ga \dl}^{(2p,a)}(i \Om_{n} \rightarrow \Om + i\eta)
= R_{\al \be, \ga \dl}^{(2p,a)}(\Om),
\]
which is Eq.~(\ref{eq:2.17a}) in Section~\ref{sec2}.

The structures of the other two 2-point functions denoted $(2p,b)$ and $(2p,c)$ are identical to one above
for $(2p,a)$. Consequently, Eqs.~(\ref{eq:2.17b}) and (\ref{eq:2.17c}) in Section~\ref{sec2} are obvious.

\subsection{Three-point functions}
Here we compute the three-point functions in the Lehmann basis. We define the time variables
$u_1 \equiv t -t_1$ and $u_2 \equiv t -t_2$. For brevity, we also define the matrix elements
$\left(X_1 \right)_{nmp} \equiv (\hv_{\al \be})_{nm} (\hv_{\ga})_{mp} (\hv_{\dl})_{pn}$ and
$\left(X_2 \right)_{nmp} \equiv (\hv_{\al \be})_{nm} (\hv_{\dl})_{mp} (\hv_{\ga})_{pn}$, without implying
summation over indices $(n, m, p)$. In terms of these $R_{\al \be, \ga, \dl}^{(3p,a)}(t, t_1, t_2)$, defined
in Eq.~(\ref{eq:2.10e}) can be written as
\begin{align}
\label{eq:A5}
&R_{\al \be, \ga, \dl}^{(3p,a)}(u_1, u_2) = \frac{1}{Z} \left[ \frac{1}{2} \theta(u_1) \theta(u_2)
\left\{ \left(X_1 \right)_{nmp} e^{i E_{pm} u_1}
\right. \right. \nonumber\\
&\times \left. e^{i E_{np} u_2}
+ \left(X_2 \right)_{nmp} e^{i E_{np} u_1} e^{i E_{pm} u_2}  \right\} e^{-\be E_p}
\nonumber\\
&- \theta(u_1) \theta(u_2 - u_1) \left\{
\left(X_1 \right)_{nmp} e^{i E_{pm} u_1} e^{i E_{np} u_2} e^{-\be E_n}
\right. \nonumber\\
&+ \left. \left. \left(X_2 \right)_{nmp} e^{i E_{np} u_1} e^{i E_{pm} u_2} e^{-\be E_m}
\right\} \right].
\end{align}
Its Fourier transform $R_{\al \be, \ga, \dl}^{(3p,a)}(\Om_1, \Om_2)$, given by Eq.~(\ref{eq:2.13d}), is
\begin{align}
\label{eq:A6}
&R_{\al \be, \ga, \dl}^{(3p,a)}(\Om_1, \Om_2) = - \frac{1}{2Z(\Om_{12} + E_{nm} + 2i\eta)}
\nonumber\\
&\times \left[ \left(X_1 \right)_{nmp}  \left\{
\frac{e^{-\be E_p}}{\Om_1 + i\eta + E_{pm}} + \frac{e^{-\be E_p} - 2e^{-\be E_n}}
{\Om_2 + i\eta + E_{np}} \right\} \right. \nonumber\\
&+ \left(X_2 \right)_{nmp} \left. \left\{
\frac{e^{-\be E_p}}{\Om_1 + i\eta + E_{np}} + \frac{e^{-\be E_p} - 2e^{-\be E_m}}
{\Om_2 + i\eta + E_{pm}} \right\}  \right],
\end{align}
where $\Om_{12} \equiv \Om_1 + \Om_2$. Thus, the symmetric combination
\begin{align}
\label{eq:A7}
&R_{\al \be, \ga, \dl}^{(3p,a)}(\Om_1, \Om_2) + R_{\al \be, \dl, \ga}^{(3p,a)}(\Om_2, \Om_1)
= \nonumber\\
&- \frac{1}{Z(\Om_{12} + E_{nm} + 2i\eta)}
\nonumber\\
&\times \left[ \left(X_1 \right)_{nmp}  \left\{
\frac{e^{-\be E_p} - e^{-\be E_m}}{\Om_1 + i\eta + E_{pm}} + \frac{e^{-\be E_p} - e^{-\be E_n}}
{\Om_2 + i\eta + E_{np}} \right\} \right. \nonumber\\
&+ \left(X_2 \right)_{nmp} \left. \left\{
\frac{e^{-\be E_p} - e^{-\be E_m}}{\Om_2 + i\eta + E_{pm}} + \frac{e^{-\be E_p} - e^{-\be E_n}}
{\Om_1 + i\eta + E_{np}} \right\}  \right].
\end{align}

Next we evaluate the imaginary time ordered correlation function
$C_{\al \be, \ga, \dl}^{(3p,a)}(\tau, \tau_1,\tau_2)$ defined by Eq.~(\ref{eq:2.15d}). We get
\begin{align}
\label{eq:A8}
&C_{\al \be, \ga, \dl}^{(3p,a)}(\tau, \tau_1,\tau_2) = \frac{1}{Z} \left[
\left(X_1 \right)_{nmp}  e^{\tau E_{nm} + \tau_1 E_{mp} + \tau_2 E_{pn}}  \right.
\nonumber\\
&\left\{ \theta(\tau - \tau_1) \theta(\tau_1 - \tau_2) e^{-\be E_n}
+ \theta(\tau_1 - \tau_2) \theta(\tau_2 - \tau) e^{-\be E_m} \right.  \nonumber\\
&+ \left.
\theta(\tau_2 - \tau) \theta(\tau - \tau_1) e^{-\be E_p} \right\} \nonumber\\
&+ \left(X_2 \right)_{nmp}  e^{\tau E_{nm} + \tau_1 E_{pn} + \tau_2 E_{mp}}
\left\{ \theta(\tau_2 - \tau_1) \theta(\tau_1 - \tau) e^{-\be E_m} \right.
\nonumber\\
&+ \left. \left. \theta(\tau_1 - \tau) \theta(\tau - \tau_2) e^{-\be E_p}
+ \theta(\tau - \tau_2) \theta(\tau_2 - \tau_1) e^{-\be E_n} \right\} \right].
\end{align}
Note, in principle $C_{\al \be, \ga, \dl}^{(3p,a)}(\tau, \tau_1,\tau_2)$ can be expressed as a function of
only two variables $s_1 \equiv \tau - \tau_1$ and $s_2 \equiv \tau - \tau_2$. However, for
$(s_1 < 0, s_2 >0)$
$C_{\al \be, \ga, \dl}^{(3p,a)}(s_1, s_2) \neq C_{\al \be, \ga, \dl}^{(3p,a)}(s_1 + \be, s_2)$,
and  for $(s_1 > 0, s_2 <0)$,
$C_{\al \be, \ga, \dl}^{(3p,a)}(s_1, s_2) \neq C_{\al \be, \ga, \dl}^{(3p,a)}(s_1, s_2 + \be)$.
In other words, the property of periodicity is lost for the three-point functions, and therefore care has
to be taken in order to define the Fourier transform that is needed to map the response function with the
correlation function. The suitable quantity, $C_{\al \be, \ga, \dl}^{(3p,a)}(i \Om_{1n}, i \Om_{2n})$
is defined in Eq.~(\ref{eq:2.16d}). In the Lehmann representation this takes the form
\begin{align}
\label{eq:A9}
&C_{\al \be, \ga, \dl}^{(3p,a)}(i \Om_{1n}, i \Om_{2n}) = \frac{\left(X_1 \right)_{nmp}}{Z}
\left( I_1 e^{-\be E_n} + I_2 e^{-\be E_m} \right. \nonumber\\
&+ \left. I_3 e^{-\be E_p} \right)     + \frac{\left(X_2 \right)_{nmp}}{Z}
\left( I_4 e^{-\be E_p} + I_5 e^{-\be E_n} \right. \nonumber\\
&+ \left. I_6 e^{-\be E_m} \right),
\end{align}
where the integrals $I_1, \cdots, I_6$ are given by
\begin{widetext}
\begin{subequations}
\label{eq:A10}
\begin{align}
I_1 &= \frac{1}{\be} \int_0^{\be} d\tau \exp[\tau(i\Om_{12n} + E_{nm})]
\int_0^{\tau} d\tau_1 \exp[\tau_1(E_{mp} - i\Om_{1n})]
\int_0^{\tau_1} d\tau_2 \exp[\tau_2(E_{pn} - i\Om_{2n})] \nonumber\\
&= \frac{1}{(i\Om_{12n} + E_{nm})(i\Om_{2n} + E_{np})}
- \frac{e^{\be E_{np}} -1}{\be (i\Om_{2n} + E_{np})^2(i\Om_{1n} + E_{pm})}
+  \frac{e^{\be E_{nm}} -1}{\be (i\Om_{12n} + E_{nm})^2(i\Om_{1n} + E_{pm})},
\label{eq:A10a} \\
I_2 &= \frac{1}{\be} \int_0^{\be} d\tau_1 \exp[\tau_1(E_{mp}- i\Om_{1n})]
\int_0^{\tau_1} d\tau_2 \exp[\tau_2(E_{pn} - i\Om_{2n})]
\int_0^{\tau_2} d\tau \exp[\tau(i\Om_{12n} + E_{nm})] \nonumber\\
&= \frac{1}{(i\Om_{12n} + E_{nm})(i\Om_{1n} + E_{pm})}
- \frac{e^{\be E_{mn}} -1}{\be (i\Om_{2n} + E_{np})(i\Om_{12n} + E_{nm})^2}
+  \frac{e^{\be E_{mp}} -1}{\be (i\Om_{1n} + E_{pm})^2(i\Om_{2n} + E_{np})},
\label{eq:A10b} \\
I_3 &= \frac{1}{\be} \int_0^{\be} d\tau_2 \exp[\tau_2(E_{pn} - i\Om_{2n})]
\int_0^{\tau_2} d\tau \exp[\tau(i\Om_{12n} + E_{nm})]
\int_0^{\tau} d\tau_1 \exp[\tau_1(E_{mp}- i\Om_{1n})] \nonumber\\
&= -\frac{1}{(i\Om_{1n} + E_{pm})(i\Om_{2n} + E_{np})}
- \frac{e^{\be E_{pn}} -1}{\be (i\Om_{2n} + E_{np})^2(i\Om_{12n} + E_{nm})}
+  \frac{e^{\be E_{pm}} -1}{\be (i\Om_{1n} + E_{pm})^2(i\Om_{12n} + E_{nm})},
\label{eq:A10c} \\
I_4 &= \frac{1}{\be} \int_0^{\be} d\tau_1 \exp[\tau_1(E_{pn}- i\Om_{1n})]
\int_0^{\tau_1} d\tau \exp[\tau(i\Om_{12n} + E_{nm})]
\int_0^{\tau} d\tau_2 \exp[\tau_2(E_{mp} - i\Om_{2n})] \nonumber\\
&= -\frac{1}{(i\Om_{1n} + E_{np})(i\Om_{2n} + E_{pm})}
- \frac{e^{\be E_{pn}} -1}{\be (i\Om_{1n} + E_{np})^2(i\Om_{12n} + E_{nm})}
+  \frac{e^{\be E_{pm}} -1}{\be (i\Om_{2n} + E_{pm})^2(i\Om_{12n} + E_{nm})},
\label{eq:A10d} \\
I_5 &= \frac{1}{\be} \int_0^{\be} d\tau \exp[\tau(i\Om_{12n} + E_{nm})]
\int_0^{\tau} d\tau_2 \exp[\tau_2(E_{mp} - i\Om_{2n})]
\int_0^{\tau_2} d\tau_1 \exp[\tau_1(E_{pn}- i\Om_{1n})] \nonumber\\
&= \frac{1}{(i\Om_{1n} + E_{np})(i\Om_{12n} + E_{nm})}
- \frac{e^{\be E_{np}} -1}{\be (i\Om_{1n} + E_{np})^2(i\Om_{2n} + E_{pm})}
+  \frac{e^{\be E_{nm}} -1}{\be (i\Om_{2n} + E_{pm})(i\Om_{12n} + E_{nm})^2},
\label{eq:A10e} \\
I_6 &= \frac{1}{\be} \int_0^{\be} d\tau_2 \exp[\tau_2(E_{mp} - i\Om_{2n})]
\int_0^{\tau_2} d\tau_1 \exp[\tau_1(E_{pn}- i\Om_{1n})]
\int_0^{\tau_1} d\tau \exp[\tau(i\Om_{12n} + E_{nm})] \nonumber\\
&= \frac{1}{(i\Om_{2n} + E_{pm})(i\Om_{12n} + E_{nm})}
- \frac{e^{\be E_{mn}} -1}{\be (i\Om_{1n} + E_{np})(i\Om_{12n} + E_{nm})^2}
+  \frac{e^{\be E_{mp}} -1}{\be (i\Om_{2n} + E_{pm})^2(i\Om_{1n} + E_{np})},
\label{eq:A10f}
\end{align}
\end{subequations}
\end{widetext}
with $i\Om_{12n} \equiv i\Om_{1n} + i\Om_{2n}$. Note, in each of the above integrals only the
first term is
useful to reconstruct the response function, while the remaining two terms are spurious. However, in
Eq.~(\ref{eq:A9}) all the spurious terms cancel and we get
\begin{align}
\label{eq:A11}
&C_{\al \be, \ga, \dl}^{(3p,a)}(i \Om_{1n}, i \Om_{2n}) = - \frac{1}{Z(i\Om_{12n} + E_{nm})}
\nonumber\\
&\times \left[ \left(X_1 \right)_{nmp}  \left\{
\frac{e^{-\be E_p} - e^{-\be E_m}}{i\Om_{1n} + E_{pm}} + \frac{e^{-\be E_p} - e^{-\be E_n}}
{i\Om_{2n} + E_{np}} \right\} \right. \nonumber\\
&+ \left(X_2 \right)_{nmp} \left. \left\{
\frac{e^{-\be E_p} - e^{-\be E_m}}{i\Om_{2n} + E_{pm}} + \frac{e^{-\be E_p} - e^{-\be E_n}}
{i\Om_{1n} + E_{np}} \right\}  \right].
\end{align}
Thus, comparing Eqs.~(\ref{eq:A7}) and (\ref{eq:A11}) we have
\begin{align}
&C_{\al \be, \ga, \dl}^{(3p,a)}(i \Om_{1n} \rightarrow \Om_1 + i\eta,
i \Om_{2n} \rightarrow \Om_2 + i\eta) \nonumber\\
&= R_{\al \be, \ga, \dl}^{(3p,a)}(\Om_{1}, \Om_{2})
+ R_{\al \be, \dl, \ga}^{(3p,a)}(\Om_{2}, \Om_{1}), \nonumber
\end{align}
which is Eq.~(\ref{eq:2.18a}) in Section~\ref{sec2}.

The algebra involving the three-point functions $(3p,b)$ is entirely analogous. For brevity we define the
matrix elements
$\left(Y_1 \right)_{nmp} \equiv (\hv_{\al})_{nm} (\hv_{\be})_{mp} (\hv_{\ga \dl})_{pn}$ and
$\left(Y_2 \right)_{nmp} \equiv (\hv_{\al})_{nm} (\hv_{\ga \dl})_{mp} (\hv_{\be})_{pn}$, without implying
summation over indices $(n, m, p)$.
In terms of these $R_{\al, \be, \ga \dl}^{(3p,b)}(t, t_1, t_2)$, defined
in Eq.~(\ref{eq:2.10f}) can be written as
\begin{align}
\label{eq:A12}
&R_{\al, \be, \ga \dl}^{(3p,b)}(u_1, u_2) = \frac{1}{2Z} \left[ \theta(u_1) \theta(u_2)
\left\{ \left(Y_1 \right)_{nmp} e^{i E_{pm} u_1}
\right. \right. \nonumber\\
&\times \left. e^{i E_{np} u_2}
+ \left(Y_2 \right)_{nmp} e^{i E_{np} u_1} e^{i E_{pm} u_2}  \right\} e^{-\be E_p}
\nonumber\\
&- \theta(u_1) \theta(u_2 - u_1) \left\{
\left(Y_1 \right)_{nmp} e^{i E_{pm} u_1} e^{i E_{np} u_2} e^{-\be E_n}
\right. \nonumber\\
&+ \left.  \left(Y_2 \right)_{nmp} e^{i E_{np} u_1} e^{i E_{pm} u_2} e^{-\be E_m}
\right\}  \nonumber\\
&- \theta(u_2) \theta(u_1 - u_2) \left\{
\left(Y_1 \right)_{nmp} e^{i E_{pm} u_1} e^{i E_{np} u_2} e^{-\be E_m}
\right. \nonumber\\
&+ \left.  \left. \left(Y_2 \right)_{nmp} e^{i E_{np} u_1} e^{i E_{pm} u_2} e^{-\be E_n}
\right\} \right].
\end{align}
It is simple to check that its Fourier transform $R_{\al, \be, \ga \dl}^{(3p,b)}(\Om_1, \Om_2)$,
given by Eq.~(\ref{eq:2.13e}), is the same as Eq.~(\ref{eq:A7}) except for a factor 1/2 and with
$(X_1, X_2) \rightarrow (Y_1, Y_2)$. In other words,
\begin{align}
\label{eq:A13}
R_{\al, \be, \ga \dl}^{(3p,b)}(\Om_1, \Om_2) &= \frac{1}{2} \left[
R_{\al \be, \ga, \dl}^{(3p,a)}(\Om_1, \Om_2) \right. \nonumber\\
&+ \left. R_{\al \be, \dl, \ga}^{(3p,a)}(\Om_2, \Om_1)
\right]_{(X_1, X_2) \rightarrow (Y_1, Y_2)}.
\end{align}

Next, comparing the definitions of the correlation functions $(3p,a)$ and $(3p,b)$ it is obvious that
\beq
\label{eq:A14}
C_{\al, \be, \ga \dl}^{(3p,b)}(\tau, \tau_1,\tau_2) =
C_{\al \be, \ga, \dl}^{(3p,a)}(\tau, \tau_1,\tau_2)_{(X_1, X_2) \rightarrow (Y_1, Y_2)},
\eeq
and
\beq
\label{eq:A15}
C_{\al, \be, \ga \dl}^{(3p,b)}(i \Om_{1n}, i \Om_{2n})
= C_{\al \be, \ga, \dl}^{(3p,a)}(i \Om_{1n}, i \Om_{2n})_{(X_1, X_2) \rightarrow (Y_1, Y_2)}.
\eeq
Thus,
\begin{align}
&C_{\al, \be, \ga \dl}^{(3p,b)}(i \Om_{1n} \rightarrow \Om_1 + i\eta,
i \Om_{2n}  \rightarrow \Om_2 + i\eta) \nonumber\\
&= 2 R_{\al, \be, \ga \dl}^{(3p,b)}(\Om_{1}, i \Om_{2}), \nonumber
\end{align}
which is Eq.~(\ref{eq:2.18b}) in Section~\ref{sec2}.

\subsection{Four-point functions}
Here we compute the four-point response and correlation functions in the Lehmann representation.
We define $u_i \equiv t - t_i$, $i = (1, 2, 3)$, and for brevity the matrix elements
\begin{align}
\left(W_1 \right)_{nmpl} &\equiv (\hv_{\al})_{nm}(\hv_{\be})_{mp}(\hv_{\ga})_{pl}(\hv_{\dl})_{ln},
\nonumber\\
\left(W_2 \right)_{nmpl} &\equiv (\hv_{\al})_{nm}(\hv_{\be})_{mp}(\hv_{\dl})_{pl}(\hv_{\ga})_{ln},
\nonumber\\
\left(W_3 \right)_{nmpl} &\equiv (\hv_{\al})_{nm}(\hv_{\ga})_{mp}(\hv_{\be})_{pl}(\hv_{\dl})_{ln},
\nonumber\\
\left(W_4 \right)_{nmpl} &\equiv (\hv_{\al})_{nm}(\hv_{\ga})_{mp}(\hv_{\dl})_{pl}(\hv_{\be})_{ln},
\nonumber\\
\left(W_5 \right)_{nmpl} &\equiv (\hv_{\al})_{nm}(\hv_{\dl})_{mp}(\hv_{\ga})_{pl}(\hv_{\be})_{ln},
\nonumber\\
\left(W_6 \right)_{nmpl} &\equiv (\hv_{\al})_{nm}(\hv_{\dl})_{mp}(\hv_{\be})_{pl}(\hv_{\ga})_{ln}.
\nonumber
\end{align}
In terms of these $R_{\al, \be, \ga, \dl}^{(4p)}(t, t_1, t_2, t_3)$, defined in Eq.~(\ref{eq:2.10g}), is given by
\begin{align}
\label{eq:A16}
&ZR_{\al, \be, \ga, \dl}^{(4p)}(u_1, u_2, u_3) = \theta(u_1) \theta(u_2 - u_1) \theta(u_3 - u_2)
\nonumber\\
&\times \left\{ \left(W_1 \right)_{nmpl} e^{i u_1 E_{pm} + i u_2 E_{lp} + i u_3 E_{nl}} e^{-\be E_n}
\right. \nonumber\\
&- \left. \left(W_5 \right)_{nmpl} e^{i u_1 E_{nl} + i u_2 E_{lp} + i u_3 E_{pm}} e^{-\be E_m} \right\}
\nonumber\\
&-  \theta(u_1) \theta(u_2) \theta(u_3 - u_2) \left\{ \left(W_5 \right)_{nmpl}
e^{i u_1 E_{nl} + i u_2 E_{pm} + i u_3 E_{lp}}  \right. \nonumber\\
&\times \left. e^{-\be E_l}
+ \left(W_2 \right)_{nmpl} e^{i u_1 E_{pm} + i u_2 E_{nl} + i u_3 E_{lp}} e^{-\be E_p} \right\}.
\end{align}
Then, the response functions in the frequency space are given by
\begin{widetext}
\begin{align}
ZR_{\al, \be, \ga, \dl}^{(4p)}(\Om_1, \Om_2, \Om_3) &=
\frac{\left(W_1 \right)_{nmpl} \exp[-\be E_n]}{(\tilde{\Om}_{123}
+ E_{nm})(\tilde{\Om}_{23} + E_{np})(\tilde{\Om}_3 + E_{nl})}
+
\frac{\left(W_2 \right)_{nmpl} \exp[-\be E_p]}{(\tilde{\Om}_{1}
+ E_{pm})(\tilde{\Om}_{23} + E_{np})(\tilde{\Om}_3 + E_{lp})} \nonumber\\
&-
\frac{\left(W_5 \right)_{nmpl} \exp[-\be E_m]}{(\tilde{\Om}_{123}
+ E_{nm})(\tilde{\Om}_{23} + E_{lm})(\tilde{\Om}_3 + E_{pm})}
-
\frac{\left(W_4 \right)_{nmpl} \exp[-\be E_l]}{(\tilde{\Om}_{1}
+ E_{nl})(\tilde{\Om}_{23} + E_{lm})(\tilde{\Om}_3 + E_{lp})} \label{eq:A17a}
\end{align}
\begin{align}
ZR_{\al, \be, \dl, \ga}^{(4p)}(\Om_1, \Om_3, \Om_2) &=
\frac{\left(W_2 \right)_{nmpl} \exp[-\be E_n]}{(\tilde{\Om}_{123}
+ E_{nm})(\tilde{\Om}_{23} + E_{np})(\tilde{\Om}_2 + E_{nl})}
+
\frac{\left(W_1 \right)_{nmpl} \exp[-\be E_p]}{(\tilde{\Om}_{1}
+ E_{pm})(\tilde{\Om}_{23} + E_{np})(\tilde{\Om}_2 + E_{lp})} \nonumber\\
&-
\frac{\left(W_4 \right)_{nmpl} \exp[-\be E_m]}{(\tilde{\Om}_{123}
+ E_{nm})(\tilde{\Om}_{23} + E_{lm})(\tilde{\Om}_2 + E_{pm})}
-
\frac{\left(W_5 \right)_{nmpl} \exp[-\be E_l]}{(\tilde{\Om}_{1}
+ E_{nl})(\tilde{\Om}_{23} + E_{lm})(\tilde{\Om}_2 + E_{lp})} \label{eq:A17b}
\end{align}
\begin{align}
ZR_{\al, \ga, \be, \dl}^{(4p)}(\Om_2, \Om_1, \Om_3) &=
\frac{\left(W_3 \right)_{nmpl} \exp[-\be E_n]}{(\tilde{\Om}_{123}
+ E_{nm})(\tilde{\Om}_{13} + E_{np})(\tilde{\Om}_3 + E_{nl})}
+
\frac{\left(W_4 \right)_{nmpl} \exp[-\be E_p]}{(\tilde{\Om}_{2}
+ E_{pm})(\tilde{\Om}_{13} + E_{np})(\tilde{\Om}_3 + E_{lp})} \nonumber\\
&-
\frac{\left(W_6 \right)_{nmpl} \exp[-\be E_m]}{(\tilde{\Om}_{123}
+ E_{nm})(\tilde{\Om}_{13} + E_{lm})(\tilde{\Om}_3 + E_{pm})}
-
\frac{\left(W_2 \right)_{nmpl} \exp[-\be E_l]}{(\tilde{\Om}_{2}
+ E_{nl})(\tilde{\Om}_{13} + E_{lm})(\tilde{\Om}_3 + E_{lp})} \label{eq:A17c}
\end{align}
\begin{align}
ZR_{\al, \ga, \dl, \be}^{(4p)}(\Om_2, \Om_3, \Om_1) &=
\frac{\left(W_4 \right)_{nmpl} \exp[-\be E_n]}{(\tilde{\Om}_{123}
+ E_{nm})(\tilde{\Om}_{13} + E_{np})(\tilde{\Om}_1 + E_{nl})}
+
\frac{\left(W_3 \right)_{nmpl} \exp[-\be E_p]}{(\tilde{\Om}_{2}
+ E_{pm})(\tilde{\Om}_{13} + E_{np})(\tilde{\Om}_1 + E_{lp})} \nonumber\\
&-
\frac{\left(W_2 \right)_{nmpl} \exp[-\be E_m]}{(\tilde{\Om}_{123}
+ E_{nm})(\tilde{\Om}_{13} + E_{lm})(\tilde{\Om}_1 + E_{pm})}
-
\frac{\left(W_6 \right)_{nmpl} \exp[-\be E_l]}{(\tilde{\Om}_{2}
+ E_{nl})(\tilde{\Om}_{13} + E_{lm})(\tilde{\Om}_1 + E_{lp})} \label{eq:A17d}
\end{align}
\begin{align}
ZR_{\al, \dl, \ga, \be}^{(4p)}(\Om_3, \Om_2, \Om_1) &=
\frac{\left(W_5 \right)_{nmpl} \exp[-\be E_n]}{(\tilde{\Om}_{123}
+ E_{nm})(\tilde{\Om}_{12} + E_{np})(\tilde{\Om}_1 + E_{nl})}
+
\frac{\left(W_6 \right)_{nmpl} \exp[-\be E_p]}{(\tilde{\Om}_{3}
+ E_{pm})(\tilde{\Om}_{12} + E_{np})(\tilde{\Om}_1 + E_{lp})} \nonumber\\
&-
\frac{\left(W_1 \right)_{nmpl} \exp[-\be E_m]}{(\tilde{\Om}_{123}
+ E_{nm})(\tilde{\Om}_{12} + E_{lm})(\tilde{\Om}_1 + E_{pm})}
-
\frac{\left(W_3 \right)_{nmpl} \exp[-\be E_l]}{(\tilde{\Om}_{3}
+ E_{nl})(\tilde{\Om}_{12} + E_{lm})(\tilde{\Om}_1 + E_{lp})} \label{eq:A17e}
\end{align}
\begin{align}
ZR_{\al, \dl, \be, \ga}^{(4p)}(\Om_3, \Om_1, \Om_2) &=
\frac{\left(W_6 \right)_{nmpl} \exp[-\be E_n]}{(\tilde{\Om}_{123}
+ E_{nm})(\tilde{\Om}_{12} + E_{np})(\tilde{\Om}_2 + E_{nl})}
+
\frac{\left(W_5 \right)_{nmpl} \exp[-\be E_p]}{(\tilde{\Om}_{3}
+ E_{pm})(\tilde{\Om}_{12} + E_{np})(\tilde{\Om}_2 + E_{lp})} \nonumber\\
&-
\frac{\left(W_3 \right)_{nmpl} \exp[-\be E_m]}{(\tilde{\Om}_{123}
+ E_{nm})(\tilde{\Om}_{12} + E_{lm})(\tilde{\Om}_2 + E_{pm})}
-
\frac{\left(W_1 \right)_{nmpl} \exp[-\be E_l]}{(\tilde{\Om}_{3}
+ E_{nl})(\tilde{\Om}_{12} + E_{lm})(\tilde{\Om}_2 + E_{lp})} \label{eq:A17f},
\end{align}
\end{widetext}
where $\tilde{\Om}_1 \equiv \Om_1 + i\eta$, etc, $\tilde{\Om}_{12} \equiv \Om_1 + \Om_2 + 2i\eta$, etc,
and $\tilde{\Om}_{123} \equiv \Om_1 + \Om_2 + \Om_3 + 3i\eta$. In the above Eq.~(\ref{eq:A17a}) is the
Fourier transform of Eq.~(\ref{eq:A16}), as defined in Eq.~(\ref{eq:2.13f}).
Eq.~(\ref{eq:A17b}) is obtained from Eq.~(\ref{eq:A17a})
by exchanging $(\ga, \Om_2) \leftrightarrow (\dl, \Om_3)$. Similarly,
Eq.~(\ref{eq:A17c}) is obtained from Eq.~(\ref{eq:A17a})
by exchanging $(\be, \Om_1) \leftrightarrow (\ga, \Om_2)$,
Eq.~(\ref{eq:A17d}) is obtained from Eq.~(\ref{eq:A17b})
by exchanging $(\be, \Om_1) \leftrightarrow (\ga, \Om_2)$,
Eq.~(\ref{eq:A17e}) is obtained from Eq.~(\ref{eq:A17d})
by exchanging $(\ga, \Om_2) \leftrightarrow (\dl, \Om_3)$, and
Eq.~(\ref{eq:A17f}) is obtained from Eq.~(\ref{eq:A17e})
by exchanging $(\be, \Om_1) \leftrightarrow (\ga, \Om_2)$.

Next we evaluate the imaginary time ordered correlation function
$C_{\al, \be, \ga, \dl}^{(4p)}(\tau, \tau_1,\tau_2,\tau_3)$ defined by Eq.~(\ref{eq:2.15f}). There are
twenty-four terms which are as follows.
\begin{widetext}
\begin{align}
\label{eq:A17}
&-Z C_{\al, \be, \ga, \dl}^{(4p)}(\tau, \tau_1,\tau_2,\tau_3) =
% 01 term
 \theta_{01} \theta_{12} \theta_{23} W_1
\exp[\tau E_{nm} + \tau_1 E_{mp} + \tau_2 E_{pl} + \tau_3 E_{ln}] e^{-\be E_n} \nonumber\\
% 02 term
&+ \theta_{10} \theta_{02} \theta_{23} W_4
\exp[\tau E_{nm} + \tau_2 E_{mp} + \tau_3 E_{pl} + \tau_1 E_{ln}] e^{-\be E_l}
% 03 term
+ \theta_{12} \theta_{20} \theta_{03} W_6
\exp[\tau E_{nm} + \tau_3 E_{mp} + \tau_1 E_{pl} + \tau_2 E_{ln}] e^{-\be E_p} \nonumber\\
% 04 term
&+ \theta_{12} \theta_{23} \theta_{30} W_1
\exp[\tau E_{nm} + \tau_1 E_{mp} + \tau_2 E_{pl} + \tau_3 E_{ln}] e^{-\be E_m}
% 05 term
+ \theta_{01} \theta_{13} \theta_{32} W_2
\exp[\tau E_{nm} + \tau_1 E_{mp} + \tau_3 E_{pl} + \tau_2 E_{ln}] e^{-\be E_n} \nonumber\\
% 06 term
&+ \theta_{10} \theta_{03} \theta_{32} W_5
\exp[\tau E_{nm} + \tau_3 E_{mp} + \tau_2 E_{pl} + \tau_1 E_{ln}] e^{-\be E_l}
% 07 term
+ \theta_{13} \theta_{30} \theta_{02} W_3
\exp[\tau E_{nm} + \tau_2 E_{mp} + \tau_1 E_{pl} + \tau_3 E_{ln}] e^{-\be E_p} \nonumber\\
% 08 term
&+ \theta_{13} \theta_{32} \theta_{20} W_2
\exp[\tau E_{nm} + \tau_1 E_{mp} + \tau_3 E_{pl} + \tau_2 E_{ln}] e^{-\be E_m}
% 09 term
+ \theta_{02} \theta_{21} \theta_{13} W_3
\exp[\tau E_{nm} + \tau_2 E_{mp} + \tau_1 E_{pl} + \tau_3 E_{ln}] e^{-\be E_n} \nonumber\\
% 10 term
&+ \theta_{20} \theta_{01} \theta_{13} W_2
\exp[\tau E_{nm} + \tau_1 E_{mp} + \tau_3 E_{pl} + \tau_2 E_{ln}] e^{-\be E_l}
% 11 term
+ \theta_{21} \theta_{10} \theta_{03} W_5
\exp[\tau E_{nm} + \tau_3 E_{mp} + \tau_2 E_{pl} + \tau_1 E_{ln}] e^{-\be E_p} \nonumber\\
% 12 term
&+ \theta_{21} \theta_{13} \theta_{30} W_3
\exp[\tau E_{nm} + \tau_2 E_{mp} + \tau_1 E_{pl} + \tau_3 E_{ln}] e^{-\be E_m}
% 13 term
+ \theta_{02} \theta_{23} \theta_{31} W_4
\exp[\tau E_{nm} + \tau_2 E_{mp} + \tau_3 E_{pl} + \tau_1 E_{ln}] e^{-\be E_n} \nonumber\\
% 14 term
&+ \theta_{20} \theta_{03} \theta_{31} W_6
\exp[\tau E_{nm} + \tau_3 E_{mp} + \tau_1 E_{pl} + \tau_2 E_{ln}] e^{-\be E_l}
% 15 term
+ \theta_{23} \theta_{30} \theta_{01} W_1
\exp[\tau E_{nm} + \tau_1 E_{mp} + \tau_2 E_{pl} + \tau_3 E_{ln}] e^{-\be E_p} \nonumber\\
% 16 term
&+ \theta_{23} \theta_{31} \theta_{10} W_4
\exp[\tau E_{nm} + \tau_2 E_{mp} + \tau_3 E_{pl} + \tau_1 E_{ln}] e^{-\be E_m}
% 17 term
+ \theta_{03} \theta_{31} \theta_{12} W_6
\exp[\tau E_{nm} + \tau_3 E_{mp} + \tau_1 E_{pl} + \tau_2 E_{ln}] e^{-\be E_n} \nonumber\\
% 18 term
&+ \theta_{30} \theta_{01} \theta_{12} W_1
\exp[\tau E_{nm} + \tau_1 E_{mp} + \tau_2 E_{pl} + \tau_3 E_{ln}] e^{-\be E_l}
% 19 term
+ \theta_{31} \theta_{10} \theta_{02} W_4
\exp[\tau E_{nm} + \tau_2 E_{mp} + \tau_3 E_{pl} + \tau_1 E_{ln}] e^{-\be E_p} \nonumber\\
% 20 term
&+ \theta_{31} \theta_{12} \theta_{20} W_6
\exp[\tau E_{nm} + \tau_3 E_{mp} + \tau_1 E_{pl} + \tau_2 E_{ln}] e^{-\be E_m}
% 21 term
+ \theta_{03} \theta_{32} \theta_{21} W_5
\exp[\tau E_{nm} + \tau_3 E_{mp} + \tau_2 E_{pl} + \tau_1 E_{ln}] e^{-\be E_n} \nonumber\\
% 22 term
&+ \theta_{30} \theta_{02} \theta_{21} W_3
\exp[\tau E_{nm} + \tau_2 E_{mp} + \tau_1 E_{pl} + \tau_3 E_{ln}] e^{-\be E_l}
% 23 term
+ \theta_{32} \theta_{20} \theta_{01} W_2
\exp[\tau E_{nm} + \tau_1 E_{mp} + \tau_3 E_{pl} + \tau_2 E_{ln}] e^{-\be E_p} \nonumber\\
% 20 term
&+ \theta_{32} \theta_{21} \theta_{10} W_5
\exp[\tau E_{nm} + \tau_3 E_{mp} + \tau_2 E_{pl} + \tau_1 E_{ln}] e^{-\be E_m}.
\end{align}
\end{widetext}
In the above $\theta_{01} \equiv \theta(\tau - \tau_1)$, $\theta_{10} \equiv \theta(\tau_1 - \tau)$,
$\theta_{12} \equiv \theta(\tau_1 - \tau_2)$, etc. We also suppressed the indices of the matrix
elements $W_i$, $i = 1, \cdots, 6$. The Fourier transform to Matsubara space
$ C_{\al, \be, \ga, \dl}^{(4p)}(i \Om_{1n}, i\Om_{2n}, i\Om_{3n})$, defined in Eq.~(\ref{eq:2.16f}),
can be written as
\begin{align}
\label{eq:A19}
&-Z C_{\al, \be, \ga, \dl}^{(4p)}(i \Om_{1n}, i\Om_{2n}, i\Om_{3n})
= \left(W_1 \right)_{nmpl} \left[ I_7 e^{-\be E_n} \right. \nonumber\\
&+ \left. I_8 e^{-\be E_m} + I_9 e^{-\be E_l} + I_{10} e^{-\be E_p} \right]
+ \cdots,
\end{align}
where the ellipsis include similar four terms involving each of the matrix elements
$W_2, \cdots, W_6$ (a total of twenty-four terms). The integrals are given by
\begin{widetext}
\begin{subequations}
\label{eq:A20}
\begin{align}
I_7 &= \frac{1}{\be} \int_0^{\be} d\tau \exp[\tau(i\Om_{123n} + E_{nm})]
\int_0^{\tau} d\tau_1 \exp[\tau_1(E_{mp} - i\Om_{1n})]
\int_0^{\tau_1} d\tau_2 \exp[\tau_2(E_{pl} - i\Om_{2n})] \nonumber\\
&\times \int_0^{\tau_2} d\tau_3 \exp[\tau_3(E_{ln} - i\Om_{3n})]
\nonumber\\
&=
%1st
\frac{1}{(E_{ln} - i\Om_{3n})(E_{pn} - i\Om_{23n})(E_{mn} - i\Om_{123n})}
%2nd
+ \frac{e^{\be E_{nl}} -1}{\be (E_{ln} - i\Om_{3n})^2(E_{pl} - i\Om_{2n})(E_{ml} - i\Om_{12n})}
\nonumber\\
%3rd
&- \frac{e^{\be E_{np}} -1}{\be (E_{pl} - i\Om_{2n})(E_{pn} - i\Om_{23n})^2(E_{mp} - i\Om_{1n})}
%4th
- \frac{e^{\be E_{nm}} -1}{\be (E_{mn} - i\Om_{123n})}
%4.1
\left\{ \frac{1}{(E_{ln} - i\Om_{3n})(E_{pl} - i\Om_{2n})(E_{ml} - i\Om_{12n})} \right.
\nonumber\\
&- \left.
%4.2
\frac{1}{(E_{ln} - i\Om_{3n})(E_{pn} - i\Om_{23n})(E_{mn} - i\Om_{123n})}
%4.3
- \frac{1}{(E_{pl} - i\Om_{2n})(E_{pn} - i\Om_{23n})(E_{mp} - i\Om_{1n})} \right\},
\label{eq:A20a}\\
I_8 &= \frac{1}{\be} \int_0^{\be} d\tau_1 \exp[\tau_1(E_{mp} - i\Om_{1n})]
\int_0^{\tau_1} d\tau_2 \exp[\tau_2(E_{pl} - i\Om_{2n})]
\int_0^{\tau_2} d\tau_3 \exp[\tau_3(E_{ln} - i\Om_{3n})] \nonumber\\
&\times \int_0^{\tau_3} d\tau \exp[\tau(i\Om_{123n} + E_{nm})]
\nonumber\\
&=
%1st
\frac{-1}{(E_{mp} - i\Om_{1n})(E_{ml} - i\Om_{12n})(E_{mn} - i\Om_{123n})}
%2nd
+ \frac{e^{\be E_{mn}} -1}{\be (E_{ln} - i\Om_{3n})(E_{pn} - i\Om_{23n})(E_{mn} - i\Om_{123n})^2}
\nonumber\\
%3rd
&- \frac{e^{\be E_{ml}} -1}{\be (E_{ln} - i\Om_{3n})(E_{ml} - i\Om_{12n})^2(E_{pl} - i\Om_{2n})}
%4th
+ \frac{e^{\be E_{mp}} -1}{\be (E_{mp} - i\Om_{1n})}
%4.1
\left\{ \frac{1}{(E_{mn} - i\Om_{123n})(E_{ml} - i\Om_{12n})(E_{mp} - i\Om_{1n})} \right.
\nonumber\\
&- \left.
%4.2
\frac{1}{(E_{mn} - i\Om_{123n})(E_{ln} - i\Om_{3n})(E_{pn} - i\Om_{23n})}
%4.3
+ \frac{1}{(E_{ln} - i\Om_{3n})(E_{ml} - i\Om_{12n})(E_{pl} - i\Om_{2n})} \right\},
\label{eq:A20b}\\
I_9 &= \frac{1}{\be} \int_0^{\be} d\tau_3 \exp[\tau_3(E_{ln} - i\Om_{3n})]
\int_0^{\tau_3} d\tau \exp[\tau(i\Om_{123n} + E_{nm})]
\int_0^{\tau} d\tau_1 \exp[\tau_1(E_{mp} - i\Om_{1n})] \nonumber\\
&\times \int_0^{\tau_1} d\tau_2 \exp[\tau_2(E_{pl} - i\Om_{2n})]
\nonumber\\
&=
%1st
\frac{-1}{(E_{pl} - i\Om_{2n})(E_{ml} - i\Om_{12n})(E_{ln} - i\Om_{3n})}
%2nd
- \frac{e^{\be E_{lp}} -1}{\be (E_{pl} - i\Om_{2n})^2(E_{mp} - i\Om_{1n})(E_{pn} - i\Om_{23n})}
\nonumber\\
%3rd
&+ \frac{e^{\be E_{lm}} -1}{\be (E_{ml} - i\Om_{12n})^2(E_{mp} - i\Om_{1n})(E_{mn} - i\Om_{123n})}
%4th
+ \frac{e^{\be E_{ln}} -1}{\be (E_{ln} - i\Om_{3n})}
%4.1
\left\{ \frac{1}{(E_{pl} - i\Om_{2n})(E_{ml} - i\Om_{12n})(E_{ln} - i\Om_{3n})} \right.
\nonumber\\
&- \left.
%4.2
\frac{1}{(E_{pl} - i\Om_{2n})(E_{mp} - i\Om_{1n})(E_{pn} - i\Om_{23n})}
%4.3
+ \frac{1}{(E_{ml} - i\Om_{12n})(E_{mp} - i\Om_{1n})(E_{mn} - i\Om_{123n})} \right\},
\label{eq:A20c}\\
I_{10} &= \frac{1}{\be} \int_0^{\be} d\tau_2 \exp[\tau_2(E_{pl} - i\Om_{2n})]
\int_0^{\tau_2} d\tau_3 \exp[\tau_3(E_{ln} - i\Om_{3n})]
\int_0^{\tau_3} d\tau \exp[\tau(i\Om_{123n} + E_{nm})] \nonumber\\
&\times \int_0^{\tau} d\tau_1 \exp[\tau_1(E_{mp} - i\Om_{1n})]
\nonumber\\
&=
%1st
\frac{1}{(E_{mp} - i\Om_{1n})(E_{pn} - i\Om_{23n})(E_{pl} - i\Om_{2n})}
%2nd
+ \frac{e^{\be E_{pm}} -1}{\be (E_{mp} - i\Om_{1n})^2(E_{mn} - i\Om_{123n})(E_{ml} - i\Om_{12n})}
\nonumber\\
%3rd
&+ \frac{e^{\be E_{pn}} -1}{\be (E_{pn} - i\Om_{23n})^2(E_{ln} - i\Om_{3n})(E_{mn} - i\Om_{123n})}
%4th
+ \frac{e^{\be E_{pl}} -1}{\be (E_{pl} - i\Om_{2n})}
%4.1
\left\{ \frac{1}{(E_{mp} - i\Om_{1n})(E_{mn} - i\Om_{123n})(E_{ml} - i\Om_{12n})} \right.
\nonumber\\
&- \left.
%4.2
\frac{1}{(E_{mp} - i\Om_{1n})(E_{pn} - i\Om_{23n})(E_{pl} - i\Om_{2n})}
%4.3
- \frac{1}{(E_{pn} - i\Om_{23n})(E_{ln} - i\Om_{3n})(E_{mn} - i\Om_{123n})} \right\},
\label{eq:A20d}
\end{align}
\end{subequations}
\end{widetext}
where $i\Om_{123n} \equiv i\Om_{1n} + i\Om_{2n} + i\Om_{3n}$, etc. In each of the above integrals only
the first term is useful to reconstruct the response function, while the remaining terms are spurious. But, as
before, all the spurious terms cancel after the summation in Eq.~(\ref{eq:A19}). It is simple to check that the
terms proportional to $\left(W_2 \right)_{nmpl}$ can be obtained from those proportional to
$\left(W_1 \right)_{nmpl} $ by exchanging $i\Om_{2n} \leftrightarrow i\Om_{3n}$. Likewise, the terms
proportional to $\left(W_3 \right)_{nmpl}$ can be obtained from those proportional to $\left(W_1 \right)_{nmpl} $
by exchanging  $i\Om_{1n} \leftrightarrow i\Om_{2n}$, and so on for $W_4, W_5$ and $W_6$.
Collecting all the terms we get
\begin{widetext}
\begin{align}
\label{eq:A21}
&Z C_{\al, \be, \ga, \dl}^{(4p)}(i \Om_{1n}, i\Om_{2n}, i\Om_{3n})
=  \nonumber\\
%1
&\left(W_1 \right)_{nmpl} \left[
% 1.1
\frac{e^{-\be E_n}}{(i\Om_{123n} + E_{nm})(i\Om_{23n} + E_{np})(i\Om_{3n} + E_{nl})}
%1.2
-  \frac{e^{-\be E_m}}{(i\Om_{123n} + E_{nm})(i\Om_{12n} + E_{lm})(i\Om_{1n} + E_{pm})}
\right. \nonumber\\
%1.3
&- \left. \frac{e^{-\be E_l}}{(i\Om_{3n} + E_{nl})(i\Om_{2n} + E_{lp})(i\Om_{12n} + E_{lm})}
%1.4
+  \frac{e^{-\be E_p}}{(i\Om_{1n} + E_{pm})(i\Om_{2n} + E_{lp})(i\Om_{23n} + E_{np})}
\right] \nonumber\\
%2
&+ \left(W_2 \right)_{nmpl} \left[
% 2.1
\frac{e^{-\be E_n}}{(i\Om_{123n} + E_{nm})(i\Om_{23n} + E_{np})(i\Om_{2n} + E_{nl})}
%2.2
-  \frac{e^{-\be E_m}}{(i\Om_{123n} + E_{nm})(i\Om_{13n} + E_{lm})(i\Om_{1n} + E_{pm})}
\right. \nonumber\\
%2.3
&- \left. \frac{e^{-\be E_l}}{(i\Om_{2n} + E_{nl})(i\Om_{3n} + E_{lp})(i\Om_{13n} + E_{lm})}
%2.4
+  \frac{e^{-\be E_p}}{(i\Om_{1n} + E_{pm})(i\Om_{3n} + E_{lp})(i\Om_{23n} + E_{np})}
\right] \nonumber\\
%3
&+ \left(W_3 \right)_{nmpl} \left[
% 3.1
\frac{e^{-\be E_n}}{(i\Om_{123n} + E_{nm})(i\Om_{13n} + E_{np})(i\Om_{3n} + E_{nl})}
%3.2
-  \frac{e^{-\be E_m}}{(i\Om_{123n} + E_{nm})(i\Om_{12n} + E_{lm})(i\Om_{2n} + E_{pm})}
\right. \nonumber\\
%3.3
&- \left. \frac{e^{-\be E_l}}{(i\Om_{3n} + E_{nl})(i\Om_{1n} + E_{lp})(i\Om_{12n} + E_{lm})}
%3.4
+  \frac{e^{-\be E_p}}{(i\Om_{2n} + E_{pm})(i\Om_{1n} + E_{lp})(i\Om_{13n} + E_{np})}
\right] \nonumber\\
%4
&+ \left(W_4 \right)_{nmpl} \left[
% 4.1
\frac{e^{-\be E_n}}{(i\Om_{123n} + E_{nm})(i\Om_{13n} + E_{np})(i\Om_{1n} + E_{nl})}
%4.2
-  \frac{e^{-\be E_m}}{(i\Om_{123n} + E_{nm})(i\Om_{23n} + E_{lm})(i\Om_{2n} + E_{pm})}
\right. \nonumber\\
%4.3
&- \left. \frac{e^{-\be E_l}}{(i\Om_{1n} + E_{nl})(i\Om_{3n} + E_{lp})(i\Om_{23n} + E_{lm})}
%4.4
+  \frac{e^{-\be E_p}}{(i\Om_{2n} + E_{pm})(i\Om_{3n} + E_{lp})(i\Om_{13n} + E_{np})}
\right] \nonumber\\
%5
&\left(W_5 \right)_{nmpl} \left[
% 5.1
\frac{e^{-\be E_n}}{(i\Om_{123n} + E_{nm})(i\Om_{12n} + E_{np})(i\Om_{1n} + E_{nl})}
%5.2
-  \frac{e^{-\be E_m}}{(i\Om_{123n} + E_{nm})(i\Om_{23n} + E_{lm})(i\Om_{3n} + E_{pm})}
\right. \nonumber\\
%5.3
&- \left. \frac{e^{-\be E_l}}{(i\Om_{1n} + E_{nl})(i\Om_{2n} + E_{lp})(i\Om_{23n} + E_{lm})}
%5.4
+  \frac{e^{-\be E_p}}{(i\Om_{3n} + E_{pm})(i\Om_{2n} + E_{lp})(i\Om_{12n} + E_{np})}
\right] \nonumber\\
%6
&\left(W_6 \right)_{nmpl} \left[
% 6.1
\frac{e^{-\be E_n}}{(i\Om_{123n} + E_{nm})(i\Om_{12n} + E_{np})(i\Om_{2n} + E_{nl})}
%6.2
-  \frac{e^{-\be E_m}}{(i\Om_{123n} + E_{nm})(i\Om_{13n} + E_{lm})(i\Om_{3n} + E_{pm})}
\right. \nonumber\\
%6.3
&- \left. \frac{e^{-\be E_l}}{(i\Om_{2n} + E_{nl})(i\Om_{1n} + E_{lp})(i\Om_{13n} + E_{lm})}
%6.4
+  \frac{e^{-\be E_p}}{(i\Om_{3n} + E_{pm})(i\Om_{1n} + E_{lp})(i\Om_{12n} + E_{np})}
\right].
\end{align}
Thus, comparing Eqs.~(\ref{eq:A17}) and (\ref{eq:A21}) we conclude that
\begin{align}
&C_{\al, \be, \ga, \dl}^{(4p)}(i \Om_{1n} \rightarrow \Om_1 + i\eta,
i\Om_{2n} \rightarrow \Om_2 + i\eta, i \Om_{3n} \rightarrow \Om_3 + i\eta) =
R_{\al, \be, \ga, \dl}^{(4p)}(\Om_1, \Om_2, \Om_3)
+ R_{\al, \be, \dl, \ga}^{(4p)}(\Om_1, \Om_3, \Om_2)  \nonumber\\
&+ R_{\al, \ga, \be, \dl}^{(4p)}(\Om_2, \Om_1, \Om_3)
+ R_{\al, \ga, \dl, \be}^{(4p)}(\Om_2, \Om_3, \Om_1)  + R_{\al, \dl, \be, \ga}^{(4p)}(\Om_3, \Om_1, \Om_2)
+ R_{\al, \dl, \ga, \be}^{(4p)}(\Om_3, \Om_2, \Om_1), \nonumber
\end{align}
%\end{widetext}
which is Eq.(\ref{eq:2.19}) in the main text.

\section{}
\label{appB}

In this appendix we provide details of Section~\ref{sec3}. The way to compute the coefficients
$Q_i(\om_1, \om_2, \om_3)_{nmpl}$, $i= 1, \cdots, 6$ is already described in the main text.
Here we simply give the final expressions.
%\begin{widetext}
\begin{align}
\label{eq:B1}
&Q_1(\om_1, \om_2, \om_3)_{nmpl} =
%line1
E_{mp} E_{pl} E_{ln} \left[ \frac{1}{\omp_{123} + E_{nm}}
\left\{ \frac{e^{-\be E_n}}{(\omp_{23} + E_{np})(\omp_3 + E_{nl})}
- \frac{e^{-\be E_m}}{(\omp_{12} + E_{lm})(\omp_1 + E_{pm})} \right\}
\right. \nonumber\\
%line2
&\left. - \frac{1}{\omp_{2} + E_{lp}}
\left\{ \frac{e^{-\be E_l}}{(\omp_{12} + E_{lm})(\omp_3 + E_{nl})}
- \frac{e^{-\be E_p}}{(\omp_{23} + E_{np})(\omp_1 + E_{pm})} \right\} \right]
%line3
- \frac{E_{pl} E_{ln}}{\omp_{23} + E_{np}} \left[
\frac{e^{-\be E_l} - e^{-\be E_p}}{\omp_2 + E_{lp}}
+ \frac{e^{-\be E_l} - e^{-\be E_n}}{\omp_3 + E_{nl}} \right]
\nonumber\\
%line4
&+ \frac{E_{mp} E_{pl}}{\omp_{12} + E_{lm}} \left[
\frac{e^{-\be E_p} - e^{-\be E_m}}{\omp_1 + E_{pm}}
+ \frac{e^{-\be E_p} - e^{-\be E_l}}{\omp_2 + E_{lp}} \right]
%line5
- \frac{E_{mp} E_{pl}}{\omp_{123} + E_{nm}} \left[
\frac{e^{-\be E_p} - e^{-\be E_m}}{\omp_1 + E_{pm}}
+ \frac{e^{-\be E_p} - e^{-\be E_n}}{\omp_{23} + E_{np}} \right]
\nonumber\\
%line6
&- \frac{E_{mp} E_{ln}}{\omp_{123} + E_{nm}} \left[
\frac{e^{-\be E_l} - e^{-\be E_m}}{\omp_{12} + E_{lm}}
+ \frac{e^{-\be E_l} - e^{-\be E_n}}{\omp_{3} + E_{nl}} \right]
%line7
+ E_{pl} \left[ \frac{e^{-\be E_n} - e^{-\be E_p}}{\omp_{23} + E_{np}} \right]
- E_{mp} \left[ \frac{e^{-\be E_l} - e^{-\be E_m}}{\omp_{12} + E_{lm}} \right]
\nonumber\\
& + E_{ln} \left[ \frac{e^{-\be E_n} - e^{-\be E_l}}{\omp_{3} + E_{nl}} \right]
- E_{pl} \left[ \frac{e^{-\be E_l} - e^{-\be E_p}}{\omp_{2} + E_{lp}} \right]
+ E_{mp} \left[ \frac{e^{-\be E_n} - e^{-\be E_m}}{\omp_{123} + E_{nm}} \right]
%line8
+ e^{-\be E_n} - e^{-\be E_l},
\end{align}
\begin{align}
\label{eq:B2}
&Q_2(\om_1, \om_2, \om_3)_{nmpl} =
%line1
E_{mp} E_{pl} E_{ln} \left[ \frac{1}{\omp_{123} + E_{nm}}
\left\{ \frac{e^{-\be E_n}}{(\omp_{23} + E_{np})(\omp_2 + E_{nl})}
- \frac{e^{-\be E_m}}{(\omp_{13} + E_{lm})(\omp_1 + E_{pm})} \right\}
\right. \nonumber\\
%line2
&\left. - \frac{1}{\omp_{3} + E_{lp}}
\left\{ \frac{e^{-\be E_l}}{(\omp_{13} + E_{lm})(\omp_2 + E_{nl})}
- \frac{e^{-\be E_p}}{(\omp_{23} + E_{np})(\omp_1 + E_{pm})} \right\} \right]
%line3
- \frac{E_{pl} E_{ln}}{\omp_{23} + E_{np}} \left[
\frac{e^{-\be E_l} - e^{-\be E_p}}{\omp_3 + E_{lp}}
+ \frac{e^{-\be E_l} - e^{-\be E_n}}{\omp_2 + E_{nl}} \right]
\nonumber\\
%line4
&+ \frac{E_{mp} E_{pl}}{\omp_{13} + E_{lm}} \left[
\frac{e^{-\be E_p} - e^{-\be E_m}}{\omp_1 + E_{pm}}
+ \frac{e^{-\be E_p} - e^{-\be E_l}}{\omp_3 + E_{lp}} \right]
%line5
+ \frac{E_{mp} E_{ln}}{\omp_{123} + E_{nm}} \left[
\frac{e^{-\be E_p} - e^{-\be E_m}}{\omp_1 + E_{pm}}
+ \frac{e^{-\be E_p} - e^{-\be E_n}}{\omp_{23} + E_{np}} \right]
\nonumber\\
%line6
&- \frac{E_{mp} E_{ln}}{\omp_{123} + E_{nm}} \left[
\frac{e^{-\be E_l} - e^{-\be E_m}}{\omp_{13} + E_{lm}}
+ \frac{e^{-\be E_l} - e^{-\be E_n}}{\omp_{2} + E_{nl}} \right]
%line7
- E_{ln} \left[ \frac{e^{-\be E_n} - e^{-\be E_p}}{\omp_{23} + E_{np}} \right]
- E_{mp} \left[ \frac{e^{-\be E_l} - e^{-\be E_m}}{\omp_{13} + E_{lm}} \right]
\nonumber\\
& - E_{pl} \left[ \frac{e^{-\be E_l} - e^{-\be E_p}}{\omp_{3} + E_{lp}} \right]
%line8
+ E_{ln} \left[ \frac{e^{-\be E_n} - e^{-\be E_l}}{\omp_{2} + E_{nl}} \right]
+ E_{mp} \left[ \frac{e^{-\be E_p} - e^{-\be E_m}}{\omp_{1} + E_{pm}} \right]
+ e^{-\be E_p} - e^{-\be E_l},
\end{align}
\begin{align}
\label{eq:B3}
&Q_3(\om_1, \om_2, \om_3)_{nmpl} =
%line1
E_{mp} E_{pl} E_{ln} \left[ \frac{1}{\omp_{123} + E_{nm}}
\left\{ \frac{e^{-\be E_n}}{(\omp_{13} + E_{np})(\omp_3 + E_{nl})}
- \frac{e^{-\be E_m}}{(\omp_{12} + E_{lm})(\omp_2 + E_{pm})} \right\}
\right. \nonumber\\
%line2
&\left. - \frac{1}{\omp_{1} + E_{lp}}
\left\{ \frac{e^{-\be E_l}}{(\omp_{12} + E_{lm})(\omp_3 + E_{nl})}
- \frac{e^{-\be E_p}}{(\omp_{13} + E_{np})(\omp_2 + E_{pm})} \right\} \right]
%line3
- \frac{E_{pl} E_{ln}}{\omp_{13} + E_{np}} \left[
\frac{e^{-\be E_l} - e^{-\be E_p}}{\omp_1 + E_{lp}}
+ \frac{e^{-\be E_l} - e^{-\be E_n}}{\omp_3 + E_{nl}} \right]
\nonumber\\
%line4
&+ \frac{E_{mp} E_{pl}}{\omp_{12} + E_{lm}} \left[
\frac{e^{-\be E_p} - e^{-\be E_m}}{\omp_2 + E_{pm}}
+ \frac{e^{-\be E_p} - e^{-\be E_l}}{\omp_1 + E_{lp}} \right]
%line5
- \frac{E_{mp} E_{pl}}{\omp_{123} + E_{nm}} \left[
\frac{e^{-\be E_p} - e^{-\be E_m}}{\omp_2 + E_{pm}}
+ \frac{e^{-\be E_p} - e^{-\be E_n}}{\omp_{13} + E_{np}} \right]
\nonumber\\
%line6
&+ \frac{E_{pl} E_{ln}}{\omp_{123} + E_{nm}} \left[
\frac{e^{-\be E_l} - e^{-\be E_m}}{\omp_{12} + E_{lm}}
+ \frac{e^{-\be E_l} - e^{-\be E_n}}{\omp_{3} + E_{nl}} \right]
%line7
+ E_{pl} \left[ \frac{e^{-\be E_n} - e^{-\be E_p}}{\omp_{13} + E_{np}} \right]
+ E_{pl} \left[ \frac{e^{-\be E_l} - e^{-\be E_m}}{\omp_{12} + E_{lm}} \right]
\nonumber\\
& - E_{pl} \left[ \frac{e^{-\be E_l} - e^{-\be E_p}}{\omp_{1} + E_{lp}} \right]
%line8
- E_{pl} \left[ \frac{e^{-\be E_n} - e^{-\be E_m}}{\omp_{123} + E_{nm}} \right],
\end{align}
\begin{align}
\label{eq:B4}
&Q_4(\om_1, \om_2, \om_3)_{nmpl} =
%line1
E_{mp} E_{pl} E_{ln} \left[ \frac{1}{\omp_{123} + E_{nm}}
\left\{ \frac{e^{-\be E_n}}{(\omp_{13} + E_{np})(\omp_1 + E_{nl})}
- \frac{e^{-\be E_m}}{(\omp_{23} + E_{lm})(\omp_2 + E_{pm})} \right\}
\right. \nonumber\\
%line2
&\left. - \frac{1}{\omp_{3} + E_{lp}}
\left\{ \frac{e^{-\be E_l}}{(\omp_{23} + E_{lm})(\omp_1 + E_{nl})}
- \frac{e^{-\be E_p}}{(\omp_{13} + E_{np})(\omp_2 + E_{pm})} \right\} \right]
%line3
+ \frac{E_{mp} E_{pl}}{\omp_{23} + E_{lm}} \left[
\frac{e^{-\be E_p} - e^{-\be E_m}}{\omp_2 + E_{pm}}
+ \frac{e^{-\be E_p} - e^{-\be E_l}}{\omp_3 + E_{lp}} \right]
\nonumber\\
%line4
&- \frac{E_{pl} E_{ln}}{\omp_{13} + E_{np}} \left[
\frac{e^{-\be E_l} - e^{-\be E_p}}{\omp_3 + E_{lp}}
+ \frac{e^{-\be E_l} - e^{-\be E_n}}{\omp_1 + E_{nl}} \right]
%line5
- \frac{E_{mp} E_{ln}}{\omp_{123} + E_{nm}} \left[
\frac{e^{-\be E_l} - e^{-\be E_m}}{\omp_{23} + E_{lm}}
+ \frac{e^{-\be E_l} - e^{-\be E_n}}{\omp_{1} + E_{nl}} \right]
\nonumber\\
%line6
&+ \frac{E_{mp} E_{ln}}{\omp_{123} + E_{nm}} \left[
\frac{e^{-\be E_p} - e^{-\be E_m}}{\omp_2 + E_{pm}}
+ \frac{e^{-\be E_p} - e^{-\be E_n}}{\omp_{13} + E_{np}} \right]
%line7
- E_{ln} \left[ \frac{e^{-\be E_n} - e^{-\be E_p}}{\omp_{13} + E_{np}} \right]
- E_{mp} \left[ \frac{e^{-\be E_l} - e^{-\be E_m}}{\omp_{23} + E_{lm}} \right]
\nonumber\\
& - E_{pl} \left[ \frac{e^{-\be E_l} - e^{-\be E_p}}{\omp_{3} + E_{lp}} \right]
%line8
+ E_{ln} \left[ \frac{e^{-\be E_n} - e^{-\be E_l}}{\omp_{1} + E_{nl}} \right]
+ E_{mp} \left[ \frac{e^{-\be E_p} - e^{-\be E_m}}{\omp_{2} + E_{pm}} \right]
+ e^{-\be E_p} - e^{-\be E_l},
\end{align}
\begin{align}
\label{eq:B5}
&Q_5(\om_1, \om_2, \om_3)_{nmpl} =
%line1
E_{mp} E_{pl} E_{ln} \left[ \frac{1}{\omp_{123} + E_{nm}}
\left\{ \frac{e^{-\be E_n}}{(\omp_{12} + E_{np})(\omp_1 + E_{nl})}
- \frac{e^{-\be E_m}}{(\omp_{23} + E_{lm})(\omp_3 + E_{pm})} \right\}
\right. \nonumber\\
%line2
&\left. - \frac{1}{\omp_{2} + E_{lp}}
\left\{ \frac{e^{-\be E_l}}{(\omp_{23} + E_{lm})(\omp_1 + E_{nl})}
- \frac{e^{-\be E_p}}{(\omp_{12} + E_{np})(\omp_3 + E_{pm})} \right\} \right]
%line3
+ \frac{E_{mp} E_{pl}}{\omp_{23} + E_{lm}} \left[
\frac{e^{-\be E_p} - e^{-\be E_m}}{\omp_3 + E_{pm}}
+ \frac{e^{-\be E_p} - e^{-\be E_l}}{\omp_2 + E_{lp}} \right]
\nonumber\\
%line4
&- \frac{E_{pl} E_{ln}}{\omp_{12} + E_{np}} \left[
\frac{e^{-\be E_l} - e^{-\be E_p}}{\omp_2 + E_{lp}}
+ \frac{e^{-\be E_l} - e^{-\be E_n}}{\omp_1 + E_{nl}} \right]
%line5
+ \frac{E_{pl} E_{ln}}{\omp_{123} + E_{nm}} \left[
\frac{e^{-\be E_l} - e^{-\be E_m}}{\omp_{23} + E_{lm}}
+ \frac{e^{-\be E_l} - e^{-\be E_n}}{\omp_{1} + E_{nl}} \right]
\nonumber\\
%line6
&+ \frac{E_{mp} E_{ln}}{\omp_{123} + E_{nm}} \left[
\frac{e^{-\be E_p} - e^{-\be E_m}}{\omp_3 + E_{pm}}
+ \frac{e^{-\be E_p} - e^{-\be E_n}}{\omp_{12} + E_{np}} \right]
%line7
- E_{ln} \left[ \frac{e^{-\be E_n} - e^{-\be E_p}}{\omp_{12} + E_{np}} \right]
+ E_{pl} \left[ \frac{e^{-\be E_l} - e^{-\be E_m}}{\omp_{23} + E_{lm}} \right]
\nonumber\\
& + E_{mp} \left[ \frac{e^{-\be E_p} - e^{-\be E_m}}{\omp_{3} + E_{pm}} \right]
%line8
- E_{pl} \left[ \frac{e^{-\be E_l} - e^{-\be E_p}}{\omp_{2} + E_{lp}} \right]
+ E_{ln} \left[ \frac{e^{-\be E_n} - e^{-\be E_m}}{\omp_{123} + E_{nm}} \right]
+ e^{-\be E_p} - e^{-\be E_m},
\end{align}
\begin{align}
\label{eq:B6}
&Q_6(\om_1, \om_2, \om_3)_{nmpl} =
%line1
E_{mp} E_{pl} E_{ln} \left[ \frac{1}{\omp_{123} + E_{nm}}
\left\{ \frac{e^{-\be E_n}}{(\omp_{12} + E_{np})(\omp_2 + E_{nl})}
- \frac{e^{-\be E_m}}{(\omp_{13} + E_{lm})(\omp_3 + E_{pm})} \right\}
\right. \nonumber\\
%line2
&\left. - \frac{1}{\omp_{1} + E_{lp}}
\left\{ \frac{e^{-\be E_l}}{(\omp_{13} + E_{lm})(\omp_2 + E_{nl})}
- \frac{e^{-\be E_p}}{(\omp_{12} + E_{np})(\omp_3 + E_{pm})} \right\} \right]
%line3
+ \frac{E_{mp} E_{pl}}{\omp_{13} + E_{lm}} \left[
\frac{e^{-\be E_p} - e^{-\be E_m}}{\omp_3 + E_{pm}}
+ \frac{e^{-\be E_p} - e^{-\be E_l}}{\omp_1 + E_{lp}} \right]
\nonumber\\
%line4
&- \frac{E_{pl} E_{ln}}{\omp_{12} + E_{np}} \left[
\frac{e^{-\be E_l} - e^{-\be E_p}}{\omp_1 + E_{lp}}
+ \frac{e^{-\be E_l} - e^{-\be E_n}}{\omp_2 + E_{nl}} \right]
%line5
+ \frac{E_{pl} E_{ln}}{\omp_{123} + E_{nm}} \left[
\frac{e^{-\be E_l} - e^{-\be E_m}}{\omp_{13} + E_{lm}}
+ \frac{e^{-\be E_l} - e^{-\be E_n}}{\omp_{2} + E_{nl}} \right]
\nonumber\\
%line6
&- \frac{E_{mp} E_{pl}}{\omp_{123} + E_{nm}} \left[
\frac{e^{-\be E_p} - e^{-\be E_m}}{\omp_3 + E_{pm}}
+ \frac{e^{-\be E_p} - e^{-\be E_n}}{\omp_{12} + E_{np}} \right]
%line7
+ E_{pl} \left[ \frac{e^{-\be E_n} - e^{-\be E_p}}{\omp_{12} + E_{np}} \right]
+ E_{pl} \left[ \frac{e^{-\be E_l} - e^{-\be E_m}}{\omp_{13} + E_{lm}} \right]
\nonumber\\
& - E_{pl} \left[ \frac{e^{-\be E_l} - e^{-\be E_p}}{\omp_{1} + E_{lp}} \right]
%line8
- E_{pl} \left[ \frac{e^{-\be E_n} - e^{-\be E_m}}{\omp_{123} + E_{nm}} \right].
\end{align}
\end{widetext}

\section{}
\label{appC}

In this appendix we provide details of the results obtained in Section~\ref{sec4}.

\subsection{Drude metal: three- and four-point correlators without vertex correction}

We first consider a correlation function of the type $(3p,a)$ such as
$C_{\al \be, \ga, \dl}^{(3p,a)}(i\om_{2n}, i\om_{3n})$ that enters in the computation of the
nonlinear electro-optical susceptibility $\Pi_{\al \be \ga \dl}^{(3)}(\om_1, \om_2, \om_3)$. From the
definition of the $(3p,a)$ correlation function given in Eq.~(\ref{eq:2.15d}), after factorization in
terms of single particle Green's functions using Wick's theorem, and in imaginary frequency space
we get
\begin{align}
& C_{\al \be, \ga, \dl}^{(3p,a)}(i\om_{2n}, i\om_{3n})
= \frac{1}{\be \vol} \sum_{\bk, \nu_n}
(v_{\bk})_{\al \be} (v_{\bk})_{\ga} (v_{\bk})_{\dl}  G_{\bk}(i\nu_n)
\nonumber \\
& G_{\bk}(i\nu_n + i\om_{2n} + i\om_{3n}) \!
\left[G_{\bk}(i\nu_n + i\om_{2n}) \! + \! G_{\bk}(i\nu_n + i\om_{3n}) \right].
\end{align}
Thus, $C_{\al \be, \ga, \dl}^{(3p,a)}(i\om_{2n}, i\om_{3n})$  is the sum of two diagrams, the
first of which is shown in Fig.~\ref{fig1} (v), while the second diagram is obtained by exchanging the
photon lines $(\om_{2n}, \ga)$ and $(\om_{3n}, \dl)$ in the first. For constant density of states we
get
\begin{align}
& C_{\al \be, \ga, \dl}^{(3p,a)}(i\om_{2n}, i\om_{3n})
= \nu_0 \la (v_{\bk})_{\al \be} (v_{\bk})_{\ga} (v_{\bk})_{\dl} \ra_{FS}
\nonumber \\
& \times \left[ \mathcal{C}^{(3p,a)}(i\om_{2n}, i\om_{3n}) +  \mathcal{C}^{(3p,a)}(i\om_{3n}, i\om_{2n})
\right],
\end{align}
where
\begin{align}
& \mathcal{C}^{(3p,a)}(i\om_{2n}, i\om_{3n}) \equiv \frac{1}{\be} \sum_{\nu_n}
\int d \ep_{\bk} G_{\bk}(i\nu_n) G_{\bk}(i\nu_n + i\om_{2n})
\nonumber \\
&\times G_{\bk}(i\nu_n + i\om_{2n} + i\om_{3n}).
\end{align}
In the above the $\ep_{\bk}$ integral can be performed by contour integration.
Since $(\om_{2n}, \om_{3n}) > 0$, the $\nu_n$ summation has non-zero contributions from
two intervals. First, for $\nu_n \in [-\om_{2n}, 0]$ the pole of the Green's function associated with the
frequency $\nu_n$ is in the lower half plane, while those associated with the frequencies
$(\nu_n + \om_{2n})$ and $(\nu_n + \om_{2n} + \om_{3n})$ are in the upper half plane.
Second, in the interval $\nu_n \in [-\om_{2n} -\om_{3n} , -\om_{2n}]$ the pole associated with
$(\nu_n + \om_{2n} + \om_{3n})$ is in the upper half plane while those of the remaining two frequencies
are in the lower half plane. Evaluating these two contributions we get
\begin{align}
\mathcal{C}^{(3p,a)}(i\om_{2n}, i\om_{3n}) &= \frac{1}{(i\om_{2n} + i\om_{3n} + i/\tau)}
\nonumber \\
&\times
\left[ \frac{i \om_{2n}}{i\om_{2n} + i/\tau} - \frac{i \om_{3n}}{i\om_{3n} + i/\tau} \right].
\end{align}
Since the above is odd under exchange of frequencies $\om_{2n}$ and $\om_{3n}$, we get
$C_{\al \be, \ga, \dl}^{(3p,a)}(i\om_{2n}, i\om_{3n})  = 0$. Using an analogous argument one
can show that correlation functions of the type $(3p,b)$ also vanish.

Next, we consider the four-point correlation function
$C_{\al, \be, \ga, \dl}^{(4p)}(i\om_{1n}, i\om_{2n}, i\om_{3n})$. It can be expressed as a
sum of six terms
\begin{align}
&C_{\al, \be, \ga, \dl}^{(4p)}(i\om_{1n}, i\om_{2n}, i\om_{3n})
= \nu_0 \la (v_{\bk})_{\al} (v_{\bk})_{\be} (v_{\bk})_{\ga} (v_{\bk})_{\dl} \ra_{FS}
\nonumber \\
&\times \left[ \mathcal{C}^{(4p)}(i\om_{1n}, i\om_{2n}, i\om_{3n})
+ \mathcal{C}^{(4p)}(i\om_{1n}, i\om_{3n}, i\om_{2n}) + \cdots \right],
\end{align}
where the ellipsis imply the remaining four terms obtained by permutation of the three external
frequencies, and
\begin{align}
& \mathcal{C}^{(4p)}(i\om_{1n}, i\om_{2n}, i\om_{3n}) \! \equiv \! \frac{1}{\be} \! \sum_{\nu_n}
\! \! \int \! \! d \ep_{\bk} G_{\bk}(i\nu_n) G_{\bk}(i\nu_n + i\om_{1n})
\nonumber \\
&\times G_{\bk}(i\nu_n + i\om_{1n} + i\om_{2n}) G_{\bk}(i\nu_n + i\om_{1n} + i\om_{2n} + i\om_{3n}).
\end{align}
The six terms can be represented diagrammatically, of which the first is shown in Fig.~\ref{fig1} (vii).
The remaining five diagrams are obtained by permuting the incoming photon lines. The energy integral
and the frequency summation can be performed as before, and we get
\begin{align}
&\mathcal{C}^{(4p)}(i\om_{1n}, i\om_{2n}, i\om_{3n})
= \frac{1}{(i\om_{1n} + i\om_{2n} + i\om_{3n} + i/\tau)}
\nonumber \\
&\times \left[
\frac{i\om_{1n}}{(i\om_{1n} + i/\tau)(i \om_{12n} + i/\tau)}
+ \frac{i\om_{3n}}{(i\om_{3n} + i/\tau)(i \om_{23n} + i/\tau)}
\right. \nonumber \\
& \left. - \frac{i\om_{2n}}{(i\om_{2n} + i/\tau)(i \om_{23n} + i/\tau)}
- \frac{i\om_{2n}}{(i\om_{2n} + i/\tau)(i \om_{12n} + i/\tau)} \right],
\end{align}
where $\om_{12n} \equiv \om_{1n} + \om_{2n}$, and so on.
From the cyclic property of the above expression it follows that
\begin{align}
\mathcal{C}^{(4p)}(i\om_{1n}, i\om_{2n}, i\om_{3n})
+ \mathcal{C}^{(4p)}(i\om_{1n}, i\om_{3n}, i\om_{2n}) + \cdots = 0,
\end{align}
which implies that $C_{\al, \be, \ga, \dl}^{(4p)}(i\om_{1n}, i\om_{2n}, i\om_{3n}) = 0$.

\subsection{Drude metal: vertex corrections}

Here we discuss the contributions to the kernel $\Pi_{\al \be \ga \dl}^{(3)}(\om_1, \om_2, \om_3)$
that involve impurity scattering induced vertex corrections. Our goal is to demonstrate that for a Drude
system, with a constant density of states, all such vertex terms vanish.
%====================
\begin{figure}[!!t]
\begin{center}
\includegraphics[width=8.7cm,trim=0 0 0 0]{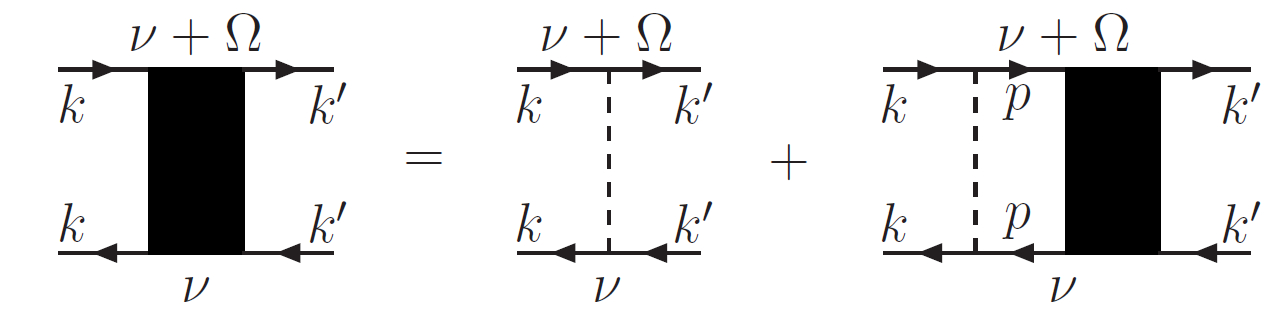}
\caption{
$P(\nu, \nu + \Om)$ is the ladder summation of repeated impurity scattering, see Eq.~(\ref{eq:C9}).
Solid lines are electron Green's functions and dashed lines imply impurity scattering of particle-hole
excitations.
}
\label{fig4}
\end{center}
\end{figure}
%====================

The treatment of vertex corrections due to weak disorder
in the diagrammatic language can be found in standard literature~\cite{Altshuler}.
One of the basic building blocks is the quantity $P(i\nu_n, i\nu_n + i\Om_n)$ shown in Fig.~\ref{fig4}
which describes one or more impurity scattering of particle-hole excitations. We get
\beq
\label{eq:C9}
 P(i\nu_n, i\nu_n + i\Om_n) = \frac{1/(2\pi \nu_0 \tau)}{1 - \Lambda(i\nu_n, i\nu_n + i\Om_n)},
 \eeq
 where
 \beq
 \Lambda(i\nu_n, i\nu_n + i\Om_n) = \frac{1}{2\pi \nu_0 \tau \vol} \sum_{\bp}
 G_{\bp}(i\nu_n)  G_{\bp}(i\nu_n + i\Om_n).
 \eeq
 The above momentum sum can be performed as a contour integral of the energy variable $\ep_{\bp}$,
 and we get
 \begin{align}
P(i\nu_n, i\nu_n + i\Om_n) &= \frac{i\Om_n + i/\tau}{2\pi i \nu_0 \tau \Om_n},
\quad \nu_n \in [ - \Om_n, 0]
\nonumber \\
&= 1/(2\pi \nu_0 \tau), \quad {\rm otherwise}.
\end{align}

We note that current operators, or combinations of them, that are odd under inversion symmetry
do not admit vertex corrections, because such terms vanish after momentum average. Thus, non-zero
vertex corrections necessarily involve momentum averages of operators such as
$(v_{\bk})_{\al} (v_{\bk})_{\be}$ or $(v_{\bk})_{\al \be}$. After momentum average such terms lead
to $\dl_{\al \be}$. This in turn implies that all the vertex terms can be grouped into three distinct
(and gauge invariant) combinations, namely those that are proportional to $\dl_{\al \be} \dl_{\ga \dl}$,
to $\dl_{\al \ga} \dl_{\be \dl}$, and to $\dl_{\al \dl} \dl_{\be \ga}$. In the following we evaluate
only the first category of vertex terms. The remaining two categories can be deduced by simply exchanging
appropriate external photon indices.
%====================
\begin{figure}[!!t]
\begin{center}
\includegraphics[width=8.7cm,trim=0 0 0 0]{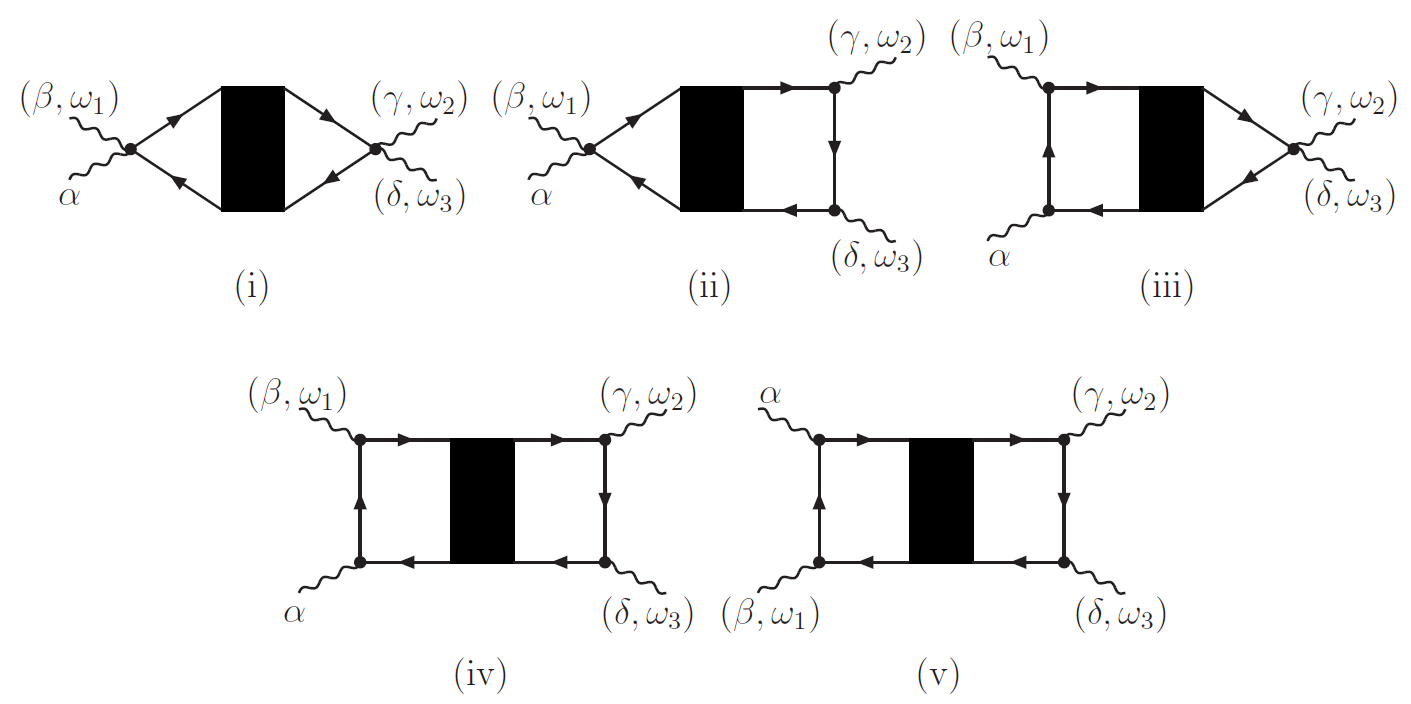}
\caption{
Impurity induced vertex correction diagrams for the computation of the kernel
$\Pi_{\al \be \ga \dl}^{(3)}(\om_1, \om_2, \om_3)$. Diagrams
(ii), (iv) and (v) have a second contribution obtained by exchanging the photon lines $(\ga, \om_2)$
and $(\dl, \om_3)$. Diagram (iii) has a second contribution obtained by exchanging photon lines
$\al$ and $(\be, \om_1)$.
}
\label{fig5}
\end{center}
\end{figure}
%====================

In total there are nine terms/diagrams that contribute to vertex corrections that are
proportional to $\dl_{\al \be} \dl_{\ga \dl}$. These are represented in Fig.~\ref{fig5},
where each of the diagrams
(ii), (iv) and (v) have a second contribution obtained by exchanging the photon lines $(\ga, \om_2)$
and $(\dl, \om_3)$, while diagram (iii) has a second contribution obtained by exchanging photon lines
$\al$ and $(\be, \om_1)$.

Diagrams of the type (i)-(iii) in Fig.~\ref{fig5} contain at least one factor of the combination
\[
\frac{1}{\vol} \sum_{\bk} (v_{\bk})_{\al \be} G_{\bk}(i\nu_n)  G_{\bk}(i\nu_n + i\Om_n).
\]
After integration by part this can be written as
\begin{align}
&- \nu_0 \la (v_{\bk})_{\al} (v_{\bk})_{\be} \ra_{FS} \int d \ep_{\bk}
G_{\bk}(i\nu_n)  G_{\bk}(i\nu_n + i\Om_n)
\nonumber \\
&\left[ G_{\bk}(i\nu_n) + G_{\bk}(i\nu_n + i\Om_n) \right] =0.
\nonumber
\end{align}
Thus, the diagrams of the type (i)-(iii) do not contribute.

Next we consider the diagram (iv). This can be expressed as
\begin{align}
(iv) &= \frac{1}{\be} \sum_{\nu_n} P(i\nu_n + i\om_{1n}, i\nu_n + i\om_{123n})
\nonumber \\
&\times L_{\al \be}(i\nu_n, i\om_{1n}, i\om_{23n})
L_{\ga \dl}(i\nu_n + i\om_{1n}, i\om_{2n}, i\om_{3n}),
\nonumber
\end{align}
where
\begin{align}
&L_{\al \be}(i\nu_n, i\om_{1n}, i\om_{23n})
\equiv \frac{1}{\vol} \sum_{\bk} (v_{\bk})_{\al} (v_{\bk})_{\be}
G_{\bk}(i\nu_n)
\nonumber \\
&\times G_{\bk}(i\nu_n + i\om_{1n}) G_{\bk}(i\nu_n + i\om_{123n}).
\end{align}
In the above $\om_{12n} \equiv \om_{1n} + \om_{2n}$, and so on.
The $\nu_n$ summation is nonzero only over the two intervals
$\nu_n \in [-\om_{12n} , - \om_{1n}]$ and $\nu_n \in [-\om_{123n} , - \om_{12n}]$.
Evaluating the contributions from these two intervals we get
\begin{align}
(iv) &= \dl_{\al \be} \dl_{\ga \dl} \frac{\nu_0 v_F^4/(d^2 \tau)}
{\om_{23n}(i \om_{123n} + i/\tau)(i \om_{23n} + i/\tau)}
\nonumber \\
&\times \left[ \frac{i \om_{3n}}{i\om_{3n} + i/\tau} -  \frac{i \om_{2n}}{i\om_{2n} + i/\tau}
\right],
\end{align}
where $d$ is the dimension. Since the partner diagram (not shown in Fig.~\ref{fig5})
of (iv) involves exchanging the frequencies $\om_{2n} \leftrightarrow \om_{3n}$, the two cancel. For the same
reason one can show that diagram (v) and its partner diagram cancel each other. Thus, overall, all the
vertex contributions drop out.

\end{document}